\documentclass[times,final]{elsarticle}
\usepackage[letterpaper,lmargin=1in,rmargin=1in,tmargin=1in,bmargin=1in]{geometry}

\global\bibsep=0pt
\usepackage{framed,multirow}

\usepackage{amsmath}
\usepackage{amssymb}
\usepackage{latexsym}
\usepackage{amsfonts}
\usepackage{stmaryrd}
\usepackage{subcaption}
\usepackage{fancyvrb}
\usepackage{bm}

\usepackage[title]{appendix}

\usepackage{xspace}
\usepackage{xifthen}

\newcommand{\curl}[1][]{\ifthenelse{\isempty{#1}}{\operatorname{\nabla{\times}}}{\operatorname{\nabla_{#1}{\times}}}}
\newcommand{\diverge}[1][]{\ifthenelse{\isempty{#1}}{\operatorname{\nabla{\cdot}}}{\operatorname{\nabla_{#1}{\cdot}}}}

\usepackage{graphicx}

\usepackage{algorithm}
\usepackage{algpseudocode}
\usepackage{xspace}
\usepackage{xifthen}

\usepackage{hyperref}

\algnewcommand{\algorithmicgoto}{\textbf{go to}}%
\algnewcommand{\Goto}{\algorithmicgoto\xspace}%
\algnewcommand{\Label}{\State\unskip}



\usepackage{amsopn}

\DeclareMathOperator*{\real}{Re}
\DeclareMathOperator*{\imag}{Im}

\DeclareMathOperator*{\ceil}{ceil}

\DeclareMathOperator{\grad}{\nabla}
\DeclareMathOperator{\erf}{erf}

\renewcommand{\vec}[1]{\ensuremath\boldsymbol{#1}}

\usepackage{listings}
\usepackage{url}
\usepackage{xcolor}
\definecolor{newcolor}{rgb}{.8,.349,.1}

\usepackage{orcidlink}

\journal{Journal of Computational Physics}

\begin{document}

\begin{frontmatter}

\title{Accelerating high-order continuum kinetic plasma simulations using multiple GPUs}%

\author[1]{Andrew Ho \orcidlink{0000-0003-0257-7113}}
\ead{ho37@llnl.gov}
\author[1]{G. V. Vogman \orcidlink{0000-0003-2566-5298}}
\ead{vogman1@llnl.gov}

\address[1]{Lawrence Livermore National Laboratory, Livermore, California, United States of America}

\begin{abstract}
  Kinetic plasma simulations solve the Vlasov-Poisson or Vlasov-Maxwell equations to evolve scalar-variable distribution functions in position-velocity phase space and vector-variable electromagnetic fields in physical space.
  The immense computational cost of evolving high-dimensional variables, and their large number of degrees of freedom, often limits the utility of continuum kinetic simulations and presents a challenge when it comes to accurately simulating real-world physical phenomena.
  To address this challenge, we developed techniques that accelerate and minimize the computational work required for a scalable Vlasov-Poisson solver.
  We present theoretical hardware compute and communication bounds required for solving a fourth-order finite-volume Vlasov-Poisson system.
  These bounds are then used to inform and evaluate the design of performance portable algorithms for a multiple graphics processing unit (GPU) accelerated version of the Vlasov-Poisson solver VCK-CPU \cite{vogman2014}.
  We demonstrate that the multi-GPU Vlasov solver implementation VCK-GPU simultaneously minimizes required inter-process data transfer while also being bounded by the machine network performance limits, leaving minimal room for theoretical performance improvements.
  This resulted in an overall strong scaling speedup per timestep of up to 40x in three-dimensional phase space (one position, two velocity coordinates) and 54x in four dimensional phase space (two position, two velocity coordinates) and a 341x increase in simulation throughput of the GPU accelerated code over the existing CPU code.
  The GPU code is also able to weak scale up to 256 compute nodes and 1024 GPUs.
  We then demonstrate how the improved compute performance can be used to explore configurations which were previously computationally infeasible such as resolving fine-scale distribution function filamentation and multi-species dynamics with realistic electron-proton mass ratios.
\end{abstract}


\end{frontmatter}

\section{Introduction}


Kinetic effects in plasmas, which are associated with particle-particle and particle-wave interactions, are important in the study of space weather\cite{palmroth2018}, electromagnetic propulsion\cite{taccogna2004}, and fusion energy applications \cite{rinderknecht2018,saarelma2017,francisquez2023}.
Such effects arise in regions where the particle mean free paths are comparable to or larger than the scale lengths of interest.
Examples of kinetic phenomena include anomalous transport in pulsed power experiments\cite{vogman2020,vogman2021,vogman2024}, laser-plasma interactions in laser inertial confinement fusion (ICF)\cite{rinderknecht2018}, and the dynamics and stability of the tokamak pedestal\cite{saarelma2017} and high field mirrors \cite{francisquez2023}.
 
Kinetic theory describes each particle species in a plasma as a distribution function in position-velocity phase space.
Evolution of the distribution function is described by the Vlasov equation, coupled to Poisson's equation or Maxwell's equations, which describe the evolution of self-consistent electromagnetic fields. 
Since in general each distribution function occupies a six-dimensional space, kinetic simulations can have a huge number of degrees of freedom and thereby significant computational cost, which limits the dynamic range that can be modeled and restricts our understanding of microphysics.
Significant numerical stiffness associated with coupling fast electron dynamics with slow ion dynamics can also limit the dynamic range of simulations.
Consequently, beyond parallelization based on domain decomposition, which is widely used in kinetic solvers, much effort has been dedicated to developing methods for reducing the computational burden of kinetic simulations.
These methods include semi-Lagrangian methods that relax stability constraints\cite{nakamura1999}, implicit solvers that facilitate larger time steps\cite{taitano2013}, sparse grid methods\cite{ricketson2017}, gyrokinetic treatment for low-frequency dynamics in the presence of strong magnetic fields\cite{dorr2018,michels2021,francisquez2023}, dynamic low-rank approximations\cite{einkemmer2024}, and particle-in-cell methods wherein distribution functions are sampled with superparticles \cite{wang2019}.

More recently, there has been significant interest in accelerating kinetic simulations by leveraging the compute power of GPU hardware, which offers significantly higher floating-point operations per second than that achievable with CPU hardware.
The maximum sustainable arithmetic intensity for CPUs is usually on the order of 2-15, while GPUs can sustain arithmetic intensities in excess of 60.
Developing algorithms that are able to make efficient use of multiple GPUs remains a challenge, however, due to the need for memory management and refactoring to minimize communication \cite{zhao2018,einkemmer2020}.
This is because many applications are strongly memory bandwidth bound on GPUs\cite{williams2009}, which calls for algorithms that focus primarily on maximizing effective memory bandwidth utilization \cite{micikevicius2009,sai2022}.
A common technique for exploiting GPUs is to re-arrange computations and fuse multiple kernels together to simultaneously minimize memory bandwidth utilization and kernel enqueue latency \cite{wang2010}.
Algorithms also have to be able to handle the unique features of kinetic simulations, which make them different from more standard hyperbolic solvers (e.g. computational fluid dynamics): namely the high dimensionality and the recurring communication between high-dimensional phase-space variables and low-dimensional physical space variables, which are related through velocity moments.
Recent efforts to refactor gyrokinetic codes for GPUs have resulted in 15x\cite{germaschewski2021} and 19x\cite{francisquez2023} speedup and scalability on GPU-accelerated supercomputers.
Extending these results to more general kinetic solvers remains an active area of research, particularly as applications seek performance-portable codes.

When it comes to accelerating continuum kinetic solvers using GPUs, semi-Lagrangian methods have received the most attention as they provide a mechanism for relaxing time step stability requirements, reducing overall computational costs.
A key challenge with these methods is the unpredictable memory access patterns, along with a significant memory overhead which limits their performance in multi-node GPU simulations \cite{sandroos2013,mehrenberger2013}.
The complex data dependencies can be simplified by utilizing splitting methods and bounding the maximum characteristic length at the cost of introducing a timestep restriction \cite{einkemmer2014}.
Semi-Lagrangian splitting methods have achieved speedups on the order of 35x while remaining scalable to multiple GPUs \cite{einkemmer2016,einkemmer2019,einkemmer2020}.

In contrast to semi-Lagrangian methods, Eulerian methods have predictable and compact data dependencies.
This feature allows straightforward decompositions into connected partitions and efficient parallel computation \cite{leveque2004,warburton2008}.
CPU-based Eulerian methods have been successfully applied to kinetic plasma methods to model electron plasma waves in stimulated Raman backscattering\cite{banks2010,banks2011}, Kelvin-Helmholtz instabilities in nonuniform low-beta plasmas \cite{vogman2020,vogman2021}, the Dory-Guest-Harris instability\cite{vogman2014,datta2021}, and plasma sheaths\cite{datta2023}.
Eulerian methods have also been demonstrated to be scalable on GPU accelerated exascale supercomputers for computational fluid dynamics \cite{PeleSoftware}.
It is expected that the same properties which make GPU accelerated Eulerian methods efficient for fluid dynamics will apply to higher dimensional continuum kinetic methods.

This paper describes a multi-GPU algorithm for a fourth-order finite-volume discretization of the Vlasov-Poisson equation system.
The algorithm exploits the efficiently scalable nature of explicit Eulerian methods, while addressing mixed-dimensionality and high-dimensionality challenges specific to kinetic solvers.
It thereby makes it feasible to perform multi-species kinetic microturbulence studies at high resolution and with realistic electron-proton mass ratios.
Section \ref{sec:discretization} presents a refactoring of the fourth-order finite-volume discretization to make it amenable to GPU computation, and presents a less restrictive CFL condition that allows for larger time steps.
Section \ref{sec:architecture} discusses the overall structure and design of the code, including handling the intricacies specific to kinetic simulations such as high dimensionality and mixed dimensionality interactions via moment integration, an effective partitioning strategy for minimizing data communication and maximizing parallel scalability, and achieving performance portability on multiple platforms.
Section \ref{sec:vv} presents benchmarks to verify and validate the correctness of the implementation in up to four phase space dimensions.
Section \ref{sec:performance} outlines the performance analysis and scaling potential achieved with our implementation.
Finally, Section \ref{sec:conclusion} presents concluding remarks and areas of future improvements.
 
\section{Fourth-order accurate finite-volume method}

\label{sec:discretization}

By utilizing a high-order accurate numerical method the number of degrees of freedom (DOFs) required to achieve the same accuracy as lower order methods is reduced \cite{warburton2008,colella2011}.
This is particularly important for continuum kinetic methods where the number of DOFs scales as $N^{D}$ where $D$ is the number of phase space dimensions, up to six total.
A downside of high order methods is that they are more complex and require more work per DOF.
As the order of accuracy increases eventually the increased complexity and cost per DOF outstrips practical considerations such as geometric fitting and parallelizability, leading to a method which is in practice slower to run.
A fourth-order accurate finite-volume spatial discretization coupled with a fourth-order accurate Runge-Kutta temporal discretization has been demonstrated to offer a reasonable balance between achieving increased convergence rates with complexity and practical runtime performance on CPUs \cite{vogman2018}.
We use the fourth-order discretization method of VCK-CPU\cite{vogman2014} as a basis for the GPU-accelerated Vlasov solver VCK-GPU.

\subsection{Spatial discretization}

\label{sec:spatial_discr}
Consider the non-dimensional magnetostatic hyperbolic Vlasov equation
\begin{gather}
  \partial_{t} f_{s}(t, \vec{r}) + \diverge[\vec{r}] (\vec{A}_{s}(t, \vec{r}) f_{s}(t, \vec{r}))  = 0\label{eq:vlasov}\\
  \vec{A}_{s} =
  \begin{bmatrix}
    \vec{v}, &
    \frac{q_{s}}{m_{s}} ((\omega_{p0} t_{0})^{2} \vec{E}(t,\vec{x}) + (\omega_{c0} t_{0}) \vec{v} \times \hat{B}) + \vec{G}
  \end{bmatrix}^{T}\\
  \begin{aligned}
    \vec{r} &=
              \begin{bmatrix}
                \vec{x}, &
                \vec{v}
              \end{bmatrix}^{T}
  \end{aligned}
\end{gather}
where $f_{s}$ is the distribution function for particle species $s$ in $\vec{x}$-$\vec{v}$ phase space with particle mass $m_{s}$ and charge $q_{s}$, $\vec{A}_{s}$ is the advection velocity along each coordinate dimension in phase space, $\vec{E}$ is the electric field, $\vec{B}$ is a constant externally applied magnetic field, $\omega_{c0} = |q_{0} B_{0} / m_{0}|$ and $\omega_{p0} = \sqrt{q_{0}^{2} n_{0} / (\epsilon_{0} m_{0})}$ are the reference cyclotron and plasma frequencies, $q_{0}$ is the elementary charge, $m_{0}$ is the reference particle mass, $t_{0}$ is the reference timescale, and $\vec{G}$ is a gravity like acceleration.
In 3D-3V Cartesian coordinates $\vec{r} = [x, y, z, v_{x}, v_{y}, v_{z}]^{T}$.
One can also consider the Vlasov equation in a subset of the six phase space dimensions so long as the number of velocity dimensions is greater than or equal to the number of spatial dimensions.
The zeroth velocity moment of the distribution function is given by
\begin{gather}
  n_{s} = \int f_{s} d\vec{v}\label{eq:number_density}
\end{gather}
where $n_{s}$ is the number density of species $s$.
The total charge density $\rho_{c}$ is defined as
\begin{gather}
  \rho_{c} = \sum_{s} q_{s} n_{s}\label{eq:charge_density}
\end{gather}
In the electrostatic limit, the non-dimensional Poisson equation for the scalar potential $\phi$ relates the distribution functions via number density and charge density to the electric field $\vec{E} = -\grad \phi$ such that
\begin{gather}
  \diverge \grad \phi = - \rho_{c}\label{eq:poisson}
\end{gather}
For the remainder of this section we will drop the species subscript $s$ and consider each species independently.
By dividing the domain into an ordered grid of cells and integrating Eq. \eqref{eq:vlasov} over a phase space cell at the multi-index $\vec{i}$, we obtain an evolution equation for the volume-integrated distribution function in terms of the flux $\vec{A} f$
\begin{gather}
  \partial_{t} \langle f\rangle_{\vec{i}} = -\oint_{\vec{i}} \vec{A} f \cdot \hat{n} dS\label{eq:finite_volume_fluxes}
\end{gather}
where the surface integral is performed over the boundary of cell $\vec{i}$ and $\langle f\rangle_{\vec{i}}$ is the volume integrated value of $f$ in cell $\vec{i}$.
Let $\vec{i} + 1/2 \vec{e}^{d}$ be the multi-index that designates the center of a cell face whose normal points in the $d$-th direction, where $\vec{e}^{d}$ denotes the unit vector in the $d$-th direction. Let $\langle A^{d} f\rangle_{\vec{i} + 1/2 \vec{e}^{d}}$ denote the surface integral over this cell face.
The fourth-order finite-volume flux over the $\vec{i} + 1/2 \vec{e}^{d}$ face is then given by\cite{colella2011}
\begin{gather}
  \langle A^{d} f\rangle_{\vec{i}+\frac{1}{2} \vec{e}^{d}} = 
  \left(
    A^{d}_{\vec{i}+\frac{1}{2} \vec{e}^{d}} f_{\vec{i}+\frac{1}{2} \vec{e}^{d}} + \frac{1}{24} \sum_{d' \neq d}  h^{2}_{d'}
    \left.
      \frac{\partial^{2} (A^{d} f)}{\partial r^{2}_{d'}}
    \right|_{\vec{i}+\frac{1}{2} \vec{e}^{d}}
  \right) \prod_{d' \neq d} h_{d'}\label{eq:fourth_fvm}
\end{gather}
where $A^{d}_{\vec{i}+1/2 \vec{e}^{d}}$ and $f_{\vec{i}+1/2 \vec{e}^{d}}$ are the face center values of $A^{d}$ and $f$ respectively.
A five point upwind stencil is used to reconstruct the distribution function at the center of each cell face\cite{vogman2014}
\begin{gather}
  \def\arraystretch{1.5}
  f_{\vec{i}+\frac{1}{2} \vec{e}^{d}} =
  \left\lbrace
    \begin{array}{l@{}l@{}l@{}l@{}l@{}r@{\quad}l@{~}l@{}}
      &\frac{2}{60} \bar{f}_{\vec{i}-2 \vec{e}^{d}} &- \frac{13}{60} \bar{f}_{\vec{i}- \vec{e}^{d}} &+ \frac{47}{60} \bar{f}_{\vec{i}} &+ \frac{27}{60} \bar{f}_{\vec{i}+ \vec{e}^{d}} &- \frac{3}{60} \bar{f}_{\vec{i}+2 \vec{e}^{d}} &\text{if} &A^{d}_{\vec{i}} > 0\\
      -&\frac{3}{60} \bar{f}_{\vec{i} - \vec{e}^{d}} &- \frac{27}{60} \bar{f}_{\vec{i}} &+ \frac{47}{60} \bar{f}_{\vec{i}+ \vec{e}^{d}} &- \frac{13}{60} \bar{f}_{\vec{i}+2  \vec{e}^{d}} &+ \frac{2}{60} \bar{f}_{\vec{i}+3 \vec{e}^{d}} &\text{if} &A^{d}_{\vec{i}} \le 0
    \end{array}
  \right.
  \label{eq:fvm_poly}
\end{gather}
where $\bar{f}_{\vec{i}} = \langle f\rangle_{\vec{i}} / V$ is the cell average value of $f$.
Using the quadrature rule in Eq. \eqref{eq:fourth_fvm} and the distribution function reconstruction in Eq. \eqref{eq:fvm_poly} to approximate the surface integral in Eq. \eqref{eq:finite_volume_fluxes} gives an evolution equation for the cell-average distribution function in the $\vec{i}$-th cell
\begin{gather}
  \partial_{t} \bar{f}_{\vec{i}} = C_{\vec{i}} - \sum_{d} \frac{A^{d}_{\vec{i}}}{h_{d}} 
  \left(
    f_{\vec{i}+\frac{1}{2} \vec{e}^{d}} - f_{\vec{i}-\frac{1}{2} \vec{e}^{d}}
  \right)\label{eq:spatial_discr}\\
  C_{\vec{i}} = \sum_{d} \frac{1}{24 h_{d}} \sum_{d' \neq d} h^{2}_{d'} 
  \left.
    \frac{\partial^{2} (\vec{A}^{d} f)}{\partial r^{2}_{d'}}
  \right|_{\vec{i} + \frac{1}{2} \vec{e}^{d}}\label{eq:spatial_corrections}
\end{gather}
where $C_{\vec{i}}$ represents the sum of all transverse derivative terms and $h_{d}$ is the cell spacing in direction $d$.
To achieve overall fourth-order accuracy, the second derivatives in $C_{\vec{i}}$ need to be at least second-order accurate\cite{vogman2018}.
Table \ref{tbl:fvm_stencils} shows the term $C_{\vec{i}}$ for the magnetostatic Vlasov system for various combinations of phase space dimensions in Cartesian coordinates.
The coefficients $c_{1}$ through $c_{5}$ in the table are given by the following expressions:
\begin{align}
  c_{1} &=
          \frac{h_{v_{x}}}{48 h_{x}}
          + \frac{(\omega_{p0} t_{0})^{2}}{96 h_{v_{x}}} \frac{q}{m}
          \left(
          E_{x,\vec{i}+\vec{e}^{1}} - E_{x,\vec{i}-\vec{e}^{1}}
          \right)\label{eq:C_k_coeff_1}\\
  c_{2} &=
          \frac{\omega_{c0} t_{0}}{48} \frac{q}{m} B_{z}
          \left(
          \frac{h_{v_{x}}}{h_{v_{y}}} - \frac{h_{v_{y}}}{h_{v_{x}}}
          \right)\label{eq:C_k_coeff_2}\\
  c_{3} &= \frac{(\omega_{p0} t_{0})^{2}}{96 h_{v_{x}}} \frac{q}{m}
          \left(
          E_{x,\vec{i} - \vec{e}^{2}} - E_{x,\vec{i} + \vec{e}^{2}}
          \right)\label{eq:C_k_coeff_3}\\
  c_{4} &= \frac{h_{v_{y}}}{48 h_{y}} + \frac{(\omega_{p0} t_{0})^{2}}{96 h_{v_{y}}} \frac{q}{m}
          \left(
          E_{y,\vec{i} + \vec{e}^{2}} - E_{y,\vec{i} - \vec{e}^{2}}
          \right)\label{eq:C_k_coeff_4}\\
  c_{5} &= \frac{(\omega_{p0} t_{0})^{2}}{96 h_{v_{y}}} \frac{q}{m}
          \left(
          E_{y,\vec{i} - \vec{e}^{1}} - E_{y,\vec{i} + \vec{e}^{1}}
          \right)\label{eq:C_k_coeff_5}
\end{align}

\begin{figure}[H]
  \centering
  \includegraphics[width=.2\textwidth]{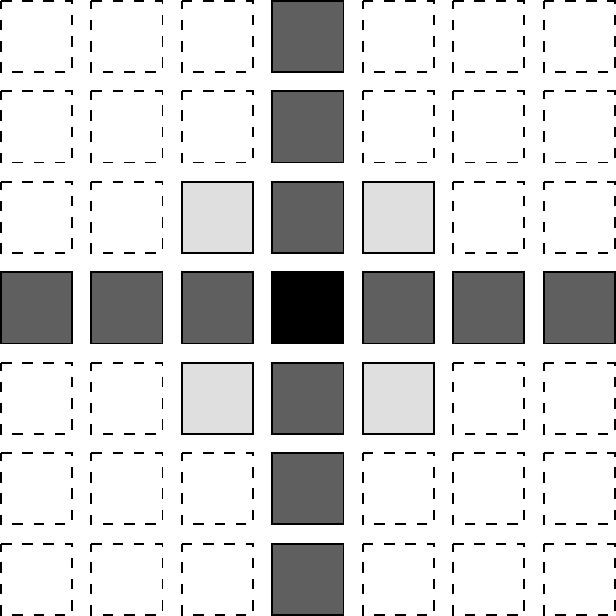}
  \caption{1D-1V fourth-order finite-volume stencil required to update the solid black cell. The black cell and a subset of the dark gray cells are used for reconstruction in Eq. \eqref{eq:fvm_poly}. Light gray cells correspond to $C_{\vec{i}}$, defined in Eq. \eqref{eq:spatial_corrections}. Dashed outline cells are not used to update the solid black cell. Dark gray and black cells are not needed for $C_{\vec{i}}$ since their contributions cancel out.}
  \label{fig:stencil_2d}
\end{figure}

Eq. \eqref{eq:spatial_discr} is amenable to efficient GPU computation because the surface integral for each cell is evaluated in one step without precalculating and storing the fluxes\cite{micikevicius2009,zhao2018,sai2022}.
This is mathematically, but not algorithmically, equivalent to the method used in VCK-CPU \cite{vogman2014}.
Note that for the Vlasov equation in Cartesian coordinates in any direction $d$ we have $A^{d}_{\vec{i}} = A^{d}_{\vec{i}+1/2 \vec{e}^{d}} = A^{d}_{\vec{i}-1/2 \vec{e}^{d}}$, allowing it to be factored out in Eq. \eqref{eq:spatial_discr}.
The high order cell center interpolation from Eq. \eqref{eq:fvm_poly} introduces a data dependency which is multiple cells wide in each direction, such that to update $\bar{f}_{\vec{i}}$ requires knowing $\bar{f}_{\vec{j}}$ such that $i_{d} \neq j_{d}$ in at most one entry and $|i_{d} - j_{d}| \le 3, \forall d$, where $i_{d}$ is the $d$-th index of the multi-index $\vec{i}$.
The term $C_{\vec{i}}$ adds edge dependencies, such that the $C_{\vec{i}}$ contribution to update $\bar{f}_{\vec{i}}$ requires knowing $\bar{f}_{\vec{j}}$ such that $|i_{d} - j_{d}| \le 1, \forall d$ when $i_{d} \neq j_{d}$ and $i_{d'} \neq j_{d'}$ for $d \neq d'$.
Figure \ref{fig:stencil_2d} shows the stencil required to update a single cell in 1D-1V phase space.

\begin{table}[H]
  \def\arraystretch{1.5}
  \centering{}
  \caption{The term $C_{\vec{i}}$ (see Eq. \eqref{eq:spatial_corrections}) in the fourth-order finite-volume discretization of the magnetostatic Vlasov equation in Cartesian coordinates. The accompanying coefficients $c_{1}$ through $c_{5}$ are given by Eqs. \eqref{eq:C_k_coeff_1}-\eqref{eq:C_k_coeff_5}.}
  \label{tbl:fvm_stencils}
  \begin{tabular}{r|c|c}
    Dimensions & $C_{\vec{i}}$ & Assumptions\\
    \hline
    $(x, v_{x})$ & \begin{tabular}{@{}l@{}}
      $c_{1}
      \left(
      \bar{f}_{\vec{i} + \vec{e}^{1} - \vec{e}^{2}} + \bar{f}_{\vec{i}-\vec{e}^{1} + \vec{e}^{2}} - \bar{f}_{\vec{i}-\vec{e}^{1}, \vec{i} - \vec{e}^{2}} - \bar{f}_{\vec{i}+\vec{e}^{1}, \vec{i}+\vec{e}^{2}}
      \right)$
    \end{tabular} & $E_{y} = E_{z} = \vec{B} = 0$\\
    \hline
    $(x, v_{x}, v_{y})$ & \begin{tabular}{@{}l@{}}$c_{1} 
      \left(
      \bar{f}_{\vec{i} - \vec{e}^{1} + \vec{e}^{2}}
      + \bar{f}_{\vec{i} + \vec{e}^{1} - \vec{e}^{2}}
      - \bar{f}_{\vec{i} - \vec{e}^{1} - \vec{e}^{2}}
      - \bar{f}_{\vec{i} + \vec{e}^{1} + \vec{e}^{2}}
      \right)$ \\
      $+c_{2}
      \left(
      \bar{f}_{\vec{i} - \vec{e}^{2} - \vec{e}^{3}}
      + \bar{f}_{\vec{i} + \vec{e}^{2} + \vec{e}^{3}}
      - \bar{f}_{\vec{i} + \vec{e}^{2} - \vec{e}^{3}}
      - \bar{f}_{\vec{i} - \vec{e}^{2} + \vec{e}^{3}}
      \right)$
    \end{tabular} & $E_{y} = E_{z} = B_{x} = B_{y} = 0$\\
    \hline
    $(x, y, v_{x}, v_{y})$ & \begin{tabular}{@{}l@{}}$c_{1} 
      \left(
      \bar{f}_{\vec{i} + \vec{e}^{1} - \vec{e}^{3}}
      + \bar{f}_{\vec{i} - \vec{e}^{1} + \vec{e}^{3}}
      - \bar{f}_{\vec{i} + \vec{e}^{1} + \vec{e}^{3}}
      - \bar{f}_{\vec{i} - \vec{e}^{1} - \vec{e}^{3}}
      \right)$ \\
      $+c_{2}
      \left(
      \bar{f}_{\vec{i} + \vec{e}^{3} + \vec{e}^{4}}
      + \bar{f}_{\vec{i} - \vec{e}^{3} - \vec{e}^{4}}
      - \bar{f}_{\vec{i} + \vec{e}^{3} - \vec{e}^{4}}
      - \bar{f}_{\vec{i} - \vec{e}^{3} + \vec{e}^{4}}
      \right)$\\
      $+c_{3}
      \left(
      \bar{f}_{\vec{i} + \vec{e}^{2} + \vec{e}^{3}}
      + \bar{f}_{\vec{i} - \vec{e}^{2} - \vec{e}^{3}}
      - \bar{f}_{\vec{i} + \vec{e}^{2} - \vec{e}^{3}}
      - \bar{f}_{\vec{i} - \vec{e}^{2} + \vec{e}^{3}}
      \right)$\\
      $+c_{4}
      \left(
      \bar{f}_{\vec{i} + \vec{e}^{2} - \vec{e}^{4}}
      + \bar{f}_{\vec{i} - \vec{e}^{2} + \vec{e}^{4}}
      - \bar{f}_{\vec{i} + \vec{e}^{2} + \vec{e}^{4}}
      - \bar{f}_{\vec{i} - \vec{e}^{2} - \vec{e}^{4}}
      \right)$\\
      $+c_{5}
      \left(
      \bar{f}_{\vec{i} + \vec{e}^{1} + \vec{e}^{4}}
      + \bar{f}_{\vec{i} - \vec{e}^{1} - \vec{e}^{4}}
      - \bar{f}_{\vec{i} + \vec{e}^{1} - \vec{e}^{4}}
      - \bar{f}_{\vec{i} - \vec{e}^{1} + \vec{e}^{4}}
      \right)$
    \end{tabular} & $E_{z} = B_{x} = B_{y} = 0$
  \end{tabular}
\end{table}

\subsection{Time stepping and stability considerations}


All explicit temporal discretizations of Eulerian methods are subject to a CFL stability condition, which depends on the combination of temporal and spatial discretization methods used \cite{courant1928}.
Choosing an appropriate time stepping method dictates both whether any stable time step exists as well as how computationally efficient it is to take a time step.
In particular, methods which have reduced memory storage/access requirements allow for efficient implementation on GPUs, where memory capacity and bandwidth are at a premium.
As before we consider each species $s$ separately and omit the $s$ subscript. The global stable timestep is then chosen as $\Delta t_{\max} = \|\Delta t_{\max,s}\|_{\infty}$.
One approach for choosing a stable time step is to perform a 1D linear Von-Neumann stability analysis\cite{chaplin2017}, then use the $L^{\infty}$ norm of the separate 1D CFL conditions for all dimensions to derive a global stable timestep \cite{vogman2014}.
We prove that it is possible to have more accurate stability bound estimates by using an $L^{1}$ norm.
This has been observed to allow on average about $20-40\%$ larger stable timesteps across full Vlasov-Poisson simulations with no extra computational work.

Performing a linear Von-Neumann stability analysis\cite{chaplin2017} on Eq. \eqref{eq:spatial_discr} for the fourth-order finite-volume method in $D$ dimensions with different constant speeds and cell spacings in each direction, the curve $\tilde{H}^{e}_{D}$ in the complex plane given by
\begin{gather}
  \tilde{H}^{e}_{D} = 
  \left\|
    \frac{\vec{A}}{60 \vec{h}}
  \right\|_{1} P\label{eq:stab_curve}\\
  P =
  2 \exp(-3j\xi) - 4 \exp(-2 j \xi) + 60 \exp(-j \xi) - 20 - 30 \exp(j \xi) + 3 \exp(2 j \xi)\label{eq:stab_curve2}
\end{gather}
must be inside the region of absolute stability for the temporal discretization in order to produce a stable numerical scheme. Here $\xi \in [0, 2\pi]$ and $j$ is the imaginary unit.
See Appendix \ref{sec:cfl_deriv} for a derivation of $\tilde{H}^{e}_{D}$.
This form is equivalent to that presented in Ref. \cite{chaplin2017} for a uniform speed $A^{d} = a$ and uniform cell size $h^{d} = h$, and permits a stable timestep size equal to or larger than that given by the max norm $\|\vec{A}/(60 \vec{h})\|_{\infty} D P$, which is used in Ref. \cite{vogman2014}.
The maximum possible increase in stable timestep size using the $L^{1}$ norm instead of the $L^{\infty}$ norm is $D$, which can be significant for high-dimensional simulations.

The curve $\tilde{H}^{e}_{D}$, defined in Eq. \eqref{eq:stab_curve}-\eqref{eq:stab_curve2} has a larger imaginary extent than real extent. For comparison, high-order discontinuous Galerkin (DG) methods and first-order finite-volume have roughly equal imaginary and real extents \cite{warburton2008}.
This means that many specialized Runge-Kutta methods developed for hyperbolic solvers, such as eSSPRK(10,4)\cite{ketcheson2008}, which have a region of absolute stability where real extent is equal to or larger than the imaginary extent, are ill-suited for the fourth-order finite-volume method under consideration.
Table \ref{tab:cfl_constant} shows the CFL $\sigma$ and effective CFL $\sigma_{\text{eff}} = \sigma / S$ for a variety of $S$-stage fourth-order accurate Runge-Kutta (RK) methods when applied in conjunction with the fourth-order finite-volume method.
Note that classical RK4 has the same region of absolute stability as the 3/8ths rule, but has different truncation error, FLOPs, and storage requirements\cite{hairer1993}.
The effective CFL gives a measure for the total number of RK stages required to advance to a fixed simulation end time, with higher values corresponding to fewer total RK stages. As computational cost scales proportional to the total number of RK stages for a full simulation, larger $\sigma_{\text{eff}}$ translates to fewer total stages and overall lower computational cost.
The effective CFL for all listed methods other than the 3/8ths rule are significantly smaller for fourth-order finite-volume than they would be for discontinuous Galerkin (not shown) or a first order finite-volume (shown in Table \ref{tab:cfl_constant} for reference) method.
The 3/8ths rule gives the largest $\sigma_{\text{eff}}$ and has the additional advantage in that it can be implemented using only the memory required to store three copies of the distribution function\cite{ho2022}.
The process used by Kubatko et al.\cite{kubatko2014} tailored for $\tilde{H}^{e}_{D}$ could produce a larger $\sigma_{\text{eff}}$, though it's unclear if the method will be as low storage and memory bandwidth efficient as the RK4 3/8ths rule.
Properly matching a temporal discretization to the spatial discretization is crucial for reducing computational costs, which is especially true for GPU implementations where reducing memory bandwidth utilization is critical.

\begin{table}[H]
  \centering{}
  \caption{CFL constant $\sigma = \Delta t_{\max} / \|\vec{A}/\vec{h}\|_{1}$ and effective CFL $\sigma_{\text{eff}}$, which is the ratio of $\sigma$ to the number of RK stages, for various RK methods when used with the fourth-order finite-volume spatial discretization described in Sec. \ref{sec:spatial_discr}. The specialized fourth-order RK methods which improve $\sigma_{\text{eff}}$ for high-order DG methods (not shown) and first-order finite-volume have a negative impact on $\sigma_{\text{eff}}$ for the fourth-order finite-volume method, and the largest $\sigma_{\text{eff}}$ of the compared methods is achieved with the 3/8ths rule.}
  \label{tab:cfl_constant}
  \begin{tabular}{l|l|l|l|l}
    Fourth-order RK Method & Num. Stages & $\sigma$ & $\sigma_{\text{eff}}$ & $\sigma_{\text{eff}}$ (first order FVM)\\
    \hline
    3/8ths rule\cite{hairer1993} & 4 & 1.73 & 0.432 & 0.348\\
    eSSPRK(5,4)\cite{spiteri2002} & 5 & 1.98 & 0.397 & 0.438\\
    SSPRK(8,4)+DG(4)\cite{kubatko2014} & 8 & 2.86 & 0.358 & 0.504\\
    eSSPRK(10,4)\cite{ketcheson2008} & 10 & 3.08 & 0.308 & 0.600
  \end{tabular}
\end{table}

\section{Code architecture overview}

\label{sec:architecture}

In this section we describe the architecture of VCK-GPU, a Vlasov-Poisson solver utilizing one GPU per MPI rank, that leverages the GPU-amenable temporal (RK4 3/8ths rule) and spatial (one-step fourth-order finite-volume) discretization in Sec. \ref{sec:discretization}.
Support for multiple compute nodes is critical for continuum kinetic simulations since simply storing the distribution function for computation can easily exceed the available memory on a single node.
The GPU-accelerated code is implemented using the Kokkos C++ template library\cite{trott2022}, which allows for a low-overhead, single-source implementation which works across various GPU architectures -- something that native tools like CUDA, HIP, and SYCL do not offer.
Furthermore, Kokkos provides a comparable level of control over parallelization strategy as these native tools.
The low-overhead and greater level of control enables near native performance that existing compiler-based options such as OpenMP and OpenACC currently cannot achieve\cite{wienke2012}.
RAJA\cite{beckingsale2019} is a similarly flexible option for writing performance portable GPU code.

The Vlasov-Poisson system, see Eqs. \eqref{eq:vlasov}-\eqref{eq:poisson}, can be broken into three distinct steps: moment integration, Poisson solve, and hyperbolic advance of the Vlasov equation, with appropriate data communication performed between steps.
The phase space computational domain is partitioned such that each GPU is assigned a unique particle species and region of phase space.
Since the number of degrees of freedom for the Poisson equation is significantly smaller than the number of degrees of freedom associated with the distribution function, a subset of the total simulation ranks is used for the Poisson solver.
See Sec. \ref{sec:partition} for details on the domain partitioning strategy, and Sec. \ref{sec:poisson} for details on the Poisson solver.
It is expected that a GPU-accelerated Vlasov-Poisson solver will be communication bandwidth limited\cite{zhao2018,einkemmer2020,francisquez2023}.
As a result, the primary focus of our design is to maximize effective bandwidth utilization.
A secondary concern which arises from the multiple problem size scales present in implementing a Vlasov-Poisson solver is ensuring the compute hardware remains fully utilized.
A common pattern for ensuring high GPU utilization for small tasks is to fuse separate operations into a single GPU compute kernel\cite{wang2010}.
This has a secondary major benefit as it allows multiple calculations to be performed on any single given data read from memory, reducing the overall volume of data which has to be read from GPU memory.
Minimizing data movement and having good data access patterns is usually the primary barrier to efficient GPU computing as the available peak arithmetic intensity of GPUs make most kernels memory bound\cite{williams2009}.
Sections \ref{sec:moment_kernel}, \ref{sec:hyper_kernel}, and \ref{sec:data_sync} describe how these guiding principles are applied to the various GPU compute kernels.

\subsection{Partitioning strategy}

\label{sec:partition}

Take $S$ number of species distribution functions to be discretized in domains with $d$ physical dimensions and $v$ velocity dimensions, and have $N_{si}$ cells along each dimension $i \in [1, ... , d, ... , d+v]$.
Each species' domain is divided with a grid of $n_{si}$ partitions along each dimension.
VCK-GPU requires that the number of cells in each physical dimension is the same for all species, such that $N_{si} = N_{i}, \forall i \in [1, d], \forall s$, and likewise the number of partitions along each physical dimension is the same for all species, such that $n_{si} = n_{i}, \forall i \in [1,d], \forall s$.
The number of cells and partitions along the velocity directions $i \in [d+1,v]$ can be different for different species.
For the purposes of estimating the communication volume we extend the requirements $N_{si} = N_{i}$ and $n_{si} = n_{i}$ to all species, $\forall i \in [1,d+v]$.
Assign $r$ partitions belonging to different species which are co-located in physical space to each rank.
For example, a 1D-2V $N_{x} \times N_{v_{x}} \times N_{v_{y}}= 1024 \times 256 \times 512$ domain could be split into $n_{x} \times n_{v_{x}} \times n_{v_{y}}= 4\times 1\times 2$ grid such that each of the eight partitions contains $256\times 256 \times 256$ cells.
The primary sources of required data communication are: global reduction for velocity moment calculation when evaluating $\rho_{c}$ in Eq. \eqref{eq:charge_density}, communication required for solving Poisson's equation given by Eq. \eqref{eq:poisson}, broadcasting appropriate pieces of the Poisson solution $\phi$, and ghost cell synchronization of $\bar{f}_{s}$.
To simplify analysis we will consider the communication required for the Poisson solver to be part of the compute time of the Poisson solver.
This is done for two reasons: the communication volume required for the Poisson solver scales differently depending on the solver used, and for problem sizes of interest the Poisson solver is sufficiently small that a single rank solver is the fastest (see Sec. \ref{sec:poisson}), which requires zero inter-rank communication.
The fourth-order finite-volume discretization requires the following number of floating point numbers to be transferred for different communication steps:
\begin{gather}
  B_{\text{reduce}} = 
  \ceil \left( \log_{2}
    \left(
      \frac{S}{r} \prod_{i=1}^{v} n_{i}
    \right)
  \right) \prod_{i=1}^{d} N_{i}\\
  B_{\phi} = B_{\text{reduce}} + 6 S
  \left(
    \prod_{i=1}^{v} n_{i}
  \right) \sum_{i=1}^{d} (n_{i} - p_{i}) \prod_{j\neq i}^{d} N_{j}\\
  B_{\text{ghost}} =
  S
  \left(
    6 \sum_{i=1}^{d+v} (n_{i} - p_{i}) \prod_{j\neq i}^{d+v} N_{j} + 2 \sum_{i=1}^{d+v} \sum_{j\neq i}^{d+v} (n_{i} - p_{i}) (n_{j} - p_{j}) \prod_{k \neq i, k \neq j}^{d+v} N_{k}
  \right)
\end{gather}
where $p_{i} = 1$ if the domain is periodic in the $i$-th direction and 0 otherwise, $B_{\text{reduce}}$ is associated with communication required for computing $\rho_{c}$, $B_{\phi}$ is associated with communication required for redistributing the Poisson solution $\phi$, and $B_{\text{ghost}}$ is associated with the communication required for ghost cell synchronization of the distribution functions (see Sec. \ref{sec:data_sync}).
Assuming $N_{i} = N$ and $n_{i} = n$, then $B_{\text{ghost}}$ scales roughly with $S n N^{d+v-1}$.
As long as $\prod_{i=1}^{d} N_{i} \ge \prod_{j=1}^{v} N_{j+d}$ there is at least four orders of magnitude more data transferred by $B_{\text{ghost}}$ than all other communication sources.
Note that $B_{\text{ghost}}$ is independent of how species are partitioned between ranks.
This means that solving one species per MPI rank requires no additional data to be transferred giving an $S$-fold increase in strong scaling.
While strictly minimizing the surface area to volume ratio can increase $B_{\text{reduce}}$, decreasing $B_{\text{ghost}}$ almost always results in a net overall reduction of all communication volume.
As data communication is the limiting factor for scalability, allowing partitioning in all dimensions while being more complex to implement is a worthwhile venture, and is the approach used here (see Sec. \ref{sec:performance}).
Another common strategy is to partition only in physical space\cite{reddell2016} as this reduces coding complexity at the cost of increased communication volume requirements.

\subsection{Moment integration}

\label{sec:moment_kernel}

\begin{figure}[H]
  \centering
  \begin{subfigure}[b]{.49\linewidth}
    \centering{}
    \includegraphics[height=.2\textheight]{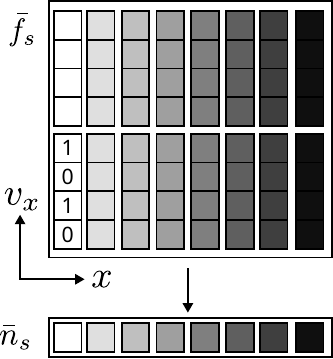}
    \caption{Algorithm L1}
    \label{algo:moment_v}
  \end{subfigure}
  \begin{subfigure}[b]{.49\linewidth}
    \centering{}
    \includegraphics[height=.2\textheight]{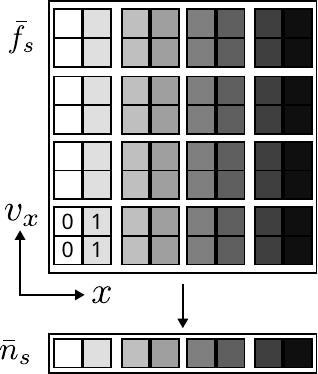}
    \caption{Algorithm L2}
    \label{algo:moment_x}
  \end{subfigure}
  \caption{Local moment integration kernel designs for different memory layouts. Algorithm L1 is optimized for memory contiguous in $\vec{v}$ while Algorithm L2 is optimized for memory contiguous in $\vec{x}$. Both algorithms are generally equally efficient so long as the layout of $\bar{f}$ matches the algorithm used.}
  \label{fig:moment_kernel}
\end{figure}

\begin{figure}[H]
  \centering{}
  \begin{subfigure}[t]{.49\linewidth}
    \centering{}
    \includegraphics[width=\textwidth]{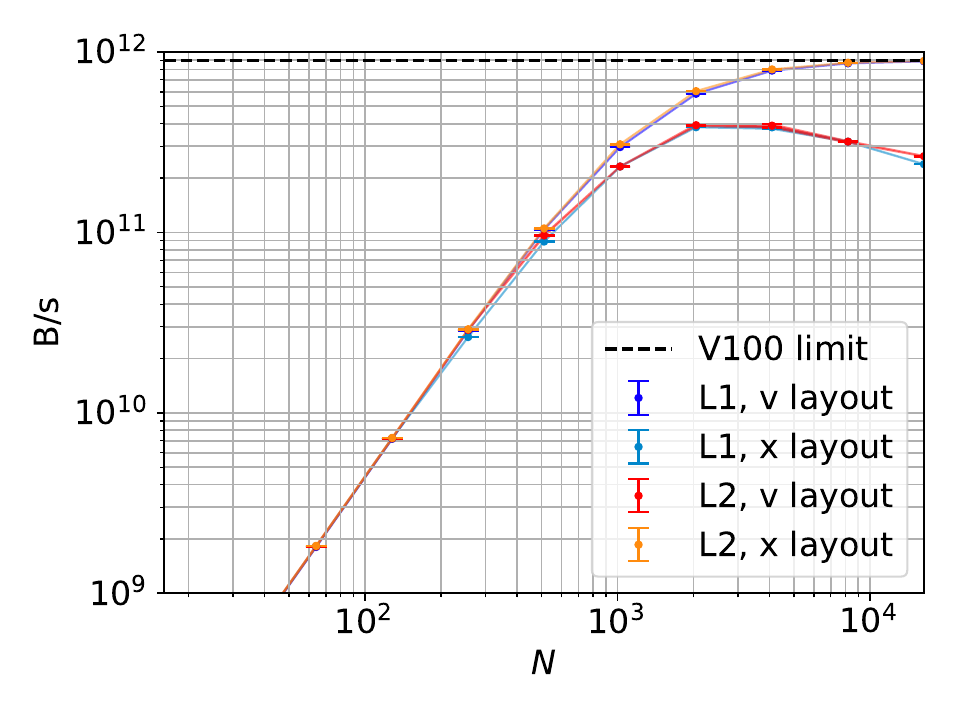}
    
    \caption{}
    \label{fig:moment0_layout_test}
  \end{subfigure}
  \begin{subfigure}[t]{.49\linewidth}
    \centering{}
    \includegraphics[width=\textwidth]{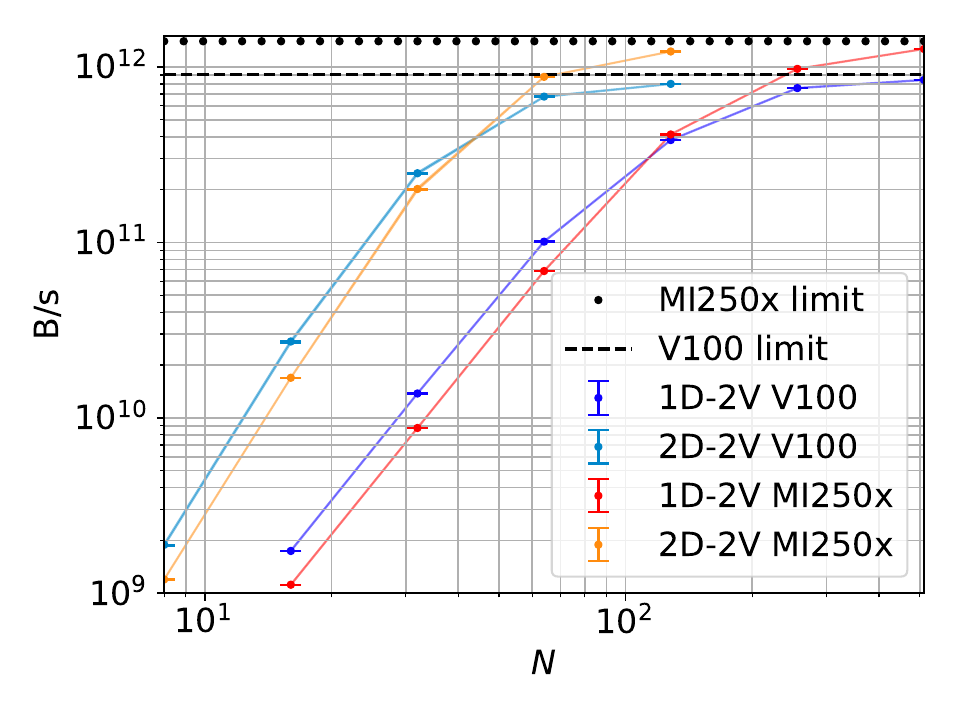}
    
    \caption{}
    \label{fig:moment0_perf_port}
  \end{subfigure}

  \caption{Local zeroth moment integration kernel performance for an $N^{d+v}$ domain. Figure \ref{fig:moment0_layout_test} demonstrates that in 1D-1V both Algorithms L1 and L2 are equally efficient so long they are matched with the correct memory layout. Figure \ref{fig:moment0_perf_port} demonstrates that Algorithm L1 can be scaled to 1D-1V and 2D-2V across multiple GPU architectures without loss of efficiency. Error bars denote the 5th, 50th, and 95th percentiles and the 95\% confidence interval is shaded. Student's t-test is used to determine the number of samples required for a 95\% confidence interval\cite{student1908}. The kernel has little noise in execution time.}
\end{figure}

The zeroth moment (see Eqs. \eqref{eq:number_density}-\eqref{eq:charge_density}) is computed by first integrating $\bar{f}_{s}$ over $\vec{v}$ in the locally assigned partition to produce a partial number density $\tilde{n}_{s}$.
MPI\_Reduce is then used on ranks which share the same region of $\vec{x}$ to get the local charge density $\rho_{c}$.
The local integration kernel is dependent on the memory data layout of $\bar{f}_{s}$.
Figure \ref{fig:moment_kernel} shows example 1D-1V parallelization strategies, which utilize four cells and two threads per team optimized for two different data layouts of $\bar{f}_{s}$.
The key guiding principle is that contiguous reads from consecutive threads in a thread team are required for peak GPU memory bandwidth.
For a layout contiguous in $v$ (labeled ``v layout'' in Fig. \ref{fig:moment0_layout_test}), Algorithm L1 achieves this by reading in a section of velocity space, performing a binary tree reduction\cite{trott2022}, and atomically adding the result to $\tilde{n}_{s}$ to safely parallelize teams over $v$.
For a layout contiguous in $x$ (labeled ``x layout'' in Fig. \ref{fig:moment0_layout_test}), Algorithm L2 achieves this by reading in a section of physical space, performing a small sequential summation in $v$, then atomically adding the result to $\tilde{n}_{s}$ to safely parallelize teams over $v$.
The number of threads per team and number of cells assigned to each team are tunable parameters dependent on the GPU model.
Figure \ref{fig:moment0_layout_test} demonstrates that using the wrong data layout results in an order of magnitude degradation of performance, however with the right combination of data layout and algorithm it is possible to achieve peak effective GPU memory bandwidth saturation.
VCK-GPU uses Algorithm L1 generalized to higher dimensions with data contiguous in $\vec{v}$ as an arbitrary choice.
Figure \ref{fig:moment0_perf_port} demonstrates performance portability of the moment integration kernel in 1D-2V and 2D-2V phase space domains, with results from the Nvidia V100 and AMD MI250x GPUs.

\subsection{Poisson solver}


\label{sec:poisson}
In order to have sufficient resolution for the high dimensional Vlasov equation there are severe limits on the number of cells along each physical space dimension\cite{vogman2020}.
This has major implications for the Poisson solver design as different algorithms are optimized for different number of degrees of freedom in the discretized Poisson system.
The smallest tested problem sizes for other applications have significantly more degrees of freedom \cite{sahasrabudhe2020}.
To test absolute performance for small system sizes two types of Poisson solvers are implemented.
The first uses sparse matrix solvers from PETSc\cite{petsc-user-ref} and HYPRE\cite{falgout2006}.
In this case, the choice of linear algebra solver and preconditioner along with the number of sub ranks used are configurable at runtime.
The second type uses FFT based methods\cite{feng2020} and is limited to solving on a single rank, though it can be GPU accelerated.
A combination of different primary solver, preconditioner, and whether the solve is performed on the CPU or GPU was tested using the Lassen supercomputer. See Table \ref{tbl:machines} for the hardware specifications.
These were tested using a slowly varying charge density on the periodic domain $\vec{x} \in [0, 1]^{d}$
with $\Delta t = 10^{-3}$, $N$ cells along each coordinate direction, and
\begin{gather}
  \rho_{c} = \sin(t) \prod_{i=1}^{D} \sin(2 \pi (x_{i} - t))
\end{gather}
across multiple solves to allow non-zero initial guesses to speed up iterative methods.
PETSc's default tolerances of $\epsilon_{a} = 10^{-10}$ and $\epsilon_{r} = 10^{-5}$ were used.
Periodic boundary conditions for the sparse matrix method are handled via null space removal techniques which preserve the original matrix properties such as definiteness and symmetry\cite{kaasschieter1988}, while the FFT solver enforces $\int \phi d\vec{x} = 0$.
The primary sparse matrix solvers tested include direct LU, GMRES, MINRES, conjugate gradients (CG), and stabilized bi-conjugate gradients (BCGS).
Preconditioner options include no preconditioning and algebraic multigrid (AMG) via BoomerAMG\cite{falgout2006}.
Note that AMG preconditioning is only done for GPU solves as PETSc+HYPRE currently does not support switching between CPU and GPU solves with a single binary.

\begin{figure}[H]
  \begin{subfigure}{.49\textwidth}
    \centering{}
    \includegraphics[width=\textwidth]{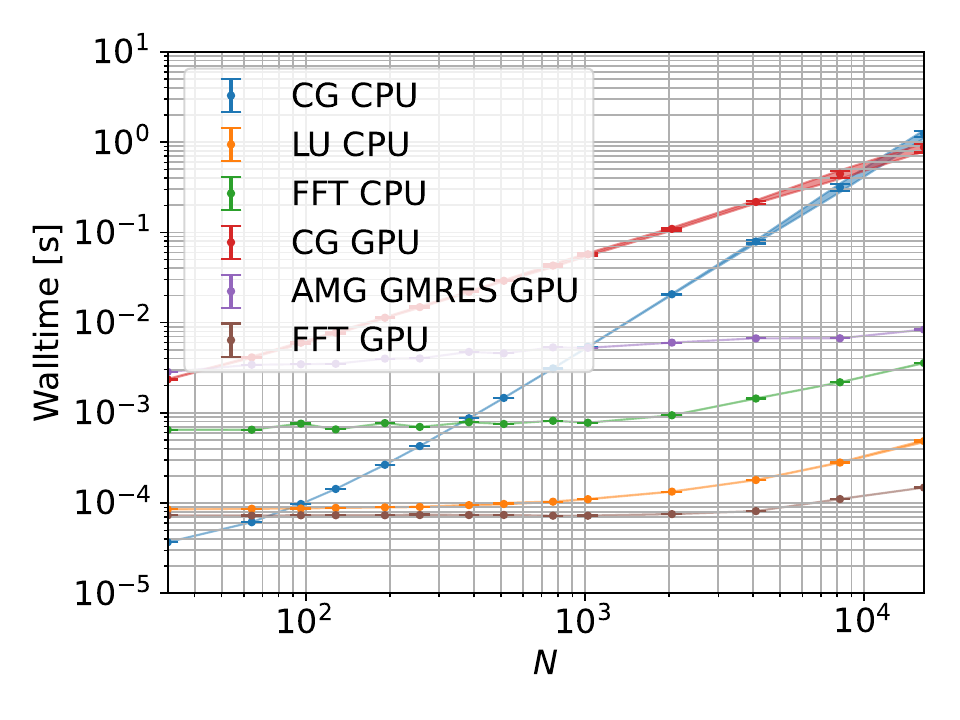}
    \caption{1D}
  \end{subfigure}
  \begin{subfigure}{.49\textwidth}
    \centering{}
    \includegraphics[width=\textwidth]{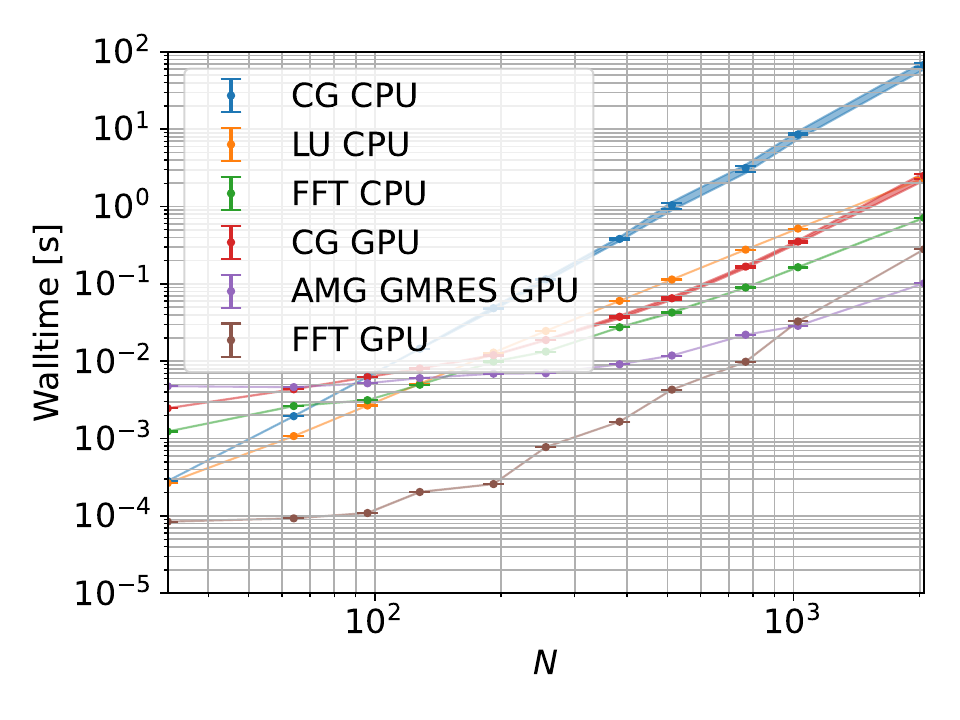}
    \caption{2D}
  \end{subfigure}
  \caption{Walltime per Poisson solve for domains with $N$ cells along each dimension. GPU-accelerated FFT solvers outperform sparse matrix solvers for small $N$, however, the better scaling performance of AMG-based methods leads eventually to AMG methods performing better. Error bars denote the 5th, 50th, and 95th percentiles and the 95\% confidence interval is shaded. Student's t-test is used to determine the number of samples required for a 95\% confidence interval\cite{student1908}.}
  \label{fig:poisson}
\end{figure}

Figure \ref{fig:poisson} shows a selection of tested combinations which performed the best in each category with the corresponding scaling trends.
For example, in 2D the combination of un-preconditioned GMRES on the GPU scales similarly to un-preconditioned CG on the GPU, however the GMRES solver had a larger constant offset resulting in longer solve times compared to CG. In this situation un-preconditioned GMRES on the GPU was omitted from the plot.
In all cases a single rank is used as multiple rank solves were all found to be slower than the single rank equivalent for the tested problem sizes.
In 1D, direct LU on the CPU or GPU-accelerated FFT solvers are at least an order of magnitude faster than all other configurations except for a few cases where for very small matrices CG CPU and MINRES CPU performed slightly faster.
In 2D GPU-accelerated FFT methods were faster than all tested sparse matrix methods up to a $1024^{2}$ domain.
At this point the better scaling performance of GPU accelerated AMG methods matches the performance of the FFT solver.
Refer to Sec. \ref{sec:performance} for a breakdown of walltime spent on the Poisson solve and how it compares to the overall solve time.
These results match similar trends as those found by others for CPU and GPU-based implementations\cite{zdenek2006,berger_vergiat2021}.
Iterative Poisson solver performance can also be improved by relaxing the solver tolerances, though this is problem dependent and it is unknown if there is any way to predict a priori how much tolerances can be relaxed.
There are potentially greater performance impacts when using non-periodic boundary conditions or non-uniform cell spacings (not including different uniform cell sizes in different dimensions) with FFT-based methods compared to sparse matrix methods\cite{feng2020}.

\subsection{Hyperbolic advance}


\label{sec:hyper_kernel}

The hyperbolic advance kernel has two primary concerns for performance: efficient memory bandwidth utilization, and reducing kernel launch latency.
Typical launch latencies are on the order of $1-10~\mu\text{s}$.
Splitting the hyperbolic advance into three steps of compute and store fluxes, sum all the surface fluxes to cells, and perform RK stage update (as done in VCK-CPU\cite{vogman2014}) not only increases the kernel enqueue latency, but also significantly increases the total amount of data which must be read/written\cite{xu2019}.
A disadvantage of fusing multiple steps into a single kernel is increased GPU register contention and reduced kernel efficiency, especially for complex PDEs.
Similarly, how the RK method is implemented can lead to different amounts of reads/writes.
For example, the 3/8ths rule\cite{hairer1993} can be implemented to require fewer kernel calls, fewer reads/writes to global GPU memory, and fewer overall memory buffers by reusing existing buffers, as shown in Table \ref{tbl:rk4_impls}.
Table \ref{tab:rk4_impl} shows the number of required global memory reads/writes of $\bar{f}$ and kernel call counts for various ways the 3/8ths rule RK4 finite-volume method could be implemented.
By fusing the entire discretized form of Eq. \eqref{eq:spatial_discr} with the fast RK4 3/8ths rule implementation into a single kernel call per RK stage there is a theoretical 2x-3x speedup over the split kernels implementation design used by VCK-CPU.
It is not expected that an implementation will achieve the full 2x-3x speedup, however this does serve as a useful indicator for identifying if a split kernel implementation could be beneficial or not, and what level of splitting to investigate.
Merging the entire RK stage into a single kernel achieves sufficient performance for many stencil-based codes that splitting is not viable\cite{micikevicius2009,sai2022}, however there are exceptions where large complex systems have other factors where splitting kernels can improve the overall performance\cite{xu2019}.

Boundary conditions at the $v_{\max}$ extents are implemented by setting velocity space ghost cells of $\bar{f}$ to have fixed initial-time values that are never updated.
The standard hyperbolic advance kernel then uses these cells as needed for the regular fourth-order finite-volume stencil.
This design choice negatively impacts conservation properties, however does offer significant performance gains.
Implementing a zero flux condition as used by VCK-CPU\cite{vogman2014} would either require additional compute kernels and/or conditional branches, both of which significantly degrade GPU compute performance.
In testing it was found that conservation problems could be sufficiently mitigated by simultaneously utilizing higher domain cell counts and extending $v_{\max}$ such that $\bar{f}$ is sufficiently small at $v_{\max}$ for all time.
Both of these conditions are already required to ensure correct physical accuracy of the fourth-order finite-volume method\cite{vogman2016}, thus this performance gain is deemed acceptable for the GPU implementation.

\begin{table}[H]
  \def\arraystretch{2}
  \centering{}
  \caption{Two ways to implement RK4 3/8ths rule. The fast form requires fewer kernel calls and global memory reads/writes compared to a direct Butcher tableau implementation.}
  \label{tbl:rk4_impls}
  \begin{tabular}{l|l}
    RK4 3/8ths rule Butcher tableau form & Fast RK4 3/8ths rule form\\
    \hline
    $\displaystyle K_{0} \gets L(f_{0})$ & $\displaystyle f_{1} \gets f_{0} + \frac{\Delta t}{3} L(f_{0})$\\
    $\displaystyle K_{1} \gets L
            \left(
            f_{0} + \frac{\Delta t}{3} K_{0}
            \right)$ & $\displaystyle f_{\text{out}} \gets 2 f_{0} - f_{1} + \Delta t L(f_{1})$\\
    $\displaystyle K_{2} \gets L
            \left(
            f_{0} + \Delta t\left(
            K_{1} - \frac{K_{0}}{3}
            \right)
            \right)$ & $\displaystyle f_{1} \gets 2 f_{0} - f_{\text{out}} + \Delta t L(f_{\text{out}})$\\
    $\displaystyle K_{3} \gets L
            \left(
            f_{0} + \Delta t (K_{0} - K_{1} + K_{2})
            \right)$ & $\displaystyle f_{\text{out}} \gets \frac{3 f_{\text{out}} - f_{0} + 3 f_{1} + \Delta t L(q_{1})}{8}$\\
    $\displaystyle f_{\text{out}} \gets f_{0} + \frac{\Delta t\left(
            K_{0} + 3 K_{1} + 3 K_{2} + K_{3}
    \right)}{8}$ &
  \end{tabular}
\end{table}

An efficient implementation for ``star shaped stencils'' (the case in Eq. \eqref{eq:spatial_discr} when $C_{\vec{i}} = 0$) in three dimensions or less is to store a sliding plane of cell values into shared memory\cite{micikevicius2009}. This has been demonstrated to be efficient since the early days of scientific GPGPU to modern hardware\cite{sai2022}.
On modern GPUs this technique is no longer required to achieve good performance, with ``nested tiling'' or ``6D-tiling'' and direct global memory access proving to be similarly fast\cite{zhao2018}.
In testing we have found that an implementation based on nested tiling works best as we can pre-compute and store into shared memory the coefficients $C_{\vec{i}}$ given by Eqs. \eqref{eq:C_k_coeff_1}-\eqref{eq:C_k_coeff_5}, which vary only in $\vec{x}$.
This reduces the demands on the quantity of shared memory compared to methods which store chunks of $\bar{f}$ into shared memory.
Storing just $C_{\vec{i}}$ into shared memory has additional benefits such as reducing arithmetic intensity and global memory I/O.
This design also scales beyond 3D at which point the sliding plane method fails to generalize gracefully.

\begin{table}[H]
  \centering
  \caption{Global memory read/write count and kernel calls for various implementations of the 3/8ths rule for the Vlasov system. Fused RHS indicates that the semi-discrete operator $L(f)$ is computed in a single kernel call. Fused stage indicates that each line in Table \ref{tbl:rk4_impls} is computed in a single kernel call. Fast RK4 indicates the Fast RK4 3/8ths rule form from Table \ref{tbl:rk4_impls} is used, while simply RK4 uses the Butcher tableau form. Numbers listed include requirements for the local moment integration kernel. Fusing together the full RK stage reduces kernel calls by half and global memory read/write counts by 2.6x.}
  \label{tab:rk4_impl}
  \begin{tabular}{c|r|r}
    Method & Global memory reads/writes & Kernel calls\\
    \hline
    Split kernels+RK4 & 42 & 16\\
    Fused RHS+RK4 & 30 & 12\\
    Fused RHS+fast RK4 & 28 & 12\\
    Fused stage+fast RK4 & 16 & 8
  \end{tabular}
\end{table}

\begin{figure}[H]
  \centering
  \includegraphics[width=.5\textwidth]{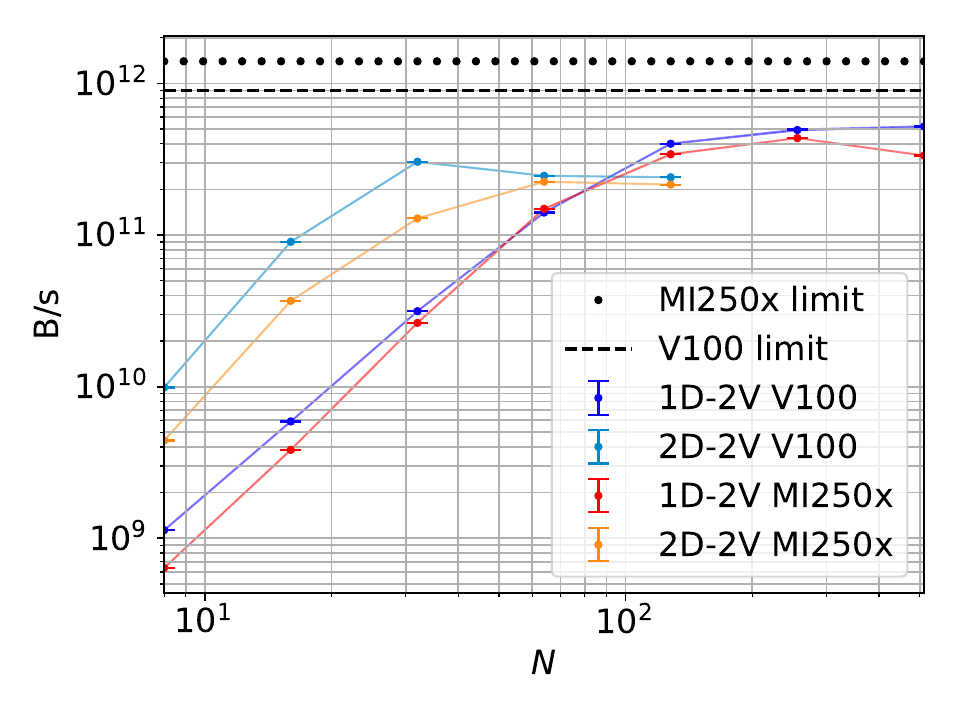}
  \caption{Effective memory bandwidth of the fused stage+fast RK4 method utilizing a nested tiling strategy for the spatial parallelization. Similar portable performance is achieved in 1D-2V while in 2D-2V the reduced caching performance of the MI250x negatively impacts overall throughput\cite{sai2022}. Error bars denote the 5th, 50th, and 95th percentiles and the 95\% confidence interval is shaded. Student's t-test is used to determine the number of samples required for a 95\% confidence interval\cite{student1908}. The kernel has little noise in execution time.}
  \label{fig:hyper_perf_port}
\end{figure}

Figure \ref{fig:hyper_perf_port} shows the effective memory bandwidth utilization achieved using the nested tiling strategy.
In 1D-2V the effective memory bandwidth is sufficiently high that there is no benefit to splitting the hyperbolic kernel on both the V100 and MI250x GPUs.
In 2D-2V this remains true for the V100, however it is less clear for the MI250x as there is available room for performance improvements, however we have been unable to find a splitting strategy which is able to realize these gains.

\subsection{Data synchronization}


\label{sec:data_sync}

Due to the wide stencil required for performing the hyperbolic advance of $\bar{f}$ (see Fig. \ref{fig:stencil_2d}), for cells which are near the edge of a partition some required cell values will belong to neighboring partitions.
A typical solution is to introduce ghost or halo cells which extend the amount of cell data the local partition stores.
These values are communicated from the appropriate neighboring ranks during a ghost/halo synchronization.
There is some overhead associated with each MPI communication send/receive pair\cite{doerfler2006}.
The fourth-order finite-volume method, described in Sec. \ref{sec:spatial_discr}, requires a three wide ghost cell region aligned with the axis for Eq. \eqref{eq:fvm_poly}, as well as some diagonal cell dependencies to compute $C_{\vec{i}}$ given by Eq. \eqref{eq:C_k_coeff_1}-\eqref{eq:C_k_coeff_5}.
Consider a partition at the center of a $3^{d+v}$ hypercube of partitions.
The number of communication pairs a rank must handle for all neighbors $N_{\text{all}}$ (naive implementation), for general fourth-order finite-volume $N_{\text{FVM}}$, and for fourth-order Vlasov-Poisson finite-volume $N_{\text{VP}}$ are given by
\begin{gather}
  N_{\text{all}} = 3^{d + v} - 1\\
  N_{\text{FVM}} = 2 (d + v) + 4 \binom{d+v}{2} = 2 (d+v)^{2}\\
  N_{\text{VP}} = 2 (d+v)^{2} - 4 \binom{d}{2} - 4 (v - d) d
\end{gather}
where $d$ is the number of physical and $v$ is the number of velocity dimensions.
The fourth order finite-volume stencil used does not require all diagonal neighbors, allowing $N_{\text{FVM}} \le N_{\text{all}}$.
$N_{\text{VP}}$ further takes advantage of cancellations specific to the Vlasov-Poisson system when computing $C_{\vec{i}}$ in Eq. \eqref{eq:spatial_corrections}, leading to $N_{\text{VP}} \le N_{\text{FVM}}$.
Both $N_{\text{FVM}}$ and $N_{\text{VP}}$ reduce the number of neighbor pairs from exponentially growing for $N_{\text{all}}$ to quadratic, a significant reduction even for $d + v = 3$.
Methods with only nearest neighbor dependencies such as DG\cite{warburton2008} can further reduce the number of neighbors to $N_{\text{NN}} = 2 (d + v)$.

\begin{figure}[H]
  \centering{}
  \includegraphics[width=.48\textwidth]{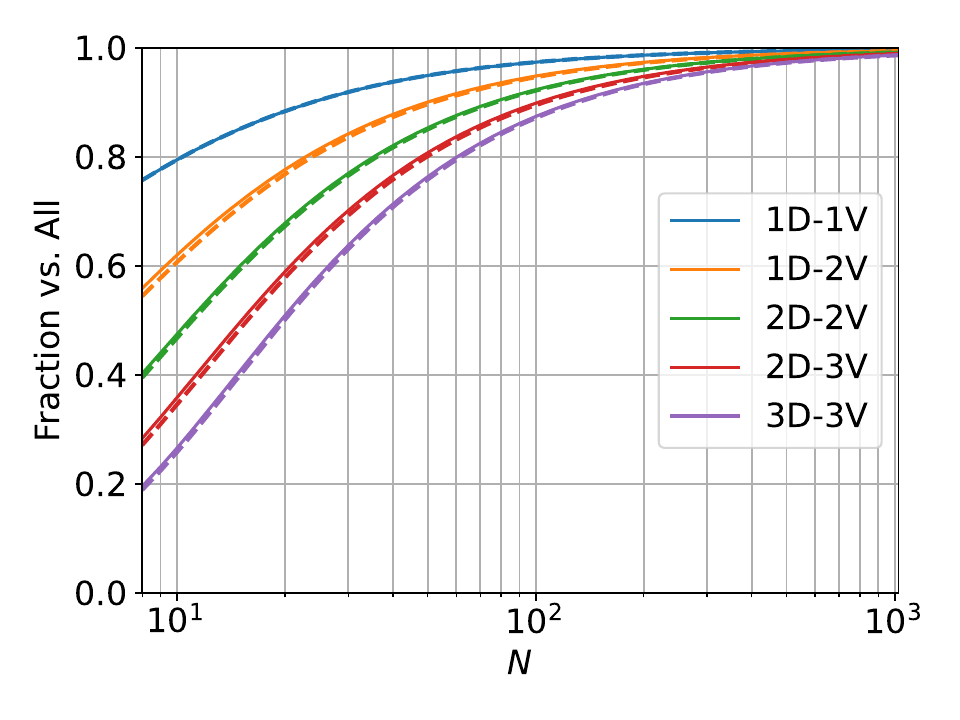}

  \caption{Fraction of ghost cell data transferred for $N_{\text{FVM}}$ and $N_{\text{VP}}$ communication-pair strategies compared to $N_{\text{all}}$ communication-pair strategy for an $N^{d+v}$ local partition. Solid line denotes $N_{\text{FVM}}$ compared to $N_{\text{all}}$ while dashed line denotes $N_{\text{VP}}$ compared to $N_{\text{all}}$. There is a significant reduction in ghost cell data volume using the $N_{\text{FVM}}$ strategy over the $N_{\text{all}}$ strategy.}
  \label{fig:ghost_cell_fraction}
\end{figure}

There is also a corresponding reduction in volume of data transferred during ghost cell synchronizations.
Figure \ref{fig:ghost_cell_fraction} shows the fraction of ghost cell data transferred for $N_{\text{FVM}}$ and $N_{\text{VP}}$ compared to $N_{\text{all}}$ for a local partition with $N$ cells along each side.
The amount of data transfer savings increases as partition sizes decrease, such that the $N_{\text{FVM}}$ communication-pair strategy reduces the data transfer volume to $60\%$ of the $N_{\text{all}}$ communication-pair strategy for a 1D-2V problem for small $N$. The savings increase with dimensionality.

\begin{figure}[H]
  \centering
  \begin{subfigure}{.49\textwidth}
    \centering{}
    \includegraphics[width=\textwidth]{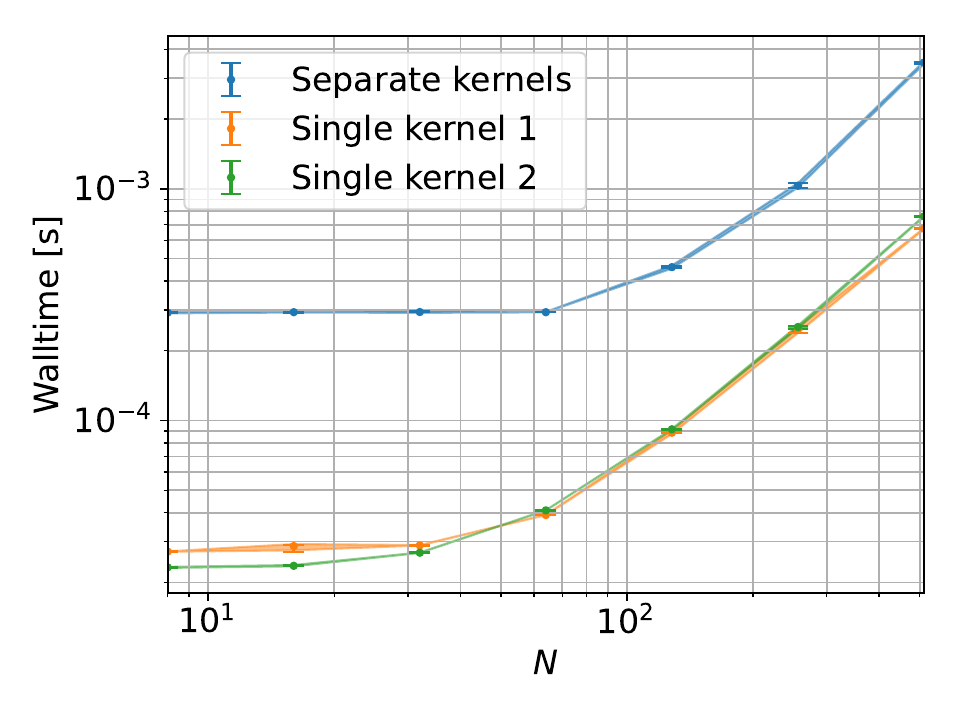}
    \caption{$d+v=3$}
  \end{subfigure}
  \begin{subfigure}{.49\textwidth}
    \centering{}
    \includegraphics[width=\textwidth]{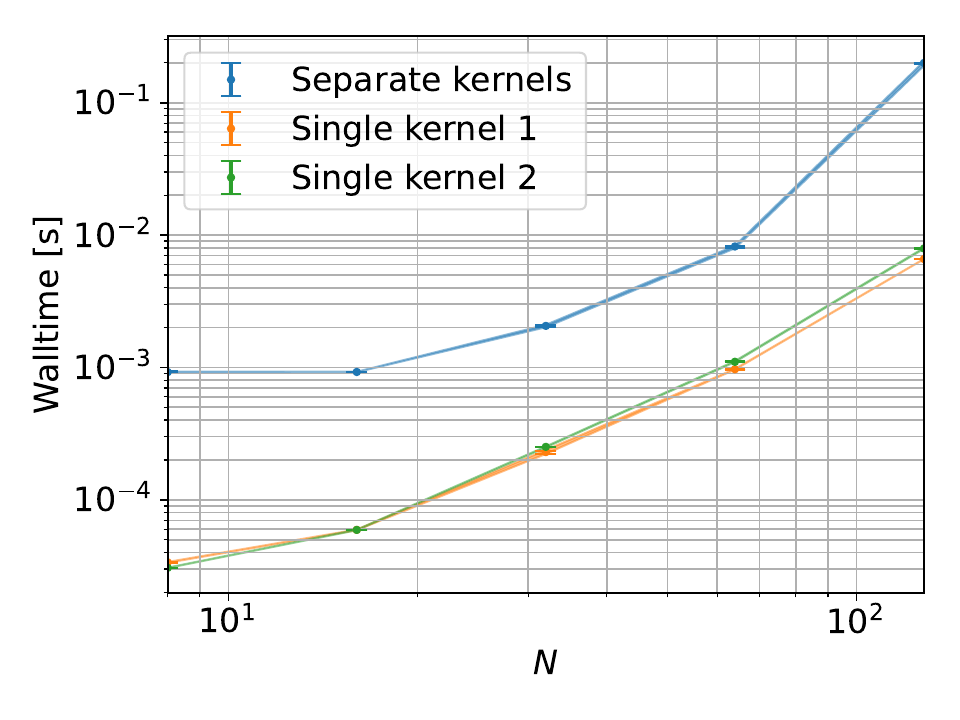}
    \caption{$d+v=4$}
  \end{subfigure}

  \caption{Comparison of methods for fusing the ghost cell pack kernel for the general fourth-order finite-volume method, for $d+v = 3$ and $d+v=4$.
    Single kernel 1 uses the indexing over flattened contiguous regions method, while single kernel 2 uses the method described in \cite{beckingsale2019}.
    Either tested method offers an order of magnitude reduction in time spent packing. Tested on an Nvidia V100 for an $N^{D}$ local partition. Error bars denote the 5th, 50th, and 95th percentiles and the 95\% confidence interval is shaded. Student's t-test is used to determine the number of samples required for a 95\% confidence interval\cite{student1908}. The kernel has little noise in execution time.}
  \label{fig:packing}
\end{figure}

In order to ensure the minimum number of send/receive pairs are used, the data must be arranged contiguously in memory per communication pair.
This can be done by design of the layout of $\bar{f}_{s}$\cite{zhao2018,reddell2016}, or by utilizing a data packing/unpacking phase.
VCK-GPU is designed with the latter approach as it allows a single compute kernel to have a simple and consistent memory access pattern.
Efficient GPU implementations of packing/unpacking utilize kernel fusion to perform the entire pack portion in a single kernel.
One method is to apply a parallel for loop over every index of the flattened contiguous region of packed data.
Each thread is responsible for identifying which cell in $\bar{f}_{s}$ its responsible for packing/unpacking.
This ensure that all GPU threads are performing useful work while utilizing the increased compute performance of the GPU to handle the extra work associated with identifying which cell to pack/unpack.
Another method to accomplish this is to specify a common team thread block size for all segments to be packed, then assign a single segment per thread team\cite{beckingsale2019}.
Figure \ref{fig:packing} demonstrates that both methods perform similarly in 3D and 4D for the fourth-order finite-volume ghost sync requirements, and offer an order of magnitude speedup of the pack/unpack phase over separate kernels per region.
In 1D-2V the overall speedup in the full simulation goes from 1.23x for large partitions to 2x for small partitions.
In 2D-2V the overall speedup ranges from 1.68x for large partitions to 4x for small partitions.
The 1D-2V speedup for large partitions is approximately double of those observed in GPU-accelerated nearest neighbor codes\cite{beckingsale2019} due to having more neighbors (18 neighbors for fourth-order finite-volume versus 6 neighbors for nearest neighbors only).

\section{Verification and validation}

\label{sec:vv}

To verify the fundamental order of accuracy of the discretization we use the grid error $L^{1}$ norm combined with Richardson extrapolation\cite{leveque2007} to estimate the numerical error using the solution $f$ at resolution $N$ and $2 N$:
\begin{gather}
  \|\hat{\epsilon}\|_{1} = \frac{1}{V} \sum^{N}_{i} |\langle f_{N}\rangle_{i} - \langle \hat{f}_{2 N}\rangle_{i}| = C h^{p}\label{eq:err_metric}
\end{gather}
where $\langle \hat{f}_{2 N}\rangle_{i}$ is an exact summation over $f_{2 N}$ to match the grid resolution of $f_{N}$ and $h$ is the cell width for $f_{N}$ and $V = \int d\vec{x} d\vec{v}$ is the volume of the phase space domain.
The distribution function $\bar{f}_{h}(t=t_{0})$ is initialized using an 8-point, 16th order accurate Gaussian quadrature rule.
This allows initialization errors to be easily separated from time advance discretization error and ensures both components are implemented correctly.
A fourth-order finite-volume quadrature can also be used in place of Gaussian quadrature\cite{vogman2014}, which is useful for initialization where initial conditions are only known at discrete points rather than an explicit analytical field\cite{vogman2019,vogman2020,vogman2021}.
To validate that the implementation is able to match theoretical predictions, a variety of plasma benchmark problems are used for which analytical dispersion relations are known: warm two-stream instability, Dory-Guest-Harris instability, acceleration-driven lower hybrid drift instability, and nonlinear Landau damping.
For single-species dynamics, we choose the reference mass $m_{0}$ to be the electron particle mass and for multi-species dynamics we choose $m_{0}$ to be the proton mass.
The reference timescale is chosen to be the plasma frequency such that $t_{0} = \omega_{p0}^{-1}$ and the characteristic length scale is problem dependent.
Instability growth rates are measured from the electric field amplitude $\|E\| = \sqrt{\int \vec{E} \cdot \vec{E} d\vec{x}}$.

\subsection{Warm two-stream instability}

\begin{figure}[H]
  \centering
  \begin{subfigure}[b]{.39\textwidth}
    \centering{}
    \includegraphics[width=\textwidth]{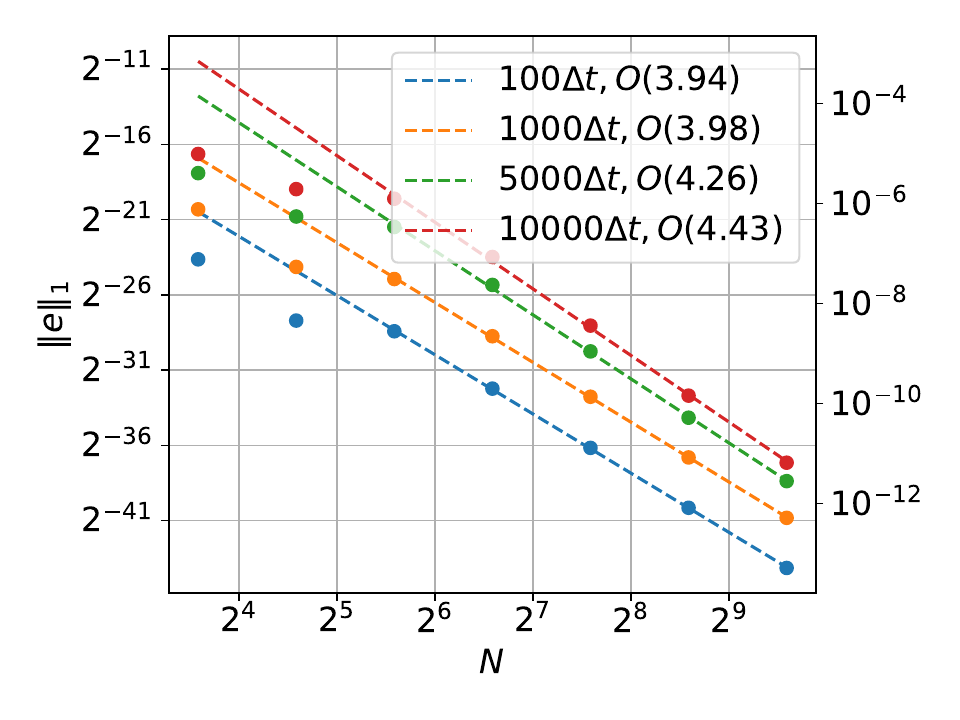}
    \caption{}
    \label{fig:two_stream_convergence}
  \end{subfigure}
    \begin{subfigure}[b]{.59\textwidth}
    \centering{}
    \includegraphics[width=\textwidth]{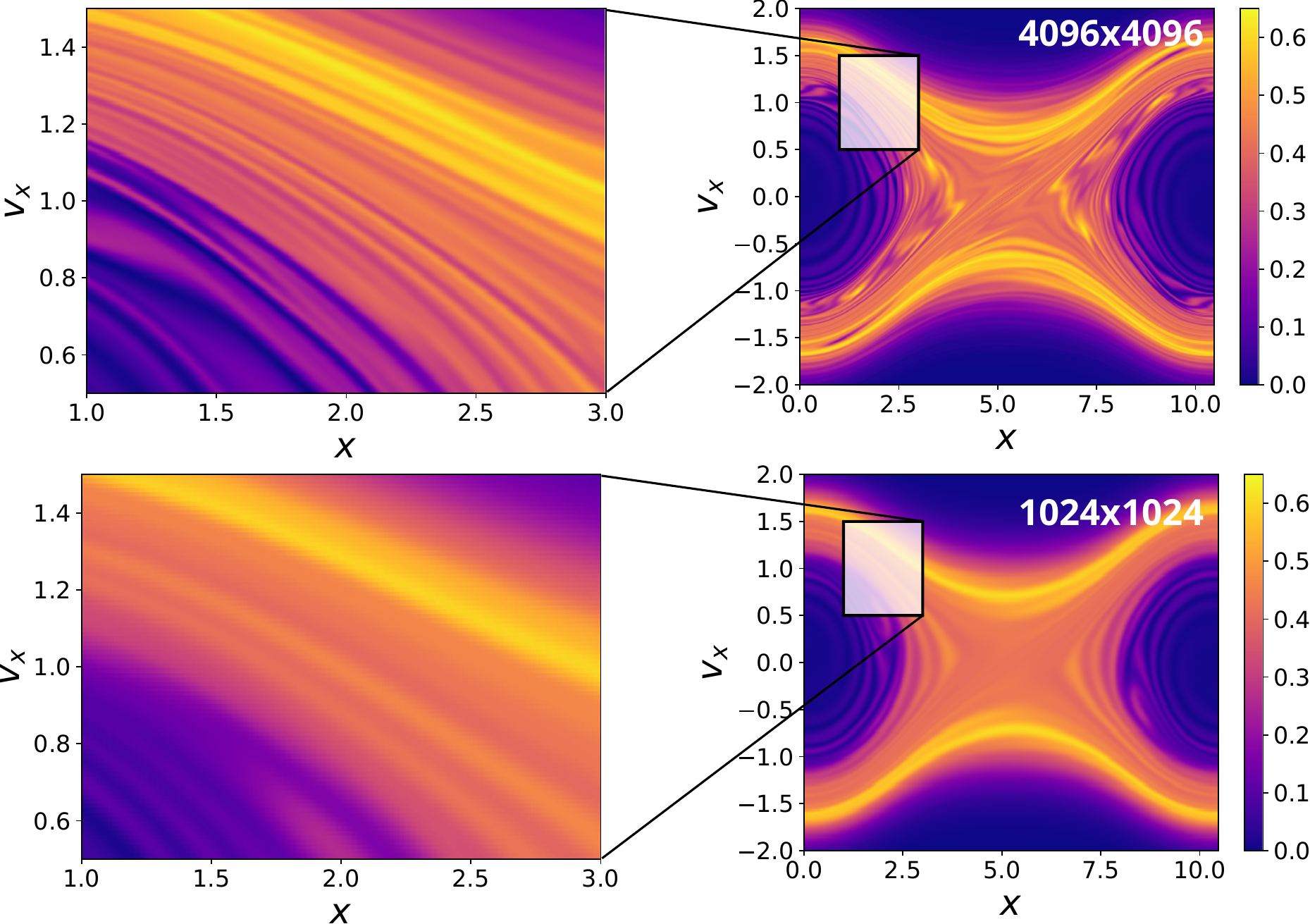}
    \caption{}
    \label{fig:two_stream_distr}
  \end{subfigure}
  \caption{The left figure shows the discretization error $\|e\|_{1}$ (see Eq. \eqref{eq:err_metric}) during the linear growth phase of the two-stream instability versus the number of cells $N$ along each dimension. A fixed timestep $\Delta t = 10^{-5}$, much smaller than the largest stable timestep, is used.
    Fourth-order numerical accuracy is maintained over multiple time steps.
    The right figure shows $f_{e}(t =200)$ for the parameter regime $v_{T}^{2}=0.1$ and $k =0.6$, with $v_{\text{max}} = \pm 6$. High order dissipation of the distribution function is greatly decreased by increasing the domain cell resolution from $1024\times 1024$ (bottom) to $4096 \times 4096$ (top).}
  \label{fig:two_stream_1}
\end{figure}

The warm two-stream instability is a 1D-1V electrostatic instability with a single dynamic electron species and a static neutralizing ion background species\cite{landau1965, kampen1955}.
The same dynamics governs the bump on tail instability\cite{filbet1979} and can be generalized to higher dimensions as well as multiple beams\cite{crews2022}.
Take an electron distribution function consisting of two counter-streaming Maxwellian beams with a perturbation in the distribution function\cite{vogman2016}
\begin{gather}
    f(x, v_{x}, t=0) = \frac{1}{v_{T} \sqrt{2 \pi}} \sum_{+,-}
                        \left(
                        \frac{1}{2} \pm \delta \sin
    \left(
      \frac{2 \pi x}{L}
    \right)
  \right) \exp
  \left(
    -\frac{\left(v_{x} \mp u\right)^{2}}{2 v_{T}^{2}}
  \right)
\end{gather}
where $v_{T}$ is the thermal velocity of each beam, $u=1$ is the beam velocity, $\delta = 10^{-5}$ is the perturbation amplitude away from the equilibrium, and the domain $x \in [0, L]$ fits a single wavelength of a given wavenumber $k = 2 \pi / L$.
The reference velocity $v_{0}$ is set to the beam velocity and the reference spatial scale is consequently $v_{0} / \omega_{p0}$.
The dispersion relation can be written in terms of the plasma dispersion function $Z$ as\cite{landau1965}
\begin{gather}
  k^{2} = - \frac{\omega_{pe}^{2}}{2 v_{T}^{2}}(1 + \zeta(1) Z(1) + \zeta(-1) Z(-1))\\
  \zeta(u) = \sqrt{\frac{1}{2 v_{T}^{2}}}
  \left(
    \frac{\omega}{|k|} - u
  \right)\\
  Z(u) = j \sqrt{\pi} (1 - \erf(-j \zeta(u))) e^{-\zeta(u)^{2}}
\end{gather}
Figure \ref{fig:two_stream_convergence} demonstrates that the  fourth-order convergence is maintained during the linear growth phase of the two-stream instability.
Higher than fourth-order initialization enables measuring error accumulated per timestep.
Figure \ref{fig:two_stream_distr} demonstrates the GPU's ability to handle high domain resolution and resolve fine-scale structures in the distribution function.

\begin{figure}[H]
  \centering
  \begin{subfigure}[t]{.49\textwidth}
    \centering{}
    \includegraphics[width=\textwidth]{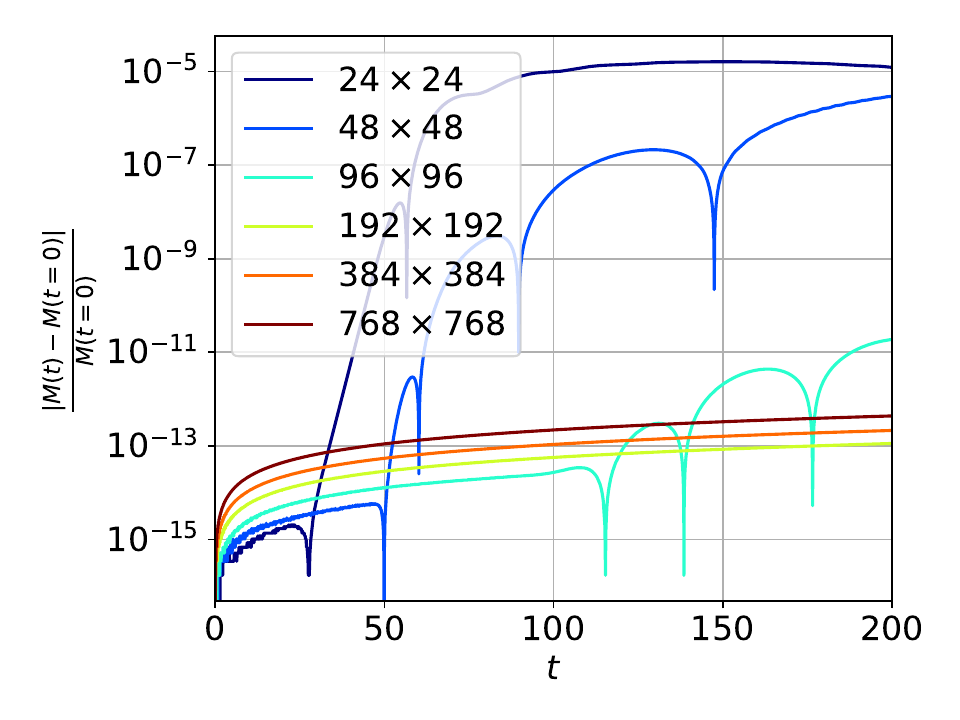}
    \caption{}
    \label{fig:two_stream_mass_cons}
  \end{subfigure}
  \begin{subfigure}[t]{.49\textwidth}
    \centering{}
    \includegraphics[width=\textwidth]{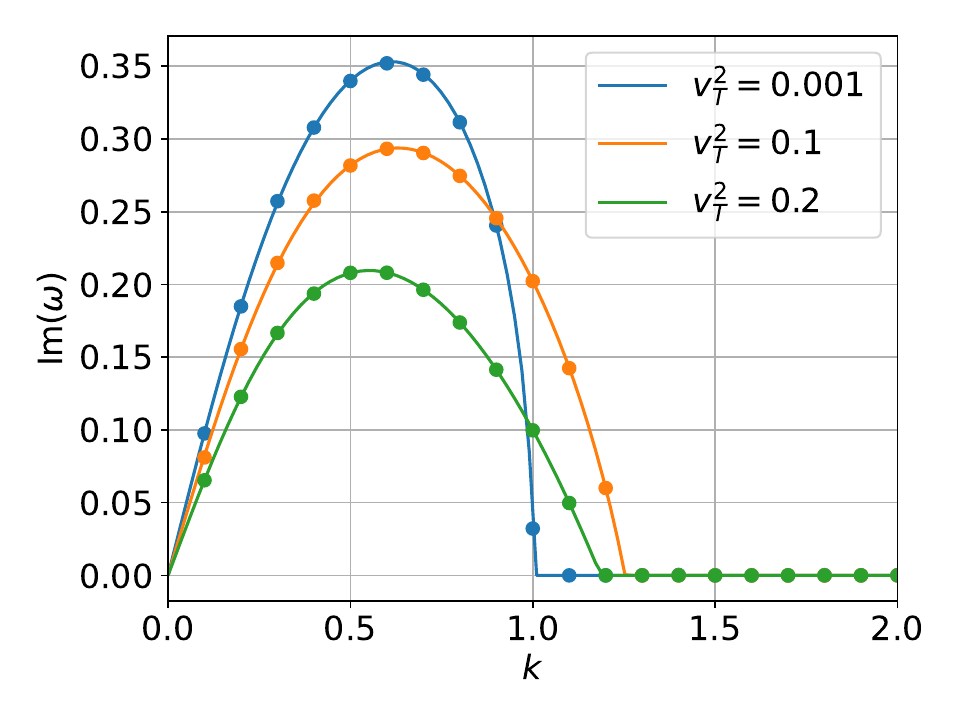}
    \caption{}
    \label{fig:two_stream_rates}
  \end{subfigure}

  \caption{The left figure shows how well total mass $M = \int f d\vec{x} d\vec{v}$ is conserved for different domain cell counts, with $v_{T}^{2} = 0.2$ and $k = 0.6$. At sufficiently high cell counts of $192\times 192$ or higher, mass loss per timestep becomes comparable to floating point truncation error. The right figure shows the measured growth rates for a variety of temperatures with circles and the theoretical growth rate in solid lines. Measured growth rates have less than $2\%$ deviation from theoretical predictions. Multiple wavenumbers are tested to show that simulations can isolate and capture fast-, slow-, and non-growing modes. A domain $2048$ cells along each direction and $v_{\max} = \pm 8$ is used.}
\end{figure}

Figure \ref{fig:two_stream_mass_cons} shows the change in total mass $M = \int f d\vec{x} d\vec{v}$ relative to the initial mass. The results demonstrate that the choice of boundary condition implementation described in Sec. \ref{sec:hyper_kernel} negatively impacts conservation of mass for low cell counts, however with sufficiently high resolution the mass loss becomes comparable to floating point truncation limits.
Zero-flux boundary conditions can ensure mass is conserved to numerical precision even at low resolution \cite{vogman2018}, however such boundary conditions reduce the efficiency of the GPU implementation.
Figure \ref{fig:two_stream_rates} validates that the correct growth rates for various beam thermal velocities $v_{T}$ are obtained.
Performance in 1D-1V is strongly limited by latency related effects, with a single GPU taking $3~\text{ms}$ per timestep with a $2048\times 2048$ cell domain. Increasing this to four GPUs decreases this to $2~\text{ms}$ per timestep due to the lack of available parallelism in 1D-1V.
This results in the tested two-stream simulations taking between 1.5 and 30 minutes, depending on the number of timesteps required.
Not having enough parallel work in 1D-1V to take full advantage of GPU acceleration results in an estimated speedup\footnote{times scaled to same number of DOFs and 40 CPU cores} over CPU codes of around 4-20x\cite{umeda2008}.

\subsection{Dory-Guest-Harris instability}


The Dory-Guest-Harris (DGH) instability is a 1D-2V magnetostatic instability\cite{dory1965} with an electromagnetic extension \cite{datta2021}. For this test we use a single dynamic electron species and a static neutralizing ion background species.
Take a perturbed electron ring distribution function of the form\cite{vogman2014}
\begin{gather}
  f(x, v_{x}, v_{y}, t=0) =  \frac{1}{\pi \ell! \alpha_{\perp}^{2}}
    \left(
      \frac{\vec{v} \cdot \vec{v}}{\alpha^{2}_{\perp}}
    \right)^{\ell} \exp
    \left(
      -\frac{\vec{v} \cdot \vec{v}}{\alpha^{2}_{\perp}}
    \right)
  \left(
    1 + \delta \sin
    \left(
      4 \tan^{-1}
      \left(
        \frac{v_{y}}{v_{x}}
      \right) - \frac{2 \pi x}{L}
    \right)
  \right)\label{eq:dgh_ic}
\end{gather}
where $\delta=10^{-4}$ is the perturbation amplitude, $\ell=4$, $\alpha_{\perp} = \sqrt{2} / 2$, and the domain $x \in [0, L]$ fits a single wavelength of a given wavenumber $k = 2 \pi / L$.
The reference velocity is set to the Alfv\`{e}n velocity $v_{0} = \sqrt{B_{0}^{2} / (\mu_{0} m_{0} n_{0})}$ and the reference length scale is consequently $v_{0} / \omega_{p0}$.
The ring distribution function is peaked at $v_{\perp 0} = \ell^{1/2} \alpha_{\perp} = \sqrt{2}$ \cite{datta2021}.
The dispersion relation of electrostatic waves propagating perpendicular to a uniform magnetic field $B_{z}$ is given by\cite{vogman2014}
\begin{gather}
  1 + \frac{\omega_{pe}^{2}}{\Omega_{e}^{2}} \int^{\pi}_{0}\frac{\sin
    \left(
      \frac{\omega}{|\Omega_{e}|} \tau
    \right)}{\sin
    \left(
      \frac{\omega}{|\Omega_{e}|} \pi
    \right)} \sin(\tau) F_{0}(\tau) d \tau = 0\\
  F_{0}(\tau) = \int_{0}^{\infty} f_{0}(v_{\perp}) J_{0}
  \left(
    2 \frac{k_{\perp} v_{\perp}}{|\Omega_{e}|} \cos\frac{\tau}{2}
  \right) 2 \pi v_{\perp} d v_{\perp}
\end{gather}
where $\Omega_{e} = q_{e} B / m_{e}$ and $J_{0}$ is the zeroth Bessel function of the first kind and $f_{0}$ is the unperturbed equilibrium electron distribution function ($\delta = 0$ in Eq. \eqref{eq:dgh_ic}).

\begin{figure}[H]
  \centering
  \begin{subfigure}[t]{.49\textwidth}
    \centering{}
    \includegraphics[width=\textwidth]{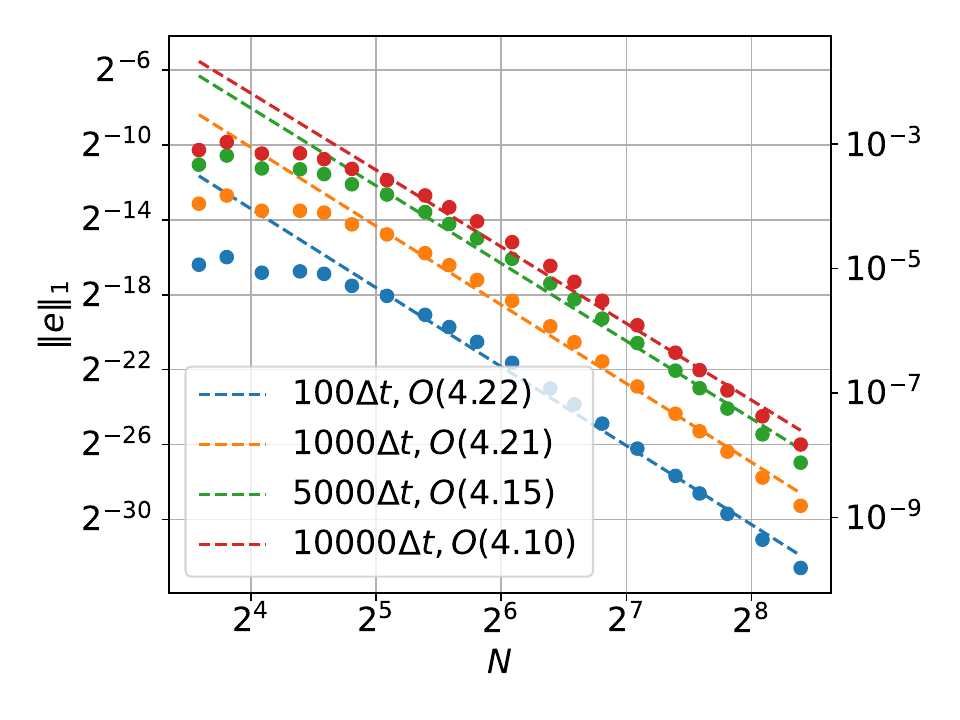}
    \caption{}
    \label{fig:dgh_error}
  \end{subfigure}
  \begin{subfigure}[t]{.49\textwidth}
    \centering{}
    \includegraphics[width=\textwidth]{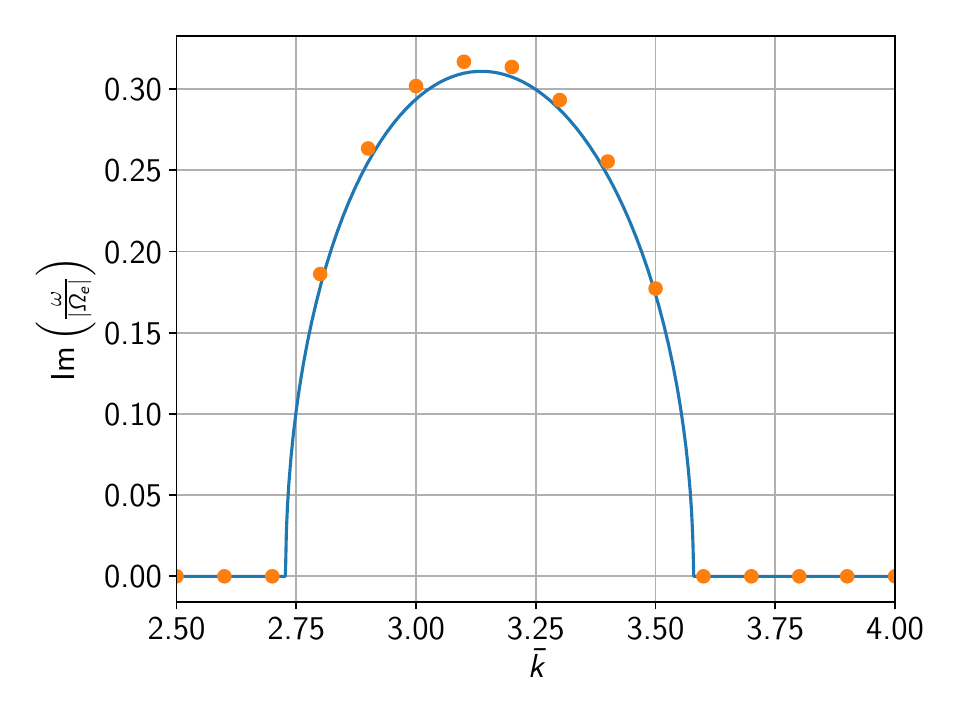}
    \caption{}
    \label{fig:dgh_growth_rates}
  \end{subfigure}
  
  \caption{The left figure shows the discretization error $\|e\|_{1}$ (see Eq. \eqref{eq:err_metric}) during the linear growth phase of the DGH instability versus the number of cells $N$ along each dimension.
    A fixed timestep $\Delta t = 10^{-3}$, much smaller than the largest stable timestep, is used.
    Fourth-order numerical accuracy is maintained over multiple time steps.
    The right figure shows the fitted growth rates with circles and the theoretical growth rates with a solid line where $|\Omega_{e} / \omega_{pe}| = 1/20$ for various $\bar{k} = k v_{\perp 0} / |\Omega_{e}|$. The fitted growth rates have a less than $5\%$ error from theoretical predictions.
    A domain resolution with $1024$ cells in each direction and $v_{\text{max}} = \pm 8$ is used.
  }
  \label{fig:dgh_convergence}
\end{figure}

\begin{figure}[H]
  \centering
  \begin{subfigure}[t]{.49\textwidth}
    \centering{}
    \includegraphics[width=\textwidth]{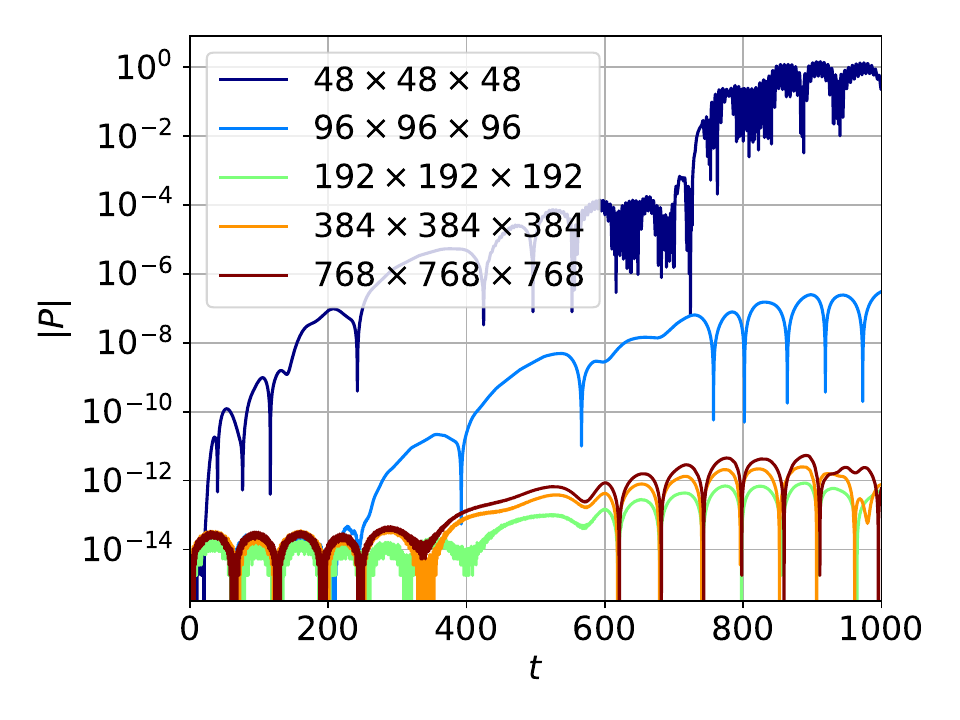}
  \end{subfigure}
  \begin{subfigure}[t]{.49\textwidth}
    \centering{}
    \includegraphics[width=\textwidth]{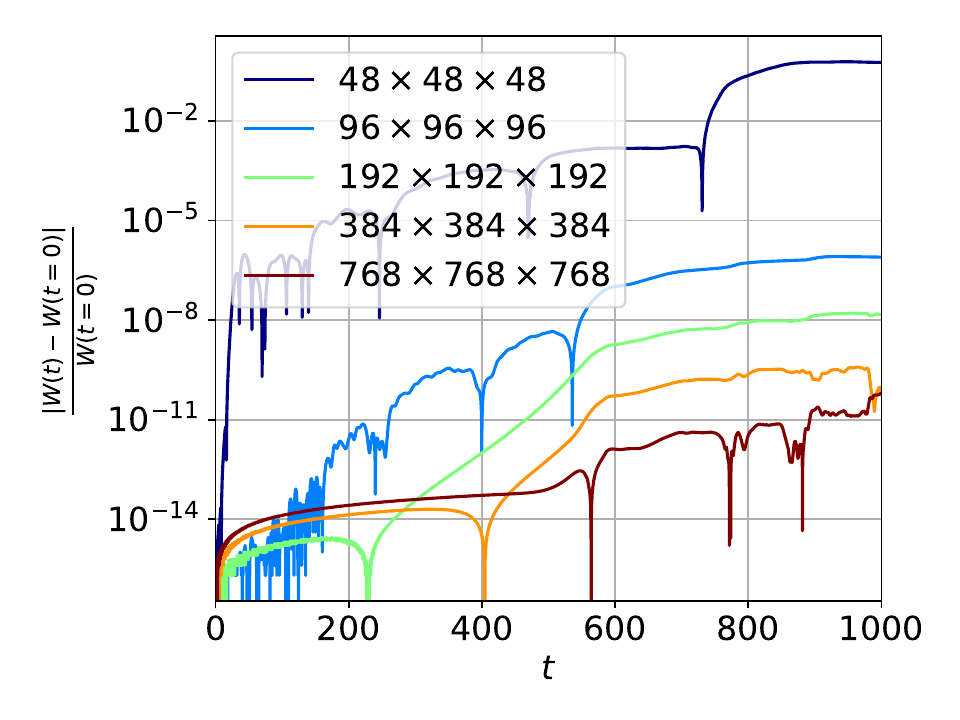}
  \end{subfigure}

  \caption{Total momentum $P = \|\int \vec{v} f d\vec{x} d\vec{v}\|$ (left) and energy $W = \int E^{2}/2 d\vec{x} + \int \vec{v} \cdot \vec{v} f / 2 d\vec{x} d\vec{v}$ (right) conservation properties for $|\Omega_{e}| / \omega_{pe} = 1/20$ and $k v_{\perp 0} / |\Omega_{e}| = 3.2$ for various domain cell counts.
    At $192\times 192\times 192$ cells momentum change per timestep becomes comparable to floating point truncation error, while energy conservation continues to improve even at the most refined tested resolution of $768 \times 768 \times 768$.
    A domain with $v_{\text{max}} = \pm 8$ is used.
  }
  \label{fig:dgh_conserve}
\end{figure}

Figure \ref{fig:dgh_error} demonstrates fourth-order convergence is maintained during the linear growth phase.
Higher than fourth-order initialization enables measuring error accumulated per timestep.
Figure \ref{fig:dgh_growth_rates} validates that the correct growth rates are obtained for various perturbations with $|\Omega_{e}| / \omega_{pe} = \omega_{c0} / \omega_{p0} = 1/20$.
Multiple wavenumbers are tested to show that simulations can isolate and capture fast-, slow-, and non-growing modes.
Figure \ref{fig:dgh_conserve} shows that the choice of boundary condition implementation described in Sec. \ref{sec:hyper_kernel} negatively impacts conservation of momentum for low cell counts, with momentum loss per timestep becoming comparable to floating point truncation error at a resolution of $192 \times 192 \times 192$.
Total energy conservation continues to improve with resolution at least to $768 \times 768 \times 768$.
These findings match and extend previously observed trends\cite{vogman2016} that while the fourth-order finite-volume method does not conserve total energy, energy conservation does improve with increasing resolution.

\subsection{Acceleration-driven lower hybrid drift instability}

\label{sec:glhdi}
\begin{figure}[H]
  \centering{}
  \begin{subfigure}[t]{.49\textwidth}
    \centering{}
    \includegraphics[width=\textwidth]{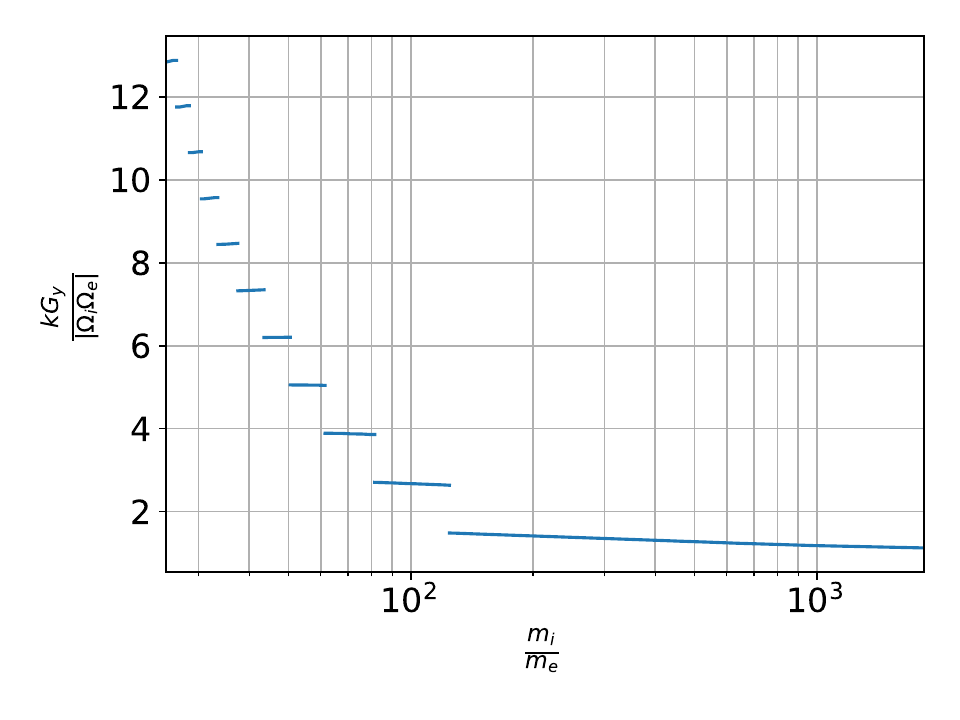}
    \caption{}
    \label{fig:glhdi_wavenumber}
  \end{subfigure}
  \begin{subfigure}[t]{.49\textwidth}
    \centering{}
    \includegraphics[width=\textwidth]{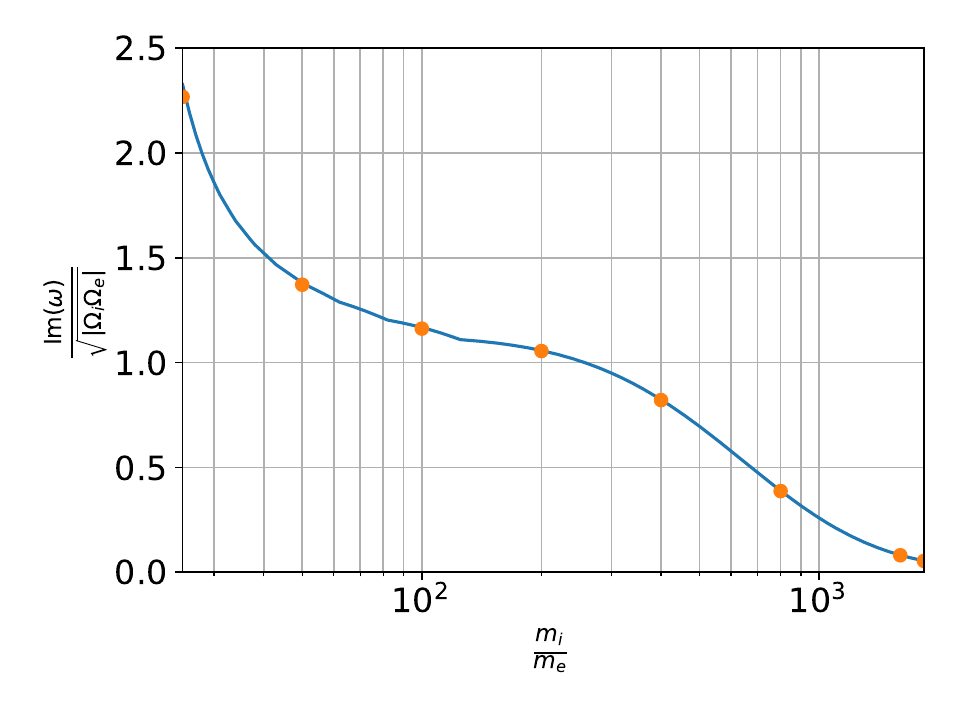}
    \caption{}
    \label{fig:glhdi_validation}
  \end{subfigure}

  \caption{The left figure shows the wavenumber corresponding to the fastest growing unstable mode for the acceleration-driven LHDI for various mass ratios for $v_{D} / v_{Ti} = 9 + 9 / m_{r}$, $|\Omega_{e} / \omega_{pe}| = 10^{-2} \sqrt{m_{r}}$, $T_{i} / T_{e} = 1$, and $\beta = 2.5 \cdot 10^{-3}$.
    The right figure shows the measured simulation growth rates with circles and the theoretical growth rates with a solid line. The measured growth rates have a less than $2.5\%$ error from theoretical predictions. VCk-GPU enables practical realistic mass ratio simulations, with VCK-CPU estimated to take 475 days walltime for the same conditions.}
\end{figure}

The acceleration-driven lower hybrid drift instability (LHDI) is a 1D-2V magnetostatic instability involving two dynamic particle species (ions and electrons) subject to an external uniform magnetic field $B_{z}$ and acceleration force $G_{y}$\cite{vogman2024}.
A sinusoidal perturbation is applied to drifting Maxwellian distributions
\begin{gather}
  f_{s}(x, v_{x}, v_{y}, t=0) = \frac{1}{2 \pi v_{Ts}^{2}} \exp
  \left(
    -\frac{
      \left(
        \vec{v} - \vec{u}_{s}
      \right)^{2}}{2 v_{Ts}^{2}}
  \right) 
  \left(
    1 + \delta_{s} \sin
    \left(
      k x
    \right)
  \right)\\
  \begin{aligned}
  u_{s,x} &= \frac{G_{y}}{\Omega_{s}} & \Omega_{s} &= \frac{q_{s} B_{z}}{m_{s}}
  \end{aligned}
\end{gather}
where $\vec{u}_{s}$ is the drift velocity of each species, $k$ is the perturbation wavenumber, and the periodic domain $x \in [0, L]$ fits one period of the perturbation $k = 2 \pi / L$.
The reference velocity is set to the Alfv\`{e}n velocity $v_{0} = \sqrt{B_{0}^{2} / (\mu_{0} m_{0} n_{0})}$ and the reference length scale is consequently $v_{0} / \omega_{p0}$.
The dispersion relation for this configuration is given by
\begin{gather}
  0 = 1 + \sum_{s} 
    \frac{\omega_{ps}^{2}}{\Omega_{s}^{2}} \int^{2\pi}_{0}
    \frac{\exp(j W_{s} \phi)}{1 - \exp(2 \pi j W_{s})} \sin \phi
    \int^{\infty}_{0} f_{0}(v_{\perp}) J_{0}
    \left(
      \frac{2 k v_{\perp}}{\Omega_{s}} \sin
      \left(
        \frac{\phi}{2}
      \right)
    \right)
    v_{\perp} d v_{\perp}      
    d\phi\\
  \begin{aligned}
  W_{s} &= \frac{\omega}{\Omega_{s}} - \frac{k G_{y}}{\Omega^{2}_{s}}
  \end{aligned}
\end{gather}
There are five key parameters of interest for acceleration driven LHDI including the drift to ion thermal velocity ratio $v_{D} / v_{Ti}$, electron cyclotron to plasma frequency ratio $|\Omega_{e} / \omega_{pe}|$, particle species mass ratio $m_{r} = m_{i} / m_{e}$, particle species temperature ratio $T_{i} / T_{e}$, and plasma pressure versus magnetic field pressure $\beta = 2 (n_{i} T_{i} + n_{e} T_{e}) / B^{2}$, where $v_{D} = |u_{i,x} - u_{e,x}|$, and $v_{Ts} = \sqrt{T_{s} / m_{s}}$\cite{vogman2024}.
Results shown below use $\delta_{e} = 10^{-3}$, $\delta_{i} = 0$, $v_{D} / v_{Ti} = 9 + 9 / m_{r}$, $|\Omega_{e}/\omega_{pe}| = 10^{-2} \sqrt{m_{r}}$, $T_{i} / T_{e} = 1$, and $\beta = 2.5 \cdot 10^{-3}$.
The wavenumber corresponding to the fastest growing unstable mode is perturbed.
This wavenumber versus particle species mass ratio is shown in Fig. \ref{fig:glhdi_wavenumber}.
A domain cell resolution of $256 \times 1024 \times 1024$ is used, with velocity boundaries chosen as
\begin{gather}
  \vec{v}_{\max,s} = \vec{u}_{s} \pm \alpha_{s} v_{Ts}
\end{gather}
with $\alpha_{i} = 12.14$ and $\alpha_{e} = 18.21$ if $m_{i}/m_{e} < 100$, or $\alpha_{e} = 6.07$ for $m_{i}/m_{e} \ge 100$ to ensure there is negligible transport of the distribution function across the $v_{\max}$ boundaries.
Figure \ref{fig:glhdi_validation} validates that the correct growth rates are obtained for various mass ratios with less than $2.5\%$ error from theoretical values.
High velocity space resolution of 1024 cells in each dimension is required to obtain correct growth rates for mass ratios $m_{r} \geq 1600$. For example, using a resolution of $256 \times 512 \times 512$ will result in no measurable instability growth.

At a mass ratio of 1836:1 VCK-GPU took about 79 hours to complete using four nodes on Lassen (16 GPUs total, see Table \ref{tbl:machines} for Lassen specifications).
Running the same conditions with VCK-CPU with four nodes on Lassen would take an estimated 475 days walltime, which implies a 144x speedup for VCK-GPU over VCK-CPU.
Increasing VCK-CPU to use 16 nodes reduces the estimated time to 119 days, with VCK-GPU still being 36x faster with a quarter of the allocated compute nodes.
The reduced compute resource requirements of VCK-GPU can also be utilized to increase simulation throughput for parametric studies.
By using just a single node per simulation on Lassen (four GPUs per simulation) a 625 simulation parametric scan was completed using 54 compute node days of allocated compute time with VCK-GPU, with individual simulations completing in approximately two hours.
The same scan run using VCK-CPU is predicted to take 50 compute node years with individual simulations completing in 11 hours.
VCK-GPU was able to facilitate a 341x increase in simulation throughput compared to VCK-CPU.

\subsection{Nonlinear Landau damping}

\begin{figure}[H]
  \centering
  \begin{subfigure}[t]{.49\textwidth}
    \centering{}
    \includegraphics[width=\textwidth]{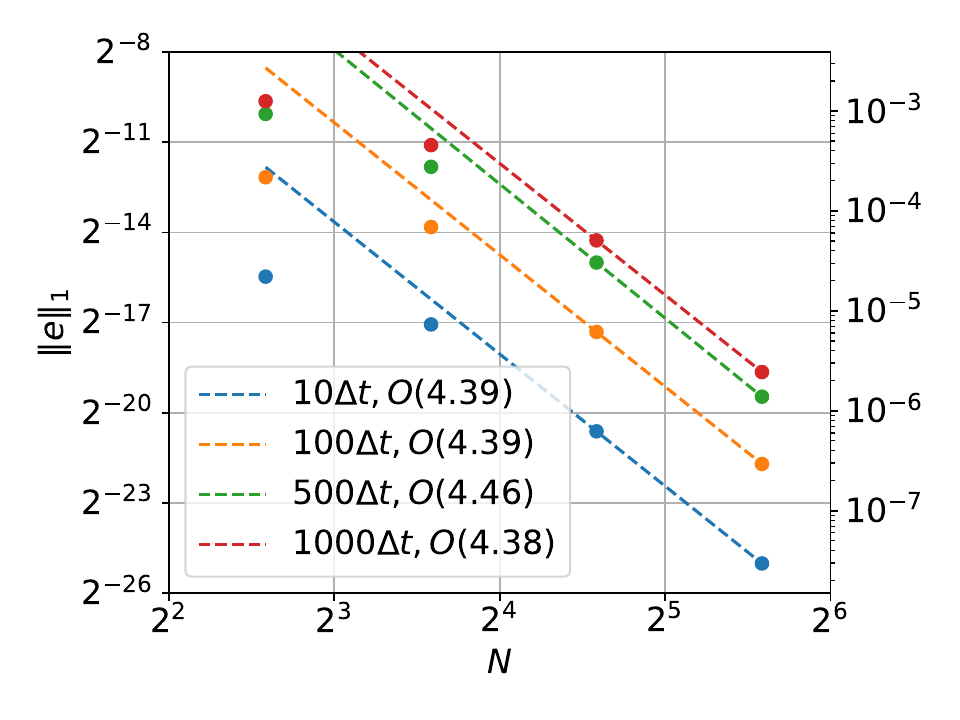}
    \caption{}
    \label{fig:landau_convergence}
  \end{subfigure}
  \begin{subfigure}[t]{.49\textwidth}
    \centering{}
    \includegraphics[width=\textwidth]{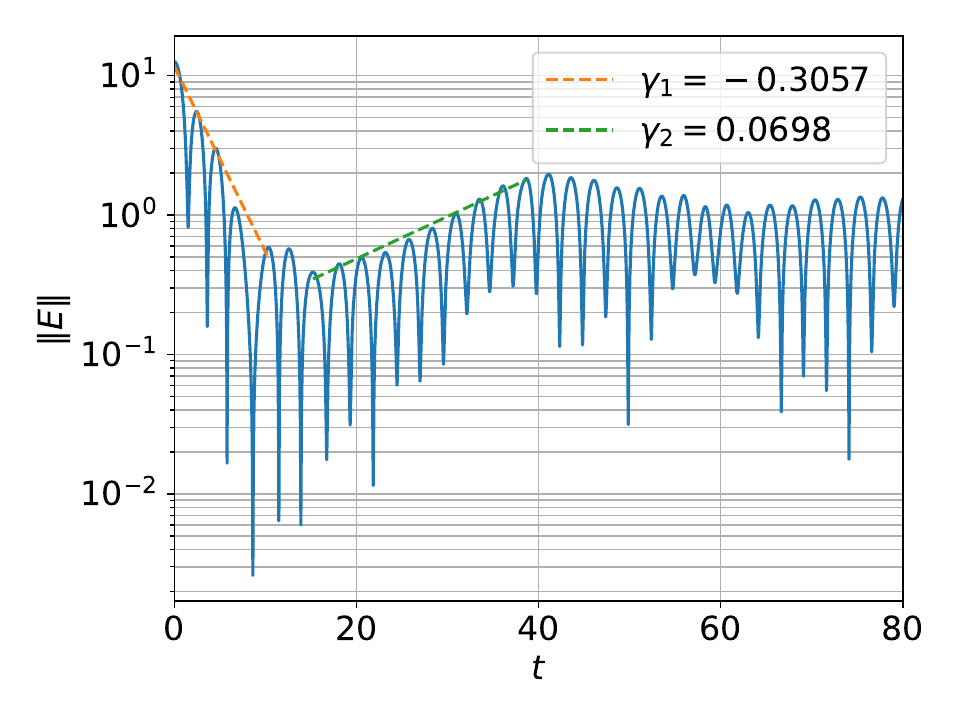}
    \caption{}
    \label{fig:landau_validation}
  \end{subfigure}
  \caption{The left figure shows the discretization error $\|e\|_{1}$ (see Eq. \eqref{eq:err_metric}) during the linear Landau damping phase versus the number of cells $N$ along each dimension. A fixed timestep $\Delta t = 10^{-3}$, much smaller than the largest stable timestep, is used.
    Fourth-order numerical accuracy is maintained over multiple time steps.
    The right figure tracks the electric field energy versus time.
    The initial 2D-2V linear strong Landau damping rate and first rebound rate match theoretical predictions. Note that presented rates are half of some literature\cite{cheng1976} which measure these using electric field energy $U_{E} = \int E^{2}/2 d\vec{x}$ rather than the electric field amplitude. A domain with $v_{\max} = \pm 8$ and 128 cells along each dimension is used.}
\end{figure}

Landau damping is an electrostatic phenomenon involving a single dynamic electron species with a uniform neutralizing static background ion species\cite{cheng1976}.
Landau damping can be extended from 1D-1V to 2D-2V and 3D-3V\cite{filbet2003}.
We consider the 2D-2V strong Landau damping version as studied by Filbet et al.\cite{filbet2003} and Einkemer\cite{einkemmer2019}.
Take an initial non-drifting perturbed Maxwellian distribution
\begin{gather}
  f(x, y, v_{x}, v_{y}, t=0) = \frac{1}{2 \pi} \exp
  \left(
    -\frac{1}{2}
      \left(
        v^{2}_{x}
        + v^{2}_{y}
      \right)
  \right) (1 + \alpha \cos(k_{x} x) + \alpha \cos(k_{y} y))
\end{gather}
with $\alpha = 0.5$, and $k_{x} = k_{y} = 0.5$ in a periodic box with side length $L_{x} = L_{y} = 4 \pi$.
The reference velocity is set to the thermal velocity $v_{0} = \sqrt{T_{0} / m_{0}}$ and the reference length scale is consequently $v_{0} / \omega_{p0}$.
Figure \ref{fig:landau_convergence} demonstrates that fourth-order numerical convergence is maintained during the initial linear phase. Higher than fourth-order initialization enables measuring error accumulated per timestep.
Figure \ref{fig:landau_validation} shows that the obtained electric field amplitude has the correct theoretical Landau damping rates, and that the non-linear phase matches those obtained by others\cite{nakamura1999, einkemmer2019, filbet2003, cheng1976, manfredi1997}.
The simulation took four V100 GPUs on a single node $92~\mathrm{ms}$ per timestep with 7740 timesteps for a total of 11.9 minutes walltime.
A $150^{4}$ cell domain, with approximately $0.5\cdot 10^{9}$ DOFs on a single V100 GPU, takes $305~\mathrm{ms}$ per timestep, which is approximately 2.3x the time compared to the $130~\mathrm{ms}$ of fourth-order SLDG run on one V100 GPU\cite{einkemmer2020}.
The time discrepancy is due to the amount of work associated with the temporal discretization.
With Strang splitting there are a total of four operations on $f$ and one Poisson solve per timestep\cite{einkemmer2016}, while RK4 requires eight operations on $f$ and four Poisson solve per timestep.
For approximately equally computationally efficient kernels on the same number of DOFs it should be expected that RK4 would take over twice as long per timestep.
As a tradeoff Strang splitting leads to a second order accurate temporal discretization\cite{einkemmer2014} while RK4 maintains fourth-order temporal accuracy.
The exact impact on simulation walltime given a desired equivalent numerical accuracy is unclear.

\section{Performance analysis}

\label{sec:performance}

Performance data was gathered using Lassen and Quartz at Lawrence Livermore National Laboratory (LLNL).
Specifications for each machine are presented in Table \ref{tbl:machines}, including the number of floating-point operations per second (FLOPS) the CPUs and GPUs theoretically can process.
Lassen has two IBM Power9 CPUs and four Nvidia V100 GPUs per compute node, with relatively high speed NVLink CPU/GPU and GPU/GPU interconnects.
Note that these specifications exclude the use of SIMD vector instructions as the CPU code does not make significant use of these.
Accepted best practices for code benchmarking are used. These include aggregating timing statistics over multiple timesteps, ranging from a few hundred up to tens of thousands, and utilizing online algorithms to ensure constant time and memory consumption during statistics collection\cite{welford1962,jain1985}.

\begin{table}[H]
  \centering{}
  \caption{HPC machine specifications used for performance benchmarks.}
  \label{tbl:machines}
  \begin{tabular}{r|r|r}
    & Lassen (L) & Quartz (Q)\\
    \hline
    \hline
    CPU model & IBM Power 9 & Intel Xeon E5-2695 v4\\
    CPU cores per node & 40\footnotemark[1] & 36\\
    CPU speed & $3.3~\text{GHz}$ & $2.6~\text{GHz}$\\
    CPU memory capacity & $256~\text{GB}$ & $128~\text{GB}$\\
    CPU memory bandwidth & $340~\text{GB/s}$ & $154~\text{GB/s}$\\
    CPU GFLOPS per node & 132 & 93.6 \\
    \hline
    GPUs per node & 4 & 0\\
    GPU model & Nvidia V100 & -\\
    GPU GFLOPS & 7000 & -\\
    GPU memory capacity & $16~\text{GB}$ & -\\
    GPU memory bandwidth & $900~\text{GB/s}$ & -\\
    GPU-CPU bandwidth & $150~\text{GB/s}$ & -\\
    \hline
    CPU peak arithmetic intensity & 3.11 & 4.86\\
    GPU peak arithmetic intensity & 62.22 & -\\
    \hline
    Network speed & $25~\text{GB/s}$ & $21~\text{GB/s}$\\
    Network topology & 1.5:1 tapered fat-tree & Two-stage fat-tree
  \end{tabular}
\end{table}
\footnotetext[1]{Four cores are reserved by the scheduler for MPI performance reasons\cite{dahm2020}, only 40 are available for compute}

\subsection{Strong scaling}

Two cases based on the two species acceleration driven LHDI problem (see Sec. \ref{sec:glhdi}) were used to test strong scaling: a 1D-2V domain with 768 cells along each dimension, and a 2D-2V domain with 128 cells along each dimension.
These domain resolutions were chosen as they were approximately the largest size problems which could fit into GPU memory with one GPU per species.
Strong scaling is then measured by increasing the amount of compute nodes used while keeping the number of degrees of freedom fixed, giving a measure of how parallelization improves time to solution.

\begin{figure}[H]
  \begin{subfigure}[t]{.49\textwidth}
    \centering{}
    \includegraphics[width=\textwidth]{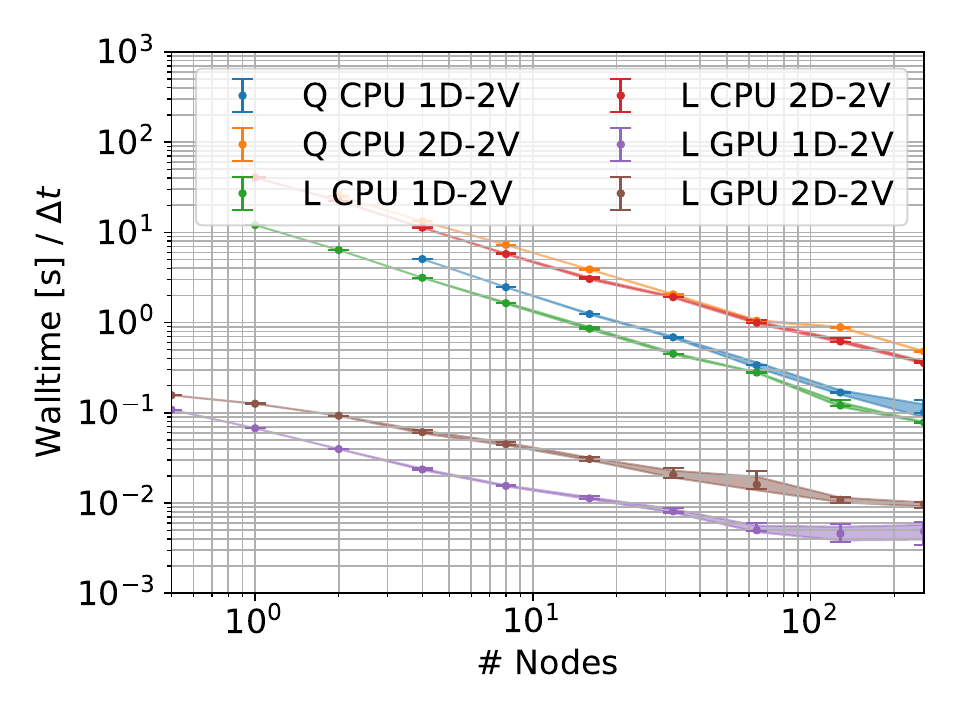}
  \end{subfigure}
  \begin{subfigure}[t]{.49\textwidth}
    \centering{}
    \includegraphics[width=\textwidth]{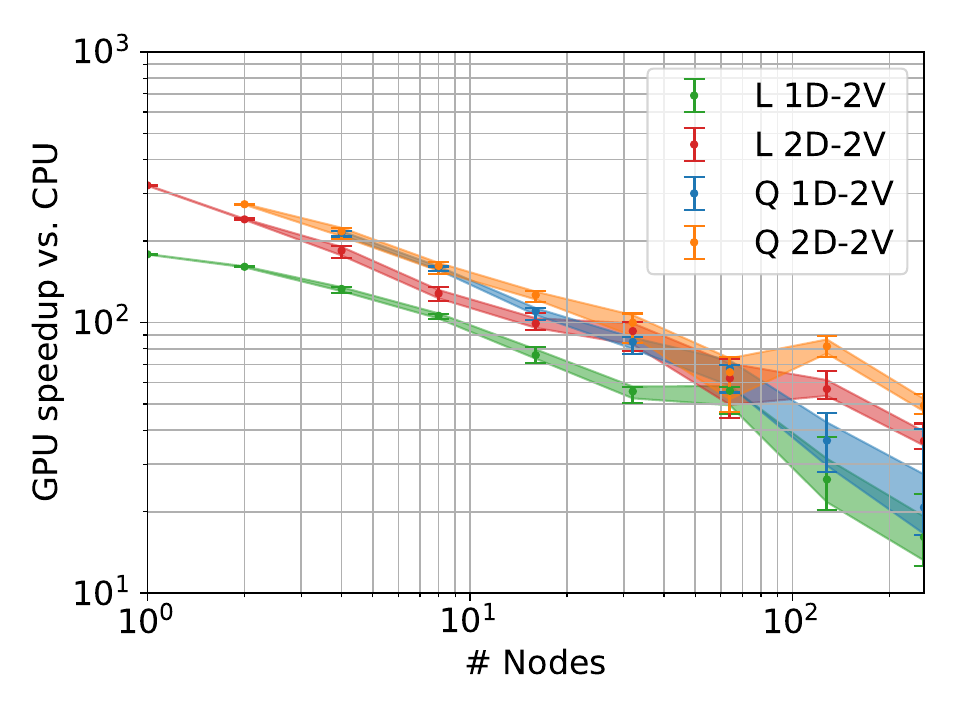}
  \end{subfigure}

  \caption{Strong scaling performance of VCK-CPU and VCK-GPU. The left plot shows walltime spent per timestep and the right plot shows the GPU code's speedup over the CPU code utilizing the same number of compute nodes. VCK-CPU has a near perfect strong scaling indicating that it is strongly compute bound, while VCK-GPU is communication bound. Error bars denote the 5th, 50th, and 95th percentiles and the 95\% confidence interval is shaded. Significant noise is introduced based on network bandwidth utilization from other concurrent jobs.}
  \label{fig:strong_speedup}
\end{figure}

\begin{figure}[H]
  \centering
  \begin{subfigure}[t]{.49\textwidth}
    \centering{}
    \includegraphics[width=\textwidth]{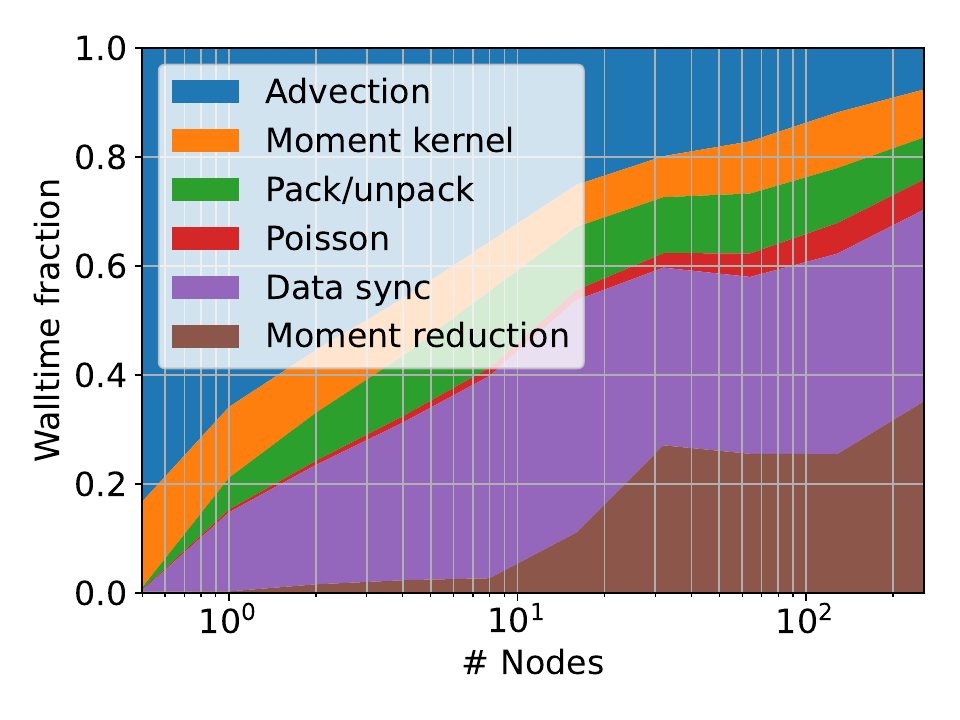}
    \caption{1D-2V}
  \end{subfigure}
  \begin{subfigure}[t]{.49\textwidth}
    \centering{}
    \includegraphics[width=\textwidth]{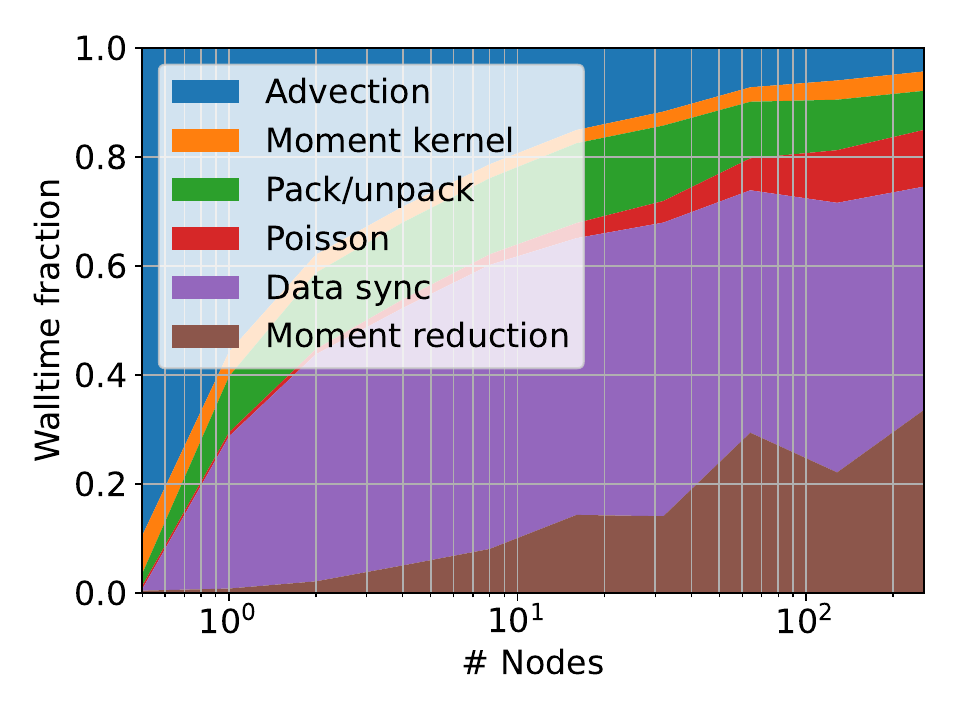}
    \caption{2D-2V}
  \end{subfigure}

  \caption{Fraction of walltime spent on each component while strong scaling. At 256 nodes communication related tasks account for approximately 70\% of the walltime.}
  \label{fig:strong_scale_fraction}
\end{figure}

Figure \ref{fig:strong_speedup} shows strong scaling performance data measured on the original VCK-CPU code as well as VCK-GPU.
Comparisons are made per allocated compute node on the respective machine, with a half node GPU run indicating that just two of the four GPUs on a Lassen node were used.
A VCK-GPU run with a single GPU per species already exceeds the performance of VCK-CPU run on 256 nodes.
The order of magnitude more time spent on computation for the CPU code hides the data communication costs from overall strong scaling trends, despite having a similar data communication volume as the GPU code.
Unlike VCK-CPU, VCK-GPU's scaling performance is strongly communication bound.
Figure \ref{fig:strong_scale_fraction} demonstrates that as the number of nodes increases, the dominant fraction of time spent rapidly shifts from compute time towards communication time for VCK-GPU.
As a result, VCK-GPU strong scales approximately proportional to the ghost sync data communication volume.
At 256 nodes in 1D-2V VCK-GPU is between 16-40x faster than VCK-CPU, while in 2D-2V VCK-GPU is between 34-54x faster.
Performing a comparison at 64 nodes where the GPU code walltime begins to saturates gives speedups of 54-96x in 1D-2V and 44-74x in 2D-2V.

\subsection{Weak scaling}


\begin{figure}[H]
  \centering
  \begin{subfigure}[t]{.49\textwidth}
    \centering{}
    \includegraphics[width=.97\textwidth]{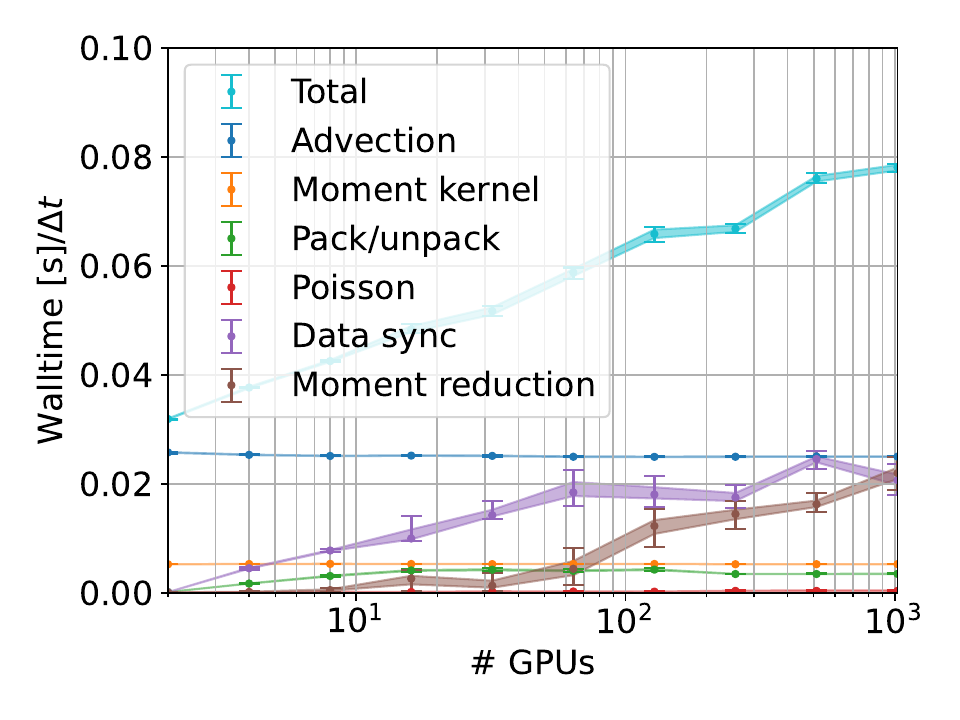}
    \caption{1D-2V}
  \end{subfigure}
  \begin{subfigure}[t]{.49\textwidth}
    \centering{}
    \includegraphics[width=.97\textwidth]{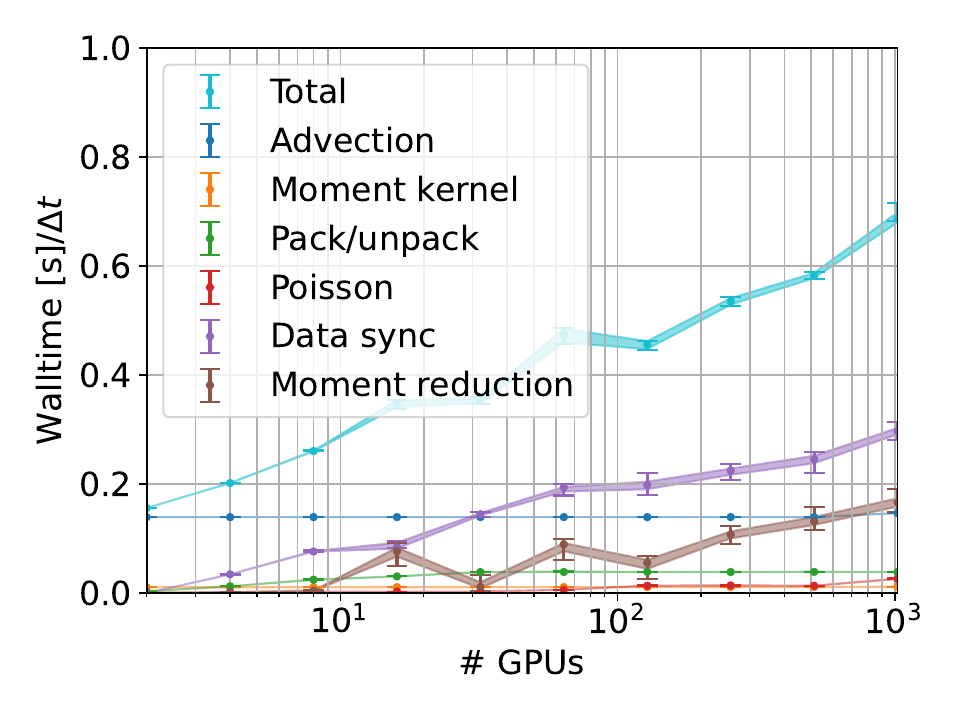}
    \caption{2D-2V}
  \end{subfigure}

  \caption{Walltime spent on each component of the Vlasov-Poisson solver by VCK-GPU while weak scaling. In 1D-2V VCK-GPU has a 2.1x increase in walltime per timestep at 256 GPUs and 2.4x increase in walltime per timestep at 1024 GPUs compared to using two GPUs.
In 2D-2V VCK-GPU has a 3.4x increase in walltime per timestep at 256 GPUs and 4.4x increase in walltime per timestep at 1024 GPUs compared to using two GPUs. Error bars denote the 5th, 50th, and 95th percentiles and the 95\% confidence interval is shaded. Significant noise is introduced based on network bandwidth utilization from other concurrent jobs.}
  \label{fig:weak_scale}
\end{figure}

Two cases based on the two-species acceleration driven LHDI problem (see Sec. \ref{sec:glhdi}) were used to test weak scaling: a 1D-2V domain with $512^{3}$ cells per GPU, and a 2D-2V domain with $128^{4}$ cells per GPU.
Weak scaling is then measured by simultaneously increasing the number of degrees of freedom with allocated compute resources, giving an indication of how large of problems parallelization enables solving.
As the number of GPUs used is doubled the global domain size is doubled in one direction at a time following the order $x$, $y$, $v_{x}$, and $v_{y}$.
Figure \ref{fig:weak_scale} shows the walltime per timestep spent on the different components while weak scaling.
In 1D-2V VCK-GPU has a 2.1x increase in walltime per timestep at 256 GPUs and 2.4x increase in walltime per timestep at 1024 GPUs compared to using two GPUs.
In 2D-2V VCK-GPU has a 3.4x increase in walltime per timestep at 256 GPUs and 4.4x increase in walltime per timestep at 1024 GPUs compared to using two GPUs.
In both cases the increase in walltime per timesteps is primarily due to the data communication cost, which has significant noise dependent on the assigned nodes and network utilization from other jobs running on the machine.
The increases in walltime is acceptable considering VCK-GPU is strongly communication bound. There is some scope to investigate why the moment reduction portion consumes a significant portion of walltime as the number of GPUs increases as there is relatively little data transferred.

\section{Conclusion}

\label{sec:conclusion}

We have shown that a fourth-order finite-volume Vlasov-Poisson solver can be implemented efficiently using multiple GPUs.
The VCK-GPU code was verified and validated to be fourth-order accurate and was demonstrated to match theoretical predictions for a variety of benchmark problems.
These include single and multi-species dynamics, and used various numbers of phase space dimensions (1D-1V, 1D-2V, and 2D-2V).
We also presented statistically rigorous strong and weak scaling code performance metrics and provided theoretical analysis on performance scaling bounds for various parts of our implementation.
The 3/8ths rule RK4 method was found to have the lowest computational costs for the fourth-order finite-volume method, simultaneously having the largest effective CFL, lowest memory footprint, and smallest effective memory bandwidth requirements.
Improved stable timestep bounds were also presented which in practice have allowed taking 20-40\% larger timesteps.

We presented near-optimal performance-portable methods for computing velocity moments of the distribution function to facilitate high-dimensional to low-dimensional coupling required for the Vlasov-Poisson system.
We also demonstrate that combining the full RK spatial-temporal hyperbolic advance into a single kernel per RK stage offers the best overall performance by minimizing data movement.
Similarly, minimizing how much ghost cell data is communicated over MPI is critical as the GPU compute time is sufficiently small that system network communication limits become the dominant factor for scaling beyond a single node.
We demonstrate that allowing partitioning in all dimensions and transmitting only ghost cells required for computations significantly improves the scalability of the code, especially for strong scaling.
The impact of minimizing ghost cell data transferred increases with total number of dimensions.
There are techniques for hiding data communication costs by partitioning the computation, such that some computations can happen concurrently with data communication\cite{reddell2016}.
Unfortunately in our testing these techniques significantly degraded the GPU utilization and kernel efficiency, leading to a gross increase of complexity for negligible performance gains.
While not explored here, the use of single precision can potentially lead to significant performance gains\cite{einkemmer2020}.

We show that VCK-GPU can perform realistic two-species proton/electron mass ratio ($m_{i} / m_{e} = 1836$) simulations of the ion dynamics in the acceleration-driven LHDI problem with a 144x speedup over VCK-CPU utilizing the same number of compute nodes or 36x speedup allowing VCK-CPU to use four times as many compute nodes.
The efficiency of the GPU code enables performing large-scale parallel parametric scans with a 341x increase in simulation throughput.
These speedups can be utilized to perform convergence studies of high dimensional kinetic problems which were not feasible using the original CPU code, as in the case of the 2D-2V Kelvin-Helmholtz instability\cite{vogman2020,vogman2021}.
Similarly the improved simulation throughput can help with characterizing the effect of microturbulence over a wide parameter space to develop anomalous transport models for complex kinetic physics\cite{vogman2024}.
 
\section*{Acknowledgments}

This work was performed under the auspices of the U.S. Department of Energy by Lawrence Livermore National Laboratory under contract DE-AC52-07NA27344. Lawrence Livermore National Security, LLC.
The authors would like to thank Sean Miller, Noah Reddell, and Uri Shumlak for illuminating discussions on performance metrics and portability tools.

\begin{appendices}
  
\section{Derivation of more accurate fourth-order finite-volume CFL stability criteria}

\label{sec:cfl_deriv}

\begin{figure}[H]
  \centering{}
  \begin{subfigure}[t]{.49\linewidth}
    \centering{}
    \includegraphics[width=.9\textwidth]{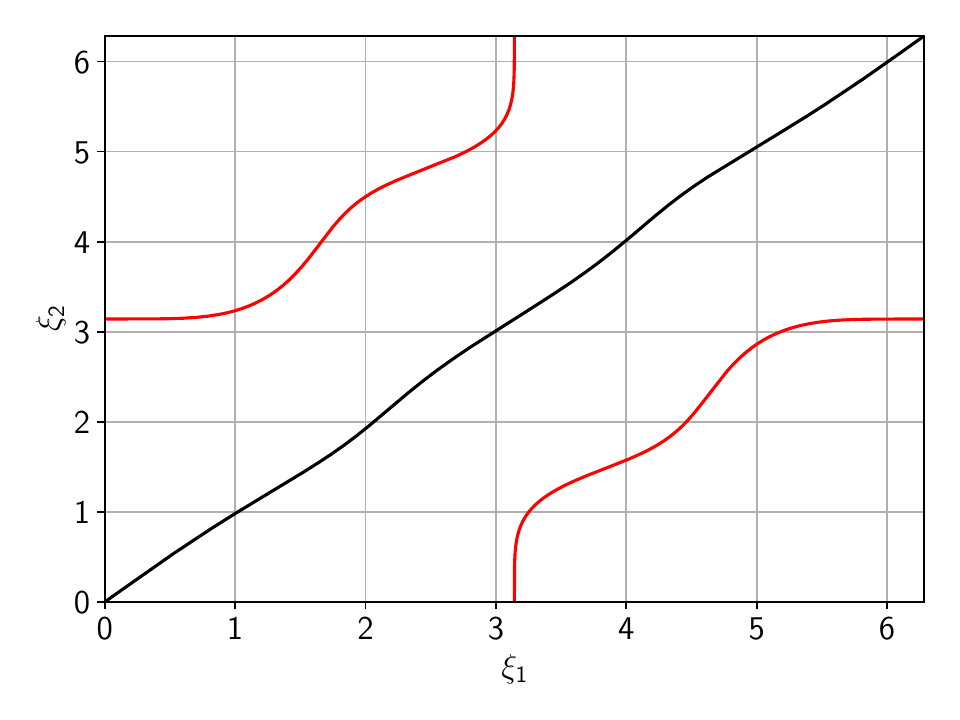}
    \caption{}
  \end{subfigure}
  \begin{subfigure}[t]{.49\linewidth}
    \centering{}
    \includegraphics[width=.9\textwidth]{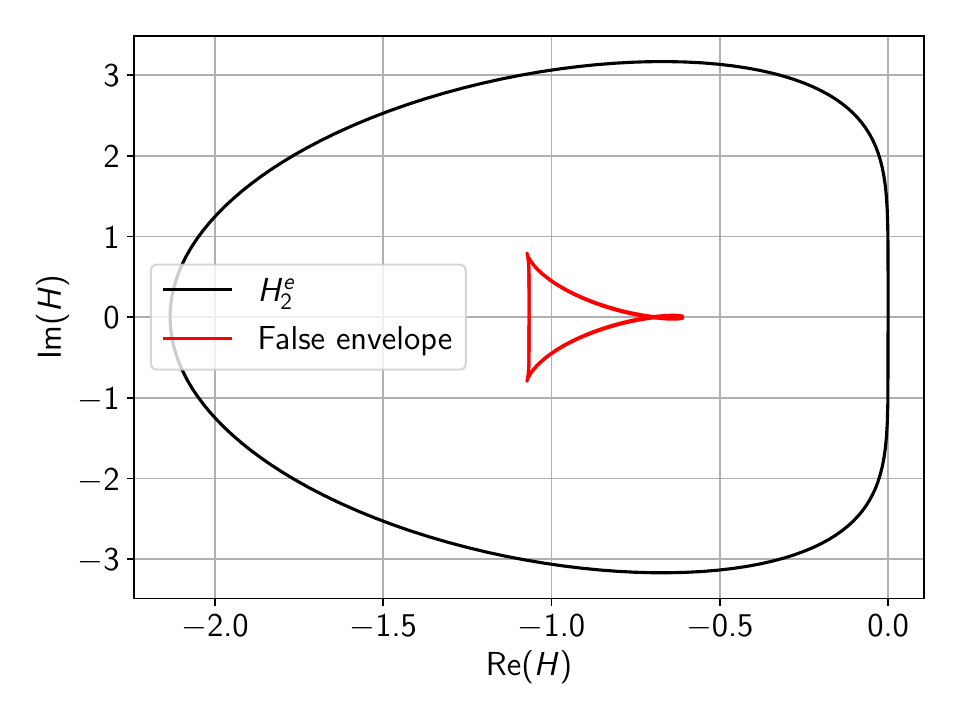}
    \caption{}
    \label{fig:envelope_enclose}
  \end{subfigure}
  \caption{The envelope $H^{e}_{2}$ for $A^{1}/h_{1} = A^{2}/h_{2} = 1$ in black and false envelope solution in red. Left figure shows all solutions $(\xi_{1}, \xi_{2})$ to Eq. \eqref{eq:envelope_cond}, while the right figure shows that the red curve is entirely enclosed by the true envelope $H^{e}_{2}$.}
  \label{fig:envelope_params}
\end{figure}

\begin{figure}[H]
  \centering
  \begin{subfigure}[t]{.49\linewidth}
    \includegraphics[width=\textwidth]{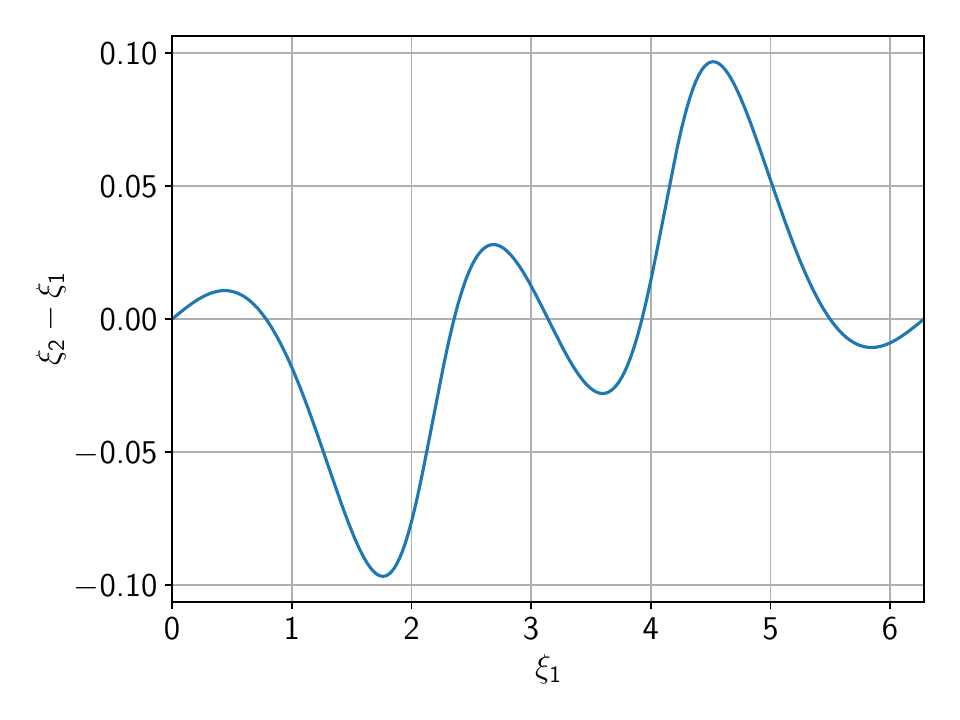}
  \end{subfigure}
  \begin{subfigure}[t]{.49\linewidth}
    \includegraphics[width=\textwidth]{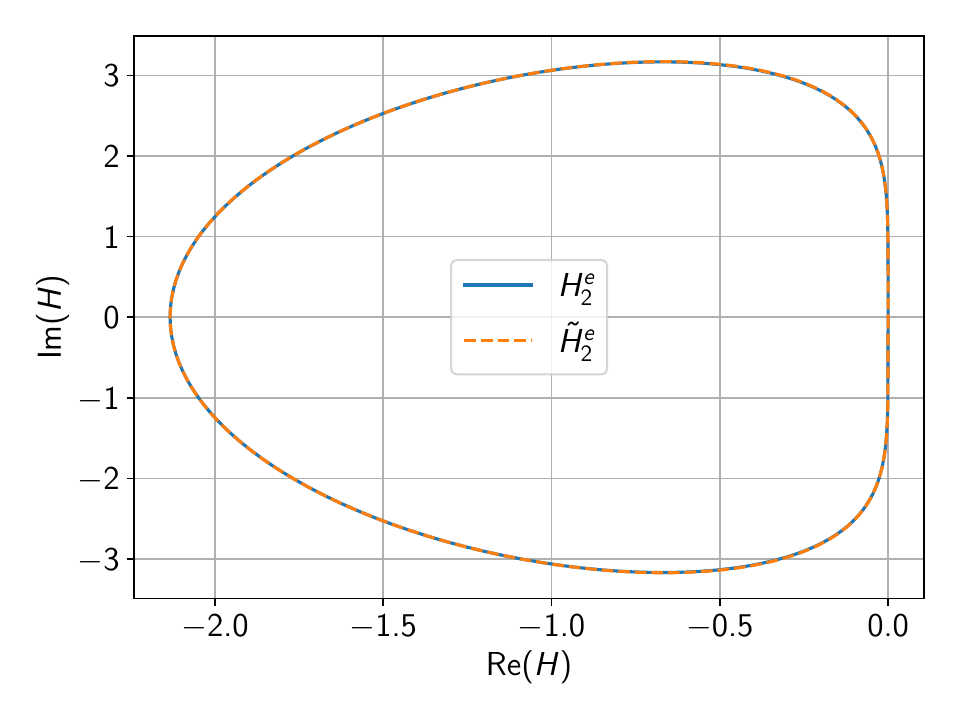}
  \end{subfigure}

  \caption{Comparison of the scaled approximate envelope $\tilde{H}^{e}_{2}$, given by Eq. \eqref{eq:approx_envelope}, with the true envelope $H^{e}_{2}$ for $a_{1}/h_{1} = a_{2}/h_{2} = 1$. The left plot shows the deviation of the true solution away from the approximation $\xi_{1} \approx \xi_{2}$. The right plot shows that the net result of this deviation on $H^{e}_{2}$ is visually indistinguishable from $\tilde{H}^{e}_{2}$ with $H^{e}_{2}$ being enclosed by $\tilde{H}^{e}_{2}$.}
  \label{fig:envelope_approx}
\end{figure}

Consider the linearized semi-discrete form of Eq. \eqref{eq:spatial_discr} such that $C_{\vec{i}} = 0$\cite{chaplin2017}.
Without loss of generality assume that the advection speed $A^{d}$ in any direction $d$ is positive.
To perform a Von-Neumann stability analysis on the semi-discrete system\cite{chaplin2017}, make the approximation that $f$ can be split into a temporal component $\Gamma(t)$ and a spatial perturbation
\begin{gather}
  f(t,\vec{r}) = \Gamma(t) \exp(j \vec{k} \cdot \vec{r})\label{eq:f_linearized}
\end{gather}
where $\vec{k}$ is the perturbation wave vector, and $\vec{r}$ is the position-velocity phase space coordinate variable.
Substituting Eq. \eqref{eq:f_linearized} into the linearized semi-discrete form of Eq. \eqref{eq:spatial_discr} gives
\begin{gather}
  \partial_{t} \Gamma(t) = \Gamma(t) \sum^{D}_{\alpha=1} P_{\alpha}(h_{\alpha} k_{\alpha})\\
  P_{\alpha}(\xi) = \frac{A^{\alpha}}{60 h_{\alpha}} P(\xi)\\
  P(\xi) = 
  2 \exp(-3j\xi) - 15 \exp(-2 j \xi) + 60 \exp(-j \xi) - 20 - 30 \exp(j \xi) + 3 \exp(2 j \xi)
\end{gather}
where $P$ is a periodic parametric curve in the complex plane as a function of $\xi \in [0,2\pi]$, $h_{\alpha}$ is the cell spacing in direction $\alpha$, $A^{\alpha}$ is the advection speed in direction $\alpha$, $k_{\alpha}$ is wavenumber in direction $\alpha$, and $j$ is the imaginary unit.
Define $H^{e}_{\beta}$ to be the envelope of dimensions one through $\beta$ of $\sum_{\alpha=1}^{\beta} P_{\alpha}(h_{\alpha} k_{\alpha})$, with $H^{e}_{1} = P_{1}$ and $1 \le \beta \le D$.
We will consider the discretized system to be stable if the interior region defined by the envelope $H^{e}_{D}$, which includes all phase space dimensions one through $D$, is within the region of absolute stability for a given temporal discretization method.
The envelope $H^{e}_{\beta+1}$ can be constructed from $H^{e}_{\beta}$ and $P_{\beta+1}$ by defining a family of curves $H_{\beta+1,\xi_{1}}(\xi_{1}, \xi_{2})$ such that the family represents translations of the curve $P_{\beta+1}(\xi_{2})$ by $H^{e}_{\beta}(\xi_{1})$.
A necessary but insufficient condition for points $(\xi_{1}, \xi_{2})$ on the envelope $H^{e}_{\beta+1}$ is that two infinitesimally adjacent curves $H_{\beta+1,\xi_{1}}$ and $H_{\beta+1,\xi_{1} + \Delta \xi_{1}}$ are tangent at $(\xi_{1}, \xi_{2})$, or\cite{yates1952}
\begin{gather}
  \partial_{\xi_{1}} \real(H_{\beta+1,\xi_{1}}(\xi_{1},\xi_{2})) \partial_{\xi_{1}} \imag(H_{\beta+1,\xi_{1}}(\xi_{1},\xi_{2})) - \partial_{\xi_{1}} \real(H_{\beta+1,\xi_{1}}(\xi_{1},\xi_{2})) \partial_{\xi_{2}} \imag(H_{\beta+1,\xi_{1}}(\xi_{1},\xi_{2})) = 0\label{eq:envelope_cond}
\end{gather}
There are two distinct curves formed which satisfy Eq. \eqref{eq:envelope_cond}, colored red and black in Fig. \ref{fig:envelope_params}. Of these two curves Fig. \ref{fig:envelope_enclose} shows that the true envelope is the black curve as it encompasses the red curve.
Despite $\xi_{1} = \xi_{2}$ not being an exact solution, we can approximate $\xi_{1} \approx \xi_{2}$ resulting in the true envelope to be enclosed by the curve
\begin{gather}
  \tilde{H}^{e}_{\beta+1} = P \sum_{\alpha=1}^{\beta+1} \frac{A^{\alpha}}{60 h_{\alpha}}
\end{gather}
where $\tilde{H}^{e}_{\beta+1}$ is a scaled version of the curve $H^{e}_{1}$.
Figure \ref{fig:envelope_approx} demonstrates the similarity between $H^{e}_{2}$ and $\tilde{H}^{e}_{2}$.
Repeating this process to computing the envelope $H^{e}_{D}$, we arrive at an alternative stability system $\partial_{t} \Gamma(t) \approx \Gamma(t) \tilde{H}^{e}_{D}$ where
\begin{gather}
  \tilde{H}^{e}_{D} = 
  \left\|
    \frac{\vec{A}}{60 \vec{h}}
  \right\|_{1} P\label{eq:approx_envelope}
\end{gather}
This stability condition is simultaneously easy to evaluate while allowing larger stable timesteps than $\|\vec{A} / (60 \vec{h})\|_{\infty} D P$\cite{vogman2014}.
Theoretically the $L^{1}$ norm used in $\tilde{H}^{e}_{D}$ could allow $D$ times larger timesteps compared to the $L^{\infty}$ norm, though in practice we have observed approximately $20-40\%$ larger timesteps.

\end{appendices}

%%Vancouver style references.
\bibliographystyle{model1-num-names}
\bibliography{refs}

\begin{thebibliography}{74}
\expandafter\ifx\csname natexlab\endcsname\relax\def\natexlab#1{#1}\fi
\providecommand{\url}[1]{\texttt{#1}}
\providecommand{\href}[2]{#2}
\providecommand{\path}[1]{#1}
\providecommand{\DOIprefix}{doi:}
\providecommand{\ArXivprefix}{arXiv:}
\providecommand{\URLprefix}{URL: }
\providecommand{\Pubmedprefix}{pmid:}
\providecommand{\doi}[1]{\href{http://dx.doi.org/#1}{\path{#1}}}
\providecommand{\Pubmed}[1]{\href{pmid:#1}{\path{#1}}}
\providecommand{\bibinfo}[2]{#2}
\ifx\xfnm\relax \def\xfnm[#1]{\unskip,\space#1}\fi
%Type = Article
\bibitem[{Vogman et~al.(2014)Vogman, Colella, and Shumlak}]{vogman2014}
\bibinfo{author}{G.~Vogman}, \bibinfo{author}{P.~Colella},
  \bibinfo{author}{U.~Shumlak},
\newblock \bibinfo{title}{Dory–guest–harris instability as a benchmark for
  continuum kinetic vlasov–poisson simulations of magnetized plasmas},
\newblock \bibinfo{journal}{Journal of Computational Physics}
  \bibinfo{volume}{277} (\bibinfo{year}{2014}) \bibinfo{pages}{101--120}.
  \DOIprefix\doi{10.1016/j.jcp.2014.08.014}.
%Type = Article
\bibitem[{Palmroth et~al.(2018)Palmroth, Ganse, Pfau-Kempf, Battarbee, Turc,
  Brito, Grandin, Hoilijoki, Sandroos, and von Alfthan}]{palmroth2018}
\bibinfo{author}{M.~Palmroth}, \bibinfo{author}{U.~Ganse},
  \bibinfo{author}{Y.~Pfau-Kempf}, \bibinfo{author}{M.~Battarbee},
  \bibinfo{author}{L.~Turc}, \bibinfo{author}{T.~Brito},
  \bibinfo{author}{M.~Grandin}, \bibinfo{author}{S.~Hoilijoki},
  \bibinfo{author}{A.~Sandroos}, \bibinfo{author}{S.~von Alfthan},
\newblock \bibinfo{title}{Vlasov methods in space physics and astrophysics},
\newblock \bibinfo{journal}{Living Reviews in Computational Astrophysics}
  \bibinfo{volume}{4} (\bibinfo{year}{2018}) \bibinfo{pages}{1}.
  \DOIprefix\doi{10.1007/s41115-018-0003-2}.
%Type = Inproceedings
\bibitem[{Taccogna et~al.(2004)Taccogna, Longo, Capitelli, and
  Schneider}]{taccogna2004}
\bibinfo{author}{F.~Taccogna}, \bibinfo{author}{S.~Longo},
  \bibinfo{author}{M.~Capitelli}, \bibinfo{author}{R.~Schneider},
\newblock \bibinfo{title}{Fully kinetic particle-in-cell simulation of a hall
  thruster},
\newblock in: \bibinfo{editor}{M.~Bubak}, \bibinfo{editor}{G.~D. van Albada},
  \bibinfo{editor}{P.~M.~A. Sloot}, \bibinfo{editor}{J.~Dongarra} (Eds.),
  \bibinfo{booktitle}{Computational Science - ICCS 2004},
  \bibinfo{publisher}{Springer Berlin Heidelberg}, \bibinfo{address}{Berlin,
  Heidelberg}, \bibinfo{year}{2004}, pp. \bibinfo{pages}{588--595}.
  \DOIprefix\doi{10.1007/978-3-540-25944-2_76}.
%Type = Article
\bibitem[{Rinderknecht et~al.(2018)Rinderknecht, Amendt, Wilks, and
  Collins}]{rinderknecht2018}
\bibinfo{author}{H.~G. Rinderknecht}, \bibinfo{author}{P.~A. Amendt},
  \bibinfo{author}{S.~C. Wilks}, \bibinfo{author}{G.~Collins},
\newblock \bibinfo{title}{Kinetic physics in icf: present understanding and
  future directions},
\newblock \bibinfo{journal}{Plasma Physics and Controlled Fusion}
  \bibinfo{volume}{60} (\bibinfo{year}{2018}) \bibinfo{pages}{064001}.
  \DOIprefix\doi{10.1088/1361-6587/aab79f}.
%Type = Article
\bibitem[{Saarelma et~al.(2017)Saarelma, Martin-Collar, Dickinson, McMillan,
  Roach, {MAST team}, and {JET Contributors}}]{saarelma2017}
\bibinfo{author}{S.~Saarelma}, \bibinfo{author}{J.~Martin-Collar},
  \bibinfo{author}{D.~Dickinson}, \bibinfo{author}{B.~F. McMillan},
  \bibinfo{author}{C.~M. Roach}, \bibinfo{author}{{MAST team}},
  \bibinfo{author}{{JET Contributors}},
\newblock \bibinfo{title}{Non-local effects on pedestal kinetic ballooning mode
  stability},
\newblock \bibinfo{journal}{Plasma Physics and Controlled Fusion}
  \bibinfo{volume}{59} (\bibinfo{year}{2017}) \bibinfo{pages}{064001}.
  \DOIprefix\doi{10.1088/1361-6587/aa66ab}.
%Type = Article
\bibitem[{Francisquez et~al.(2023)Francisquez, Rosen, Mandell, Hakim, Forest,
  and Hammett}]{francisquez2023}
\bibinfo{author}{M.~Francisquez}, \bibinfo{author}{M.~H. Rosen},
  \bibinfo{author}{N.~R. Mandell}, \bibinfo{author}{A.~Hakim},
  \bibinfo{author}{C.~B. Forest}, \bibinfo{author}{G.~W. Hammett},
\newblock \bibinfo{title}{{Toward continuum gyrokinetic study of high-field
  mirrors}},
\newblock \bibinfo{journal}{Physics of Plasmas} \bibinfo{volume}{30}
  (\bibinfo{year}{2023}) \bibinfo{pages}{102504}.
  \DOIprefix\doi{10.1063/5.0152440}.
%Type = Article
\bibitem[{Vogman et~al.(2020)Vogman, Hammer, Shumlak, and Farmer}]{vogman2020}
\bibinfo{author}{G.~V. Vogman}, \bibinfo{author}{J.~H. Hammer},
  \bibinfo{author}{U.~Shumlak}, \bibinfo{author}{W.~A. Farmer},
\newblock \bibinfo{title}{{Two-fluid and kinetic transport physics of
  Kelvin–Helmholtz instabilities in nonuniform low-beta plasmas}},
\newblock \bibinfo{journal}{Physics of Plasmas} \bibinfo{volume}{27}
  (\bibinfo{year}{2020}) \bibinfo{pages}{102109}.
  \DOIprefix\doi{10.1063/5.0014489}.
%Type = Article
\bibitem[{Vogman and Hammer(2021)}]{vogman2021}
\bibinfo{author}{G.~V. Vogman}, \bibinfo{author}{J.~H. Hammer},
\newblock \bibinfo{title}{High-fidelity kinetic modeling of instabilities and
  gyromotion physics in nonuniform low-beta plasmas},
\newblock \bibinfo{journal}{Physics of Plasmas} \bibinfo{volume}{28}
  (\bibinfo{year}{2021}) \bibinfo{pages}{062103}.
  \DOIprefix\doi{10.1063/5.0045983}.
%Type = Article
\bibitem[{Vogman and Hammer(2024)}]{vogman2024}
\bibinfo{author}{G.~V. Vogman}, \bibinfo{author}{J.~H. Hammer},
\newblock \bibinfo{title}{Complete quasilinear model for the
  acceleration-driven lower hybrid drift instability and a computational
  assessment of its validity},
\newblock \bibinfo{journal}{Phys. Rev. E} \bibinfo{volume}{110}
  (\bibinfo{year}{2024}) \bibinfo{pages}{025201}.
  \DOIprefix\doi{10.1103/PhysRevE.110.025201}.
%Type = Article
\bibitem[{Nakamura and Yabe(1999)}]{nakamura1999}
\bibinfo{author}{T.~Nakamura}, \bibinfo{author}{T.~Yabe},
\newblock \bibinfo{title}{Cubic interpolated propagation scheme for solving the
  hyper-dimensional vlasov—poisson equation in phase space},
\newblock \bibinfo{journal}{Computer Physics Communications}
  \bibinfo{volume}{120} (\bibinfo{year}{1999}) \bibinfo{pages}{122--154}.
  \DOIprefix\doi{10.1016/S0010-4655(99)00247-7}.
%Type = Article
\bibitem[{Taitano et~al.(2013)Taitano, Knoll, Chac\'{o}n, and
  Chen}]{taitano2013}
\bibinfo{author}{W.~T. Taitano}, \bibinfo{author}{D.~A. Knoll},
  \bibinfo{author}{L.~Chac\'{o}n}, \bibinfo{author}{G.~Chen},
\newblock \bibinfo{title}{Development of a consistent and stable fully implicit
  moment method for vlasov--ampère particle in cell (pic) system},
\newblock \bibinfo{journal}{SIAM Journal on Scientific Computing}
  \bibinfo{volume}{35} (\bibinfo{year}{2013}) \bibinfo{pages}{S126--S149}.
  \DOIprefix\doi{10.1137/120881385}.
%Type = Article
\bibitem[{Ricketson and Cerfon(2016)}]{ricketson2017}
\bibinfo{author}{L.~F. Ricketson}, \bibinfo{author}{A.~J. Cerfon},
\newblock \bibinfo{title}{Sparse grid techniques for particle-in-cell schemes},
\newblock \bibinfo{journal}{Plasma Physics and Controlled Fusion}
  \bibinfo{volume}{59} (\bibinfo{year}{2016}) \bibinfo{pages}{024002}.
  \DOIprefix\doi{10.1088/1361-6587/59/2/024002}.
%Type = Article
\bibitem[{Dorr et~al.(2018)Dorr, Colella, Dorf, Ghosh, Hittinger, and
  Schwartz}]{dorr2018}
\bibinfo{author}{M.~R. Dorr}, \bibinfo{author}{P.~Colella},
  \bibinfo{author}{M.~A. Dorf}, \bibinfo{author}{D.~Ghosh},
  \bibinfo{author}{J.~A. Hittinger}, \bibinfo{author}{P.~O. Schwartz},
\newblock \bibinfo{title}{High-order discretization of a gyrokinetic vlasov
  model in edge plasma geometry},
\newblock \bibinfo{journal}{Journal of Computational Physics}
  \bibinfo{volume}{373} (\bibinfo{year}{2018}) \bibinfo{pages}{605--630}.
  \DOIprefix\doi{10.1016/j.jcp.2018.07.008}.
%Type = Article
\bibitem[{Michels et~al.(2021)Michels, Stegmeir, Ulbl, Jarema, and
  Jenko}]{michels2021}
\bibinfo{author}{D.~Michels}, \bibinfo{author}{A.~Stegmeir},
  \bibinfo{author}{P.~Ulbl}, \bibinfo{author}{D.~Jarema},
  \bibinfo{author}{F.~Jenko},
\newblock \bibinfo{title}{Gene-x: A full-f gyrokinetic turbulence code based on
  the flux-coordinate independent approach},
\newblock \bibinfo{journal}{Computer Physics Communications}
  \bibinfo{volume}{264} (\bibinfo{year}{2021}) \bibinfo{pages}{107986}.
  \DOIprefix\doi{10.1016/j.cpc.2021.107986}.
%Type = Article
\bibitem[{Einkemmer(2024)}]{einkemmer2024}
\bibinfo{author}{L.~Einkemmer},
\newblock \bibinfo{title}{Accelerating the simulation of kinetic shear alfvén
  waves with a dynamical low-rank approximation},
\newblock \bibinfo{journal}{Journal of Computational Physics}
  \bibinfo{volume}{501} (\bibinfo{year}{2024}) \bibinfo{pages}{112757}.
  \DOIprefix\doi{10.1016/j.jcp.2024.112757}.
%Type = Article
\bibitem[{Wang et~al.(2019)Wang, Ethier, Tang, Ibrahim, Madduri, Williams, and
  Oliker}]{wang2019}
\bibinfo{author}{B.~Wang}, \bibinfo{author}{S.~Ethier},
  \bibinfo{author}{W.~Tang}, \bibinfo{author}{K.~Z. Ibrahim},
  \bibinfo{author}{K.~Madduri}, \bibinfo{author}{S.~Williams},
  \bibinfo{author}{L.~Oliker},
\newblock \bibinfo{title}{Modern gyrokinetic particle-in-cell simulation of
  fusion plasmas on top supercomputers},
\newblock \bibinfo{journal}{The International Journal of High Performance
  Computing Applications} \bibinfo{volume}{33} (\bibinfo{year}{2019})
  \bibinfo{pages}{169--188}. \DOIprefix\doi{10.1177/1094342017712059}.
%Type = Inproceedings
\bibitem[{Zhao et~al.(2018)Zhao, Williams, Hall, and Johansen}]{zhao2018}
\bibinfo{author}{T.~Zhao}, \bibinfo{author}{S.~Williams},
  \bibinfo{author}{M.~Hall}, \bibinfo{author}{H.~Johansen},
\newblock \bibinfo{title}{Delivering performance-portable stencil computations
  on cpus and gpus using bricks},
\newblock in: \bibinfo{booktitle}{2018 IEEE/ACM International Workshop on
  Performance, Portability and Productivity in HPC (P3HPC)},
  \bibinfo{year}{2018}, pp. \bibinfo{pages}{59--70}.
  \DOIprefix\doi{10.1109/P3HPC.2018.00009}.
%Type = Article
\bibitem[{Einkemmer(2020)}]{einkemmer2020}
\bibinfo{author}{L.~Einkemmer},
\newblock \bibinfo{title}{Semi-lagrangian vlasov simulation on gpus},
\newblock \bibinfo{journal}{Computer Physics Communications}
  \bibinfo{volume}{254} (\bibinfo{year}{2020}) \bibinfo{pages}{107351}.
  \DOIprefix\doi{10.1016/j.cpc.2020.107351}.
%Type = Article
\bibitem[{Williams et~al.(2009)Williams, Waterman, and
  Patterson}]{williams2009}
\bibinfo{author}{S.~Williams}, \bibinfo{author}{A.~Waterman},
  \bibinfo{author}{D.~Patterson},
\newblock \bibinfo{title}{Roofline: an insightful visual performance model for
  multicore architectures},
\newblock \bibinfo{journal}{Communications of the ACM} \bibinfo{volume}{52}
  (\bibinfo{year}{2009}) \bibinfo{pages}{65–76}.
  \DOIprefix\doi{10.1145/1498765.1498785}.
%Type = Inproceedings
\bibitem[{Micikevicius(2009)}]{micikevicius2009}
\bibinfo{author}{P.~Micikevicius},
\newblock \bibinfo{title}{3d finite difference computation on gpus using cuda},
\newblock in: \bibinfo{booktitle}{Proceedings of 2nd Workshop on General
  Purpose Processing on Graphics Processing Units}, GPGPU-2,
  \bibinfo{publisher}{Association for Computing Machinery},
  \bibinfo{address}{New York, NY, USA}, \bibinfo{year}{2009}, p.
  \bibinfo{pages}{79–84}. \DOIprefix\doi{10.1145/1513895.1513905}.
%Type = Article
\bibitem[{Sai et~al.(2022)Sai, Mellor-Crummey, Meng, Zhou, Araya-Polo, and
  Meng}]{sai2022}
\bibinfo{author}{R.~Sai}, \bibinfo{author}{J.~Mellor-Crummey},
  \bibinfo{author}{X.~Meng}, \bibinfo{author}{K.~Zhou},
  \bibinfo{author}{M.~Araya-Polo}, \bibinfo{author}{J.~Meng},
\newblock \bibinfo{title}{Accelerating high-order stencils on gpus},
\newblock \bibinfo{journal}{Concurrency and Computation: Practice and
  Experience} \bibinfo{volume}{34} (\bibinfo{year}{2022})
  \bibinfo{pages}{e6467}. \DOIprefix\doi{10.1002/cpe.6467}.
%Type = Inproceedings
\bibitem[{Wang et~al.(2010)Wang, Lin, and Yi}]{wang2010}
\bibinfo{author}{G.~Wang}, \bibinfo{author}{Y.~Lin}, \bibinfo{author}{W.~Yi},
\newblock \bibinfo{title}{Kernel fusion: An effective method for better power
  efficiency on multithreaded gpu},
\newblock in: \bibinfo{booktitle}{2010 IEEE/ACM Int'l Conference on Green
  Computing and Communications \& Int'l Conference on Cyber, Physical and
  Social Computing}, \bibinfo{year}{2010}, pp. \bibinfo{pages}{344--350}.
  \DOIprefix\doi{10.1109/GreenCom-CPSCom.2010.102}.
%Type = Article
\bibitem[{Germaschewski et~al.(2021)Germaschewski, Allen, Dannert, Hrywniak,
  Donaghy, Merlo, Ethier, D'Azevedo, Jenko, and
  Bhattacharjee}]{germaschewski2021}
\bibinfo{author}{K.~Germaschewski}, \bibinfo{author}{B.~Allen},
  \bibinfo{author}{T.~Dannert}, \bibinfo{author}{M.~Hrywniak},
  \bibinfo{author}{J.~Donaghy}, \bibinfo{author}{G.~Merlo},
  \bibinfo{author}{S.~Ethier}, \bibinfo{author}{E.~D'Azevedo},
  \bibinfo{author}{F.~Jenko}, \bibinfo{author}{A.~Bhattacharjee},
\newblock \bibinfo{title}{{Toward exascale whole-device modeling of fusion
  devices: Porting the GENE gyrokinetic microturbulence code to GPU}},
\newblock \bibinfo{journal}{Physics of Plasmas} \bibinfo{volume}{28}
  (\bibinfo{year}{2021}) \bibinfo{pages}{062501}.
  \DOIprefix\doi{10.1063/5.0046327}.
%Type = Article
\bibitem[{Sandroos et~al.(2013)Sandroos, Honkonen, {von Alfthan}, and
  Palmroth}]{sandroos2013}
\bibinfo{author}{A.~Sandroos}, \bibinfo{author}{I.~Honkonen},
  \bibinfo{author}{S.~{von Alfthan}}, \bibinfo{author}{M.~Palmroth},
\newblock \bibinfo{title}{Multi-gpu simulations of vlasov’s equation using
  vlasiator},
\newblock \bibinfo{journal}{Parallel Computing} \bibinfo{volume}{39}
  (\bibinfo{year}{2013}) \bibinfo{pages}{306--318}.
  \DOIprefix\doi{10.1016/j.parco.2013.05.001}.
%Type = Article
\bibitem[{Mehrenberger et~al.(2013)Mehrenberger, Steiner, Marradi, Crouseilles,
  Sonnendrücker, and Afeyan}]{mehrenberger2013}
\bibinfo{author}{M.~Mehrenberger}, \bibinfo{author}{C.~Steiner},
  \bibinfo{author}{L.~Marradi}, \bibinfo{author}{N.~Crouseilles},
  \bibinfo{author}{E.~Sonnendrücker}, \bibinfo{author}{B.~Afeyan},
\newblock \bibinfo{title}{Vlasov on gpu (vog project)},
\newblock \bibinfo{journal}{ESAIM: Proc.} \bibinfo{volume}{43}
  (\bibinfo{year}{2013}) \bibinfo{pages}{37--58}.
  \DOIprefix\doi{10.1051/proc/201343003}.
%Type = Article
\bibitem[{Einkemmer and Ostermann(2014)}]{einkemmer2014}
\bibinfo{author}{L.~Einkemmer}, \bibinfo{author}{A.~Ostermann},
\newblock \bibinfo{title}{Convergence analysis of a discontinuous
  galerkin/strang splitting approximation for the vlasov--poisson equations},
\newblock \bibinfo{journal}{SIAM Journal on Numerical Analysis}
  \bibinfo{volume}{52} (\bibinfo{year}{2014}) \bibinfo{pages}{757--778}.
  \DOIprefix\doi{10.1137/120898620}.
%Type = Article
\bibitem[{Einkemmer(2016)}]{einkemmer2016}
\bibinfo{author}{L.~Einkemmer},
\newblock \bibinfo{title}{High performance computing aspects of a dimension
  independent semi-lagrangian discontinuous galerkin code},
\newblock \bibinfo{journal}{Computer Physics Communications}
  \bibinfo{volume}{202} (\bibinfo{year}{2016}) \bibinfo{pages}{326--336}.
  \DOIprefix\doi{10.1016/j.cpc.2016.01.012}.
%Type = Article
\bibitem[{Einkemmer(2019)}]{einkemmer2019}
\bibinfo{author}{L.~Einkemmer},
\newblock \bibinfo{title}{A performance comparison of semi-lagrangian
  discontinuous galerkin and spline based vlasov solvers in four dimensions},
\newblock \bibinfo{journal}{Journal of Computational Physics}
  \bibinfo{volume}{376} (\bibinfo{year}{2019}) \bibinfo{pages}{937--951}.
  \DOIprefix\doi{10.1016/j.jcp.2018.10.012}.
%Type = Book
\bibitem[{LeVeque(2004)}]{leveque2004}
\bibinfo{author}{R.~J. LeVeque}, \bibinfo{title}{Finite Volume Methods for
  Hyperbolic Problems}, \bibinfo{publisher}{Cambridge Texts in Applied
  Mathematics}, \bibinfo{year}{2004}. \DOIprefix\doi{10.1017/CBO9780511791253}.
%Type = Article
\bibitem[{Warburton and Hagstrom(2008)}]{warburton2008}
\bibinfo{author}{T.~Warburton}, \bibinfo{author}{T.~Hagstrom},
\newblock \bibinfo{title}{Taming the cfl number for discontinuous galerkin
  methods on structured meshes},
\newblock \bibinfo{journal}{SIAM Journal on Numerical Analysis}
  \bibinfo{volume}{46} (\bibinfo{year}{2008}) \bibinfo{pages}{3151--3180}.
  \DOIprefix\doi{10.1137/060672601}.
%Type = Article
\bibitem[{Banks and Hittinger(2010)}]{banks2010}
\bibinfo{author}{J.~W. Banks}, \bibinfo{author}{J.~A.~F. Hittinger},
\newblock \bibinfo{title}{A new class of nonlinear finite-volume methods for
  vlasov simulation},
\newblock \bibinfo{journal}{IEEE Transactions on Plasma Science}
  \bibinfo{volume}{38} (\bibinfo{year}{2010}) \bibinfo{pages}{2198--2207}.
  \DOIprefix\doi{10.1109/TPS.2010.2056937}.
%Type = Article
\bibitem[{Banks et~al.(2011)Banks, Berger, Brunner, Cohen, and
  Hittinger}]{banks2011}
\bibinfo{author}{J.~W. Banks}, \bibinfo{author}{R.~L. Berger},
  \bibinfo{author}{S.~Brunner}, \bibinfo{author}{B.~I. Cohen},
  \bibinfo{author}{J.~A.~F. Hittinger},
\newblock \bibinfo{title}{{Two-dimensional Vlasov simulation of electron plasma
  wave trapping, wavefront bowing, self-focusing, and sideloss}},
\newblock \bibinfo{journal}{Physics of Plasmas} \bibinfo{volume}{18}
  (\bibinfo{year}{2011}) \bibinfo{pages}{052102}.
  \DOIprefix\doi{10.1063/1.3577784}.
%Type = Article
\bibitem[{Datta et~al.(2021)Datta, Crews, and Shumlak}]{datta2021}
\bibinfo{author}{I.~A.~M. Datta}, \bibinfo{author}{D.~W. Crews},
  \bibinfo{author}{U.~Shumlak},
\newblock \bibinfo{title}{{Electromagnetic extension of the
  Dory–Guest–Harris instability as a benchmark for Vlasov–Maxwell
  continuum kinetic simulations of magnetized plasmas}},
\newblock \bibinfo{journal}{Physics of Plasmas} \bibinfo{volume}{28}
  (\bibinfo{year}{2021}) \bibinfo{pages}{072112}.
  \DOIprefix\doi{10.1063/5.0057230}.
%Type = Article
\bibitem[{Datta and Shumlak(2023)}]{datta2023}
\bibinfo{author}{I.~Datta}, \bibinfo{author}{U.~Shumlak},
\newblock \bibinfo{title}{Computationally efficient high-fidelity plasma
  simulations by coupling multi-species kinetic and multi-fluid models on
  decomposed domains},
\newblock \bibinfo{journal}{Journal of Computational Physics}
  \bibinfo{volume}{483} (\bibinfo{year}{2023}) \bibinfo{pages}{112073}.
  \DOIprefix\doi{10.1016/j.jcp.2023.112073}.
%Type = Article
\bibitem[{{Henry de Frahan} et~al.(2024){Henry de Frahan}, Esclapez, Rood,
  Wimer, Mullowney, Perry, Owen, Sitaraman, Yellapantula, Hassanaly, Rahimi,
  Martin, Doronina, A., Rieth, Ge, Sankaran, Almgren, Zhang, Bell, Grout, Day,
  and Chen}]{PeleSoftware}
\bibinfo{author}{M.~T. {Henry de Frahan}}, \bibinfo{author}{L.~Esclapez},
  \bibinfo{author}{J.~Rood}, \bibinfo{author}{N.~T. Wimer},
  \bibinfo{author}{P.~Mullowney}, \bibinfo{author}{B.~A. Perry},
  \bibinfo{author}{L.~Owen}, \bibinfo{author}{H.~Sitaraman},
  \bibinfo{author}{S.~Yellapantula}, \bibinfo{author}{M.~Hassanaly},
  \bibinfo{author}{M.~J. Rahimi}, \bibinfo{author}{M.~J. Martin},
  \bibinfo{author}{O.~A. Doronina}, \bibinfo{author}{S.~N. A.},
  \bibinfo{author}{M.~Rieth}, \bibinfo{author}{W.~Ge},
  \bibinfo{author}{R.~Sankaran}, \bibinfo{author}{A.~S. Almgren},
  \bibinfo{author}{W.~Zhang}, \bibinfo{author}{J.~B. Bell},
  \bibinfo{author}{R.~Grout}, \bibinfo{author}{M.~S. Day},
  \bibinfo{author}{J.~H. Chen},
\newblock \bibinfo{title}{The pele simulation suite for reacting flows at
  exascale},
\newblock \bibinfo{journal}{Proceedings of the 2024 SIAM Conference on Parallel
  Processing for Scientific Computing}  (\bibinfo{year}{2024})
  \bibinfo{pages}{13--25}. \DOIprefix\doi{10.1137/1.9781611977967.2}.
%Type = Article
\bibitem[{Colella et~al.(2011)Colella, Dorr, Hittinger, and
  Martin}]{colella2011}
\bibinfo{author}{P.~Colella}, \bibinfo{author}{M.~Dorr},
  \bibinfo{author}{J.~Hittinger}, \bibinfo{author}{D.~Martin},
\newblock \bibinfo{title}{High-order, finite-volume methods in mapped
  coordinates},
\newblock \bibinfo{journal}{Journal of Computational Physics}
  \bibinfo{volume}{230} (\bibinfo{year}{2011}) \bibinfo{pages}{2952--2976}.
  \DOIprefix\doi{10.1016/j.jcp.2010.12.044}.
%Type = Article
\bibitem[{Vogman et~al.(2018)Vogman, Shumlak, and Colella}]{vogman2018}
\bibinfo{author}{G.~Vogman}, \bibinfo{author}{U.~Shumlak},
  \bibinfo{author}{P.~Colella},
\newblock \bibinfo{title}{Conservative fourth-order finite-volume
  vlasov–poisson solver for axisymmetric plasmas in cylindrical
  (r,$v_r$,$v_{\theta}$) phase space coordinates},
\newblock \bibinfo{journal}{Journal of Computational Physics}
  \bibinfo{volume}{373} (\bibinfo{year}{2018}) \bibinfo{pages}{877--899}.
  \DOIprefix\doi{10.1016/j.jcp.2018.07.029}.
%Type = Article
\bibitem[{Courant et~al.(1928)Courant, Friedrichs, and Lewy}]{courant1928}
\bibinfo{author}{R.~Courant}, \bibinfo{author}{K.~Friedrichs},
  \bibinfo{author}{H.~Lewy},
\newblock \bibinfo{title}{{\"U}ber die partiellen differenzengleichungen der
  mathematischen physik},
\newblock \bibinfo{journal}{Mathematische Annalen} \bibinfo{volume}{100}
  (\bibinfo{year}{1928}) \bibinfo{pages}{32--74}.
  \DOIprefix\doi{10.1007/BF01448839}.
%Type = Article
\bibitem[{Chaplin and Colella(2017)}]{chaplin2017}
\bibinfo{author}{C.~Chaplin}, \bibinfo{author}{P.~Colella},
\newblock \bibinfo{title}{{A single-stage flux-corrected transport algorithm
  for high-order finite-volume methods}},
\newblock \bibinfo{journal}{Communications in Applied Mathematics and
  Computational Science} \bibinfo{volume}{12} (\bibinfo{year}{2017})
  \bibinfo{pages}{1 -- 24}. \DOIprefix\doi{10.2140/camcos.2017.12.1}.
%Type = Article
\bibitem[{Ketcheson(2008)}]{ketcheson2008}
\bibinfo{author}{D.~I. Ketcheson},
\newblock \bibinfo{title}{Highly efficient strong stability-preserving
  runge–kutta methods with low-storage implementations},
\newblock \bibinfo{journal}{SIAM Journal on Scientific Computing}
  \bibinfo{volume}{30} (\bibinfo{year}{2008}) \bibinfo{pages}{2113--2136}.
  \DOIprefix\doi{10.1137/07070485X}.
%Type = Book
\bibitem[{Hairer et~al.(1993)Hairer, Wanner, and N{\o}rsett}]{hairer1993}
\bibinfo{author}{E.~Hairer}, \bibinfo{author}{G.~Wanner},
  \bibinfo{author}{S.~P. N{\o}rsett}, \bibinfo{title}{Solving Ordinary
  Differential Equations I}, \bibinfo{edition}{2} ed.,
  \bibinfo{publisher}{Springer Berlin, Heidelberg}, \bibinfo{year}{1993}.
  \DOIprefix\doi{10.1007/978-3-540-78862-1}.
%Type = Phdthesis
\bibitem[{Ho(2022)}]{ho2022}
\bibinfo{author}{A.~Ho}, \bibinfo{title}{Modeling plasma systems using a
  domain-hybridized physical model}, Ph.D. thesis, University of Washington,
  \bibinfo{year}{2022}. \URLprefix \url{http://hdl.handle.net/1773/48801}.
%Type = Article
\bibitem[{Kubatko et~al.(2014)Kubatko, Yeager, and Ketcheson}]{kubatko2014}
\bibinfo{author}{E.~J. Kubatko}, \bibinfo{author}{B.~A. Yeager},
  \bibinfo{author}{D.~I. Ketcheson},
\newblock \bibinfo{title}{Optimal strong-stability-preserving runge--kutta time
  discretizations for discontinuous galerkin methods},
\newblock \bibinfo{journal}{Journal of Scientific Computing}
  \bibinfo{volume}{60} (\bibinfo{year}{2014}) \bibinfo{pages}{313--344}.
  \DOIprefix\doi{10.1007/s10915-013-9796-7}.
%Type = Article
\bibitem[{Spiteri and Ruuth(2002)}]{spiteri2002}
\bibinfo{author}{R.~J. Spiteri}, \bibinfo{author}{S.~J. Ruuth},
\newblock \bibinfo{title}{A new class of optimal high-order
  strong-stability-preserving time discretization methods},
\newblock \bibinfo{journal}{SIAM Journal on Numerical Analysis}
  \bibinfo{volume}{40} (\bibinfo{year}{2002}) \bibinfo{pages}{469--491}.
  \DOIprefix\doi{10.1137/S0036142901389025}.
%Type = Article
\bibitem[{Trott et~al.(2022)Trott, Lebrun-Grandié, Arndt, Ciesko, Dang,
  Ellingwood, Gayatri, Harvey, Hollman, Ibanez, Liber, Madsen, Miles,
  Poliakoff, Powell, Rajamanickam, Simberg, Sunderland, Turcksin, and
  Wilke}]{trott2022}
\bibinfo{author}{C.~R. Trott}, \bibinfo{author}{D.~Lebrun-Grandié},
  \bibinfo{author}{D.~Arndt}, \bibinfo{author}{J.~Ciesko},
  \bibinfo{author}{V.~Dang}, \bibinfo{author}{N.~Ellingwood},
  \bibinfo{author}{R.~Gayatri}, \bibinfo{author}{E.~Harvey},
  \bibinfo{author}{D.~S. Hollman}, \bibinfo{author}{D.~Ibanez},
  \bibinfo{author}{N.~Liber}, \bibinfo{author}{J.~Madsen},
  \bibinfo{author}{J.~Miles}, \bibinfo{author}{D.~Poliakoff},
  \bibinfo{author}{A.~Powell}, \bibinfo{author}{S.~Rajamanickam},
  \bibinfo{author}{M.~Simberg}, \bibinfo{author}{D.~Sunderland},
  \bibinfo{author}{B.~Turcksin}, \bibinfo{author}{J.~Wilke},
\newblock \bibinfo{title}{Kokkos 3: Programming model extensions for the
  exascale era},
\newblock \bibinfo{journal}{IEEE Transactions on Parallel and Distributed
  Systems} \bibinfo{volume}{33} (\bibinfo{year}{2022})
  \bibinfo{pages}{805--817}. \DOIprefix\doi{10.1109/TPDS.2021.3097283}.
%Type = Inproceedings
\bibitem[{Wienke et~al.(2012)Wienke, Springer, Terboven, and
  an~Mey}]{wienke2012}
\bibinfo{author}{S.~Wienke}, \bibinfo{author}{P.~Springer},
  \bibinfo{author}{C.~Terboven}, \bibinfo{author}{D.~an~Mey},
\newblock \bibinfo{title}{Openacc --- first experiences with real-world
  applications},
\newblock in: \bibinfo{editor}{C.~Kaklamanis},
  \bibinfo{editor}{T.~Papatheodorou}, \bibinfo{editor}{P.~G. Spirakis} (Eds.),
  \bibinfo{booktitle}{Euro-Par 2012 Parallel Processing},
  \bibinfo{publisher}{Springer Berlin Heidelberg}, \bibinfo{address}{Berlin,
  Heidelberg}, \bibinfo{year}{2012}, pp. \bibinfo{pages}{859--870}.
%Type = Inproceedings
\bibitem[{Beckingsale et~al.(2019)Beckingsale, Burmark, Hornung, Jones,
  Killian, Kunen, Pearce, Robinson, Ryujin, and Scogland}]{beckingsale2019}
\bibinfo{author}{D.~A. Beckingsale}, \bibinfo{author}{J.~Burmark},
  \bibinfo{author}{R.~Hornung}, \bibinfo{author}{H.~Jones},
  \bibinfo{author}{W.~Killian}, \bibinfo{author}{A.~J. Kunen},
  \bibinfo{author}{O.~Pearce}, \bibinfo{author}{P.~Robinson},
  \bibinfo{author}{B.~S. Ryujin}, \bibinfo{author}{T.~R. Scogland},
\newblock \bibinfo{title}{Raja: Portable performance for large-scale scientific
  applications},
\newblock in: \bibinfo{booktitle}{2019 IEEE/ACM International Workshop on
  Performance, Portability and Productivity in HPC (P3HPC)},
  \bibinfo{year}{2019}, pp. \bibinfo{pages}{71--81}.
  \DOIprefix\doi{10.1109/P3HPC49587.2019.00012}.
%Type = Phdthesis
\bibitem[{Reddell(2016)}]{reddell2016}
\bibinfo{author}{N.~Reddell}, \bibinfo{title}{A Kinetic Vlasov Model for Plasma
  Simulation Using Discontinuous Galerkin Method on Many-Core Architectures},
  Ph.D. thesis, University of Washington, \bibinfo{address}{Seattle, WA},
  \bibinfo{year}{2016}. \URLprefix \url{http://hdl.handle.net/1773/36473}.
%Type = Article
\bibitem[{Student(1908)}]{student1908}
\bibinfo{author}{Student},
\newblock \bibinfo{title}{The probable error of a mean},
\newblock \bibinfo{journal}{Biometrika} \bibinfo{volume}{6}
  (\bibinfo{year}{1908}) \bibinfo{pages}{1--25}.
%Type = Inproceedings
\bibitem[{Sahasrabudhe and Berzins(2020)}]{sahasrabudhe2020}
\bibinfo{author}{D.~Sahasrabudhe}, \bibinfo{author}{M.~Berzins},
\newblock \bibinfo{title}{Improving performance of the hypre iterative solver
  for uintah combustion codes on manycore architectures using mpi endpoints and
  kernel consolidation},
\newblock in: \bibinfo{editor}{V.~V. Krzhizhanovskaya},
  \bibinfo{editor}{G.~Z{\'a}vodszky}, \bibinfo{editor}{M.~H. Lees},
  \bibinfo{editor}{J.~J. Dongarra}, \bibinfo{editor}{P.~M.~A. Sloot},
  \bibinfo{editor}{S.~Brissos}, \bibinfo{editor}{J.~Teixeira} (Eds.),
  \bibinfo{booktitle}{Computational Science -- ICCS 2020},
  \bibinfo{publisher}{Springer International Publishing},
  \bibinfo{address}{Cham}, \bibinfo{year}{2020}, pp. \bibinfo{pages}{175--190}.
%Type = Techreport
\bibitem[{Balay et~al.(2024)Balay, Abhyankar, Adams, Benson, Brown, Brune,
  Buschelman, Constantinescu, Dalcin, Dener, Eijkhout, Faibussowitsch, Gropp,
  Hapla, Isaac, Jolivet, Karpeev, Kaushik, Knepley, Kong, Kruger, May, McInnes,
  Mills, Mitchell, Munson, Roman, Rupp, Sanan, Sarich, Smith, Zampini, Zhang,
  Zhang, and Zhang}]{petsc-user-ref}
\bibinfo{author}{S.~Balay}, \bibinfo{author}{S.~Abhyankar},
  \bibinfo{author}{M.~F. Adams}, \bibinfo{author}{S.~Benson},
  \bibinfo{author}{J.~Brown}, \bibinfo{author}{P.~Brune},
  \bibinfo{author}{K.~Buschelman}, \bibinfo{author}{E.~Constantinescu},
  \bibinfo{author}{L.~Dalcin}, \bibinfo{author}{A.~Dener},
  \bibinfo{author}{V.~Eijkhout}, \bibinfo{author}{J.~Faibussowitsch},
  \bibinfo{author}{W.~D. Gropp}, \bibinfo{author}{V.~Hapla},
  \bibinfo{author}{T.~Isaac}, \bibinfo{author}{P.~Jolivet},
  \bibinfo{author}{D.~Karpeev}, \bibinfo{author}{D.~Kaushik},
  \bibinfo{author}{M.~G. Knepley}, \bibinfo{author}{F.~Kong},
  \bibinfo{author}{S.~Kruger}, \bibinfo{author}{D.~A. May},
  \bibinfo{author}{L.~C. McInnes}, \bibinfo{author}{R.~T. Mills},
  \bibinfo{author}{L.~Mitchell}, \bibinfo{author}{T.~Munson},
  \bibinfo{author}{J.~E. Roman}, \bibinfo{author}{K.~Rupp},
  \bibinfo{author}{P.~Sanan}, \bibinfo{author}{J.~Sarich},
  \bibinfo{author}{B.~F. Smith}, \bibinfo{author}{S.~Zampini},
  \bibinfo{author}{H.~Zhang}, \bibinfo{author}{H.~Zhang},
  \bibinfo{author}{J.~Zhang}, \bibinfo{title}{{PETSc/TAO} Users Manual},
  \bibinfo{type}{Technical Report} \bibinfo{number}{ANL-21/39 - Revision 3.21},
  Argonne National Laboratory, \bibinfo{year}{2024}.
  \DOIprefix\doi{10.2172/2205494}.
%Type = Inproceedings
\bibitem[{Falgout et~al.(2006)Falgout, Jones, and Yang}]{falgout2006}
\bibinfo{author}{R.~D. Falgout}, \bibinfo{author}{J.~E. Jones},
  \bibinfo{author}{U.~M. Yang},
\newblock \bibinfo{title}{The design and implementation of hypre, a library of
  parallel high performance preconditioners},
\newblock in: \bibinfo{editor}{A.~M. Bruaset}, \bibinfo{editor}{A.~Tveito}
  (Eds.), \bibinfo{booktitle}{Numerical Solution of Partial Differential
  Equations on Parallel Computers}, \bibinfo{publisher}{Springer Berlin
  Heidelberg}, \bibinfo{address}{Berlin, Heidelberg}, \bibinfo{year}{2006}, pp.
  \bibinfo{pages}{267--294}.
%Type = Article
\bibitem[{Feng and Zhao(2020)}]{feng2020}
\bibinfo{author}{H.~Feng}, \bibinfo{author}{S.~Zhao},
\newblock \bibinfo{title}{Fft-based high order central difference schemes for
  three-dimensional poisson's equation with various types of boundary
  conditions},
\newblock \bibinfo{journal}{Journal of Computational Physics}
  \bibinfo{volume}{410} (\bibinfo{year}{2020}) \bibinfo{pages}{109391}.
  \DOIprefix\doi{10.1016/j.jcp.2020.109391}.
%Type = Article
\bibitem[{Kaasschieter(1988)}]{kaasschieter1988}
\bibinfo{author}{E.~Kaasschieter},
\newblock \bibinfo{title}{Preconditioned conjugate gradients for solving
  singular systems},
\newblock \bibinfo{journal}{Journal of Computational and Applied Mathematics}
  \bibinfo{volume}{24} (\bibinfo{year}{1988}) \bibinfo{pages}{265--275}.
  \DOIprefix\doi{10.1016/0377-0427(88)90358-5}.
%Type = Inproceedings
\bibitem[{Zdenek and Rudolf(2006)}]{zdenek2006}
\bibinfo{author}{B.~Zdenek}, \bibinfo{author}{H.~Rudolf},
\newblock \bibinfo{title}{A comparison of advanced poisson equation solvers
  applied to the particle-in-cell plasma model},
\newblock in: \bibinfo{booktitle}{WDS'06 Proceedings of Contributed Papers,
  Part III- Physics}, \bibinfo{year}{2006}, pp. \bibinfo{pages}{187--192}.
%Type = Misc
\bibitem[{Berger-Vergiat et~al.(2021)Berger-Vergiat, Kelley, Rajamanickam, Hu,
  Swirydowicz, Mullowney, Thomas, and Yamazaki}]{berger_vergiat2021}
\bibinfo{author}{L.~Berger-Vergiat}, \bibinfo{author}{B.~Kelley},
  \bibinfo{author}{S.~Rajamanickam}, \bibinfo{author}{J.~Hu},
  \bibinfo{author}{K.~Swirydowicz}, \bibinfo{author}{P.~Mullowney},
  \bibinfo{author}{S.~Thomas}, \bibinfo{author}{I.~Yamazaki},
  \bibinfo{title}{Two-stage gauss--seidel preconditioners and smoothers for
  krylov solvers on a gpu cluster}, \bibinfo{year}{2021}.
  \href{http://arxiv.org/abs/2104.01196}{\tt arXiv:2104.01196}.
%Type = Article
\bibitem[{Xu et~al.(2019)Xu, Fu, Luk, Gan, Shi, Xue, Yang, Jiang, He, and
  Yang}]{xu2019}
\bibinfo{author}{J.~Xu}, \bibinfo{author}{H.~Fu}, \bibinfo{author}{W.~Luk},
  \bibinfo{author}{L.~Gan}, \bibinfo{author}{W.~Shi}, \bibinfo{author}{W.~Xue},
  \bibinfo{author}{C.~Yang}, \bibinfo{author}{Y.~Jiang},
  \bibinfo{author}{C.~He}, \bibinfo{author}{G.~Yang},
\newblock \bibinfo{title}{Optimizing finite volume method solvers on nvidia
  gpus},
\newblock \bibinfo{journal}{IEEE Transactions on Parallel and Distributed
  Systems} \bibinfo{volume}{30} (\bibinfo{year}{2019})
  \bibinfo{pages}{2790--2805}. \DOIprefix\doi{10.1109/TPDS.2019.2926084}.
%Type = Phdthesis
\bibitem[{Vogman(2016)}]{vogman2016}
\bibinfo{author}{G.~V. Vogman}, \bibinfo{title}{Fourth-order conservative
  Vlasov-Maxwell solver for Cartesian and cylindrical phase space coordinates},
  Ph.D. thesis, University of California, Berkeley, \bibinfo{year}{2016}.
  \URLprefix \url{https://escholarship.org/uc/item/1c49t97t}.
%Type = Inproceedings
\bibitem[{Doerfler and Brightwell(2006)}]{doerfler2006}
\bibinfo{author}{D.~Doerfler}, \bibinfo{author}{R.~Brightwell},
\newblock \bibinfo{title}{Measuring mpi send and receive overhead and
  application availability in high performance network interfaces},
\newblock in: \bibinfo{editor}{B.~Mohr}, \bibinfo{editor}{J.~L. Tr{\"a}ff},
  \bibinfo{editor}{J.~Worringen}, \bibinfo{editor}{J.~Dongarra} (Eds.),
  \bibinfo{booktitle}{Recent Advances in Parallel Virtual Machine and Message
  Passing Interface}, \bibinfo{publisher}{Springer Berlin Heidelberg},
  \bibinfo{address}{Berlin, Heidelberg}, \bibinfo{year}{2006}, pp.
  \bibinfo{pages}{331--338}. \DOIprefix\doi{10.1007/11846802_46}.
%Type = Book
\bibitem[{LeVeque(2007)}]{leveque2007}
\bibinfo{author}{R.~J. LeVeque}, \bibinfo{title}{Finite Difference Methods for
  Ordinary and Partial Differential Equations}, \bibinfo{publisher}{SIAM},
  \bibinfo{year}{2007}. \DOIprefix\doi{10.1137/1.9780898717839}.
%Type = Article
\bibitem[{Vogman et~al.(2019)Vogman, Hammer, and Farmer}]{vogman2019}
\bibinfo{author}{G.~V. Vogman}, \bibinfo{author}{J.~H. Hammer},
  \bibinfo{author}{W.~A. Farmer},
\newblock \bibinfo{title}{Customizable two-species kinetic equilibria for
  nonuniform low-beta plasmas},
\newblock \bibinfo{journal}{Physics of Plasmas} \bibinfo{volume}{26}
  (\bibinfo{year}{2019}) \bibinfo{pages}{042119}.
  \DOIprefix\doi{10.1063/1.5089465}.
%Type = Incollection
\bibitem[{Landau(1965)}]{landau1965}
\bibinfo{author}{L.~D. Landau},
\newblock \bibinfo{title}{61 - on the vibrations of the electronic plasma},
\newblock in: \bibinfo{editor}{D.~{TER HAAR}} (Ed.),
  \bibinfo{booktitle}{Collected Papers of L.D. Landau},
  \bibinfo{publisher}{Pergamon}, \bibinfo{year}{1965}, pp.
  \bibinfo{pages}{445--460}.
  \DOIprefix\doi{10.1016/B978-0-08-010586-4.50066-3}.
%Type = Article
\bibitem[{{Van Kampen}(1955)}]{kampen1955}
\bibinfo{author}{N.~{Van Kampen}},
\newblock \bibinfo{title}{On the theory of stationary waves in plasmas},
\newblock \bibinfo{journal}{Physica} \bibinfo{volume}{21}
  (\bibinfo{year}{1955}) \bibinfo{pages}{949--963}.
  \DOIprefix\doi{10.1016/S0031-8914(55)93068-8}.
%Type = Article
\bibitem[{Filbert and Kellogg(1979)}]{filbet1979}
\bibinfo{author}{P.~C. Filbert}, \bibinfo{author}{P.~J. Kellogg},
\newblock \bibinfo{title}{Electrostatic noise at the plasma frequency beyond
  the earth's bow shock},
\newblock \bibinfo{journal}{Journal of Geophysical Research: Space Physics}
  \bibinfo{volume}{84} (\bibinfo{year}{1979}) \bibinfo{pages}{1369--1381}.
  \DOIprefix\doi{10.1029/JA084iA04p01369}.
%Type = Phdthesis
\bibitem[{Crews(2022)}]{crews2022}
\bibinfo{author}{D.~W. Crews}, \bibinfo{title}{Numerical simulation of
  collisionless kinetic plasma turbulence}, Ph.D. thesis, University of
  Washington, \bibinfo{year}{2022}. \URLprefix
  \url{http://hdl.handle.net/1773/48798}.
%Type = Article
\bibitem[{Umeda(2008)}]{umeda2008}
\bibinfo{author}{T.~Umeda},
\newblock \bibinfo{title}{A conservative and non-oscillatory scheme for vlasov
  code simulations},
\newblock \bibinfo{journal}{Earth, Planets and Space} \bibinfo{volume}{60}
  (\bibinfo{year}{2008}) \bibinfo{pages}{773--779}.
  \DOIprefix\doi{10.1186/BF03352826}.
%Type = Article
\bibitem[{Dory et~al.(1965)Dory, Guest, and Harris}]{dory1965}
\bibinfo{author}{R.~A. Dory}, \bibinfo{author}{G.~E. Guest},
  \bibinfo{author}{E.~G. Harris},
\newblock \bibinfo{title}{Unstable electrostatic plasma waves propagating
  perpendicular to a magnetic field},
\newblock \bibinfo{journal}{Phys. Rev. Lett.} \bibinfo{volume}{14}
  (\bibinfo{year}{1965}) \bibinfo{pages}{131--133}.
  \DOIprefix\doi{10.1103/PhysRevLett.14.131}.
%Type = Article
\bibitem[{Cheng and Knorr(1976)}]{cheng1976}
\bibinfo{author}{C.~Cheng}, \bibinfo{author}{G.~Knorr},
\newblock \bibinfo{title}{The integration of the vlasov equation in
  configuration space},
\newblock \bibinfo{journal}{Journal of Computational Physics}
  \bibinfo{volume}{22} (\bibinfo{year}{1976}) \bibinfo{pages}{330--351}.
  \DOIprefix\doi{10.1016/0021-9991(76)90053-X}.
%Type = Article
\bibitem[{Filbet and Sonnendrücker(2003)}]{filbet2003}
\bibinfo{author}{F.~Filbet}, \bibinfo{author}{E.~Sonnendrücker},
\newblock \bibinfo{title}{Comparison of eulerian vlasov solvers},
\newblock \bibinfo{journal}{Computer Physics Communications}
  \bibinfo{volume}{150} (\bibinfo{year}{2003}) \bibinfo{pages}{247--266}.
  \DOIprefix\doi{10.1016/S0010-4655(02)00694-X}.
%Type = Article
\bibitem[{Manfredi(1997)}]{manfredi1997}
\bibinfo{author}{G.~Manfredi},
\newblock \bibinfo{title}{Long-time behavior of nonlinear landau damping},
\newblock \bibinfo{journal}{Phys. Rev. Lett.} \bibinfo{volume}{79}
  (\bibinfo{year}{1997}) \bibinfo{pages}{2815--2818}.
  \DOIprefix\doi{10.1103/PhysRevLett.79.2815}.
%Type = Article
\bibitem[{Welford(1962)}]{welford1962}
\bibinfo{author}{B.~P. Welford},
\newblock \bibinfo{title}{Note on a method for calculating corrected sums of
  squares and products},
\newblock \bibinfo{journal}{Technometrics} \bibinfo{volume}{4}
  (\bibinfo{year}{1962}) \bibinfo{pages}{419--420}.
  \DOIprefix\doi{10.1080/00401706.1962.10490022}.
%Type = Article
\bibitem[{Jain and Chlamtac(1985)}]{jain1985}
\bibinfo{author}{R.~Jain}, \bibinfo{author}{I.~Chlamtac},
\newblock \bibinfo{title}{The p2 algorithm for dynamic calculation of quantiles
  and histograms without storing observations},
\newblock \bibinfo{journal}{Commun. ACM} \bibinfo{volume}{28}
  (\bibinfo{year}{1985}) \bibinfo{pages}{1076–1085}.
  \DOIprefix\doi{10.1145/4372.4378}.
%Type = Article
\bibitem[{Dahm et~al.(2020)Dahm, Richards, Black, Bertsch, Grinberg, Karlin,
  Kokkila-Schumacher, León, Neely, Pankajakshan, and Pearce}]{dahm2020}
\bibinfo{author}{J.~P. Dahm}, \bibinfo{author}{D.~F. Richards},
  \bibinfo{author}{A.~Black}, \bibinfo{author}{A.~D. Bertsch},
  \bibinfo{author}{L.~Grinberg}, \bibinfo{author}{I.~Karlin},
  \bibinfo{author}{S.~Kokkila-Schumacher}, \bibinfo{author}{E.~A. León},
  \bibinfo{author}{J.~R. Neely}, \bibinfo{author}{R.~Pankajakshan},
  \bibinfo{author}{O.~Pearce},
\newblock \bibinfo{title}{Sierra center of excellence: Lessons learned},
\newblock \bibinfo{journal}{IBM Journal of Research and Development}
  \bibinfo{volume}{64} (\bibinfo{year}{2020}) \bibinfo{pages}{2:1--2:14}.
  \DOIprefix\doi{10.1147/JRD.2019.2961069}.
%Type = Book
\bibitem[{Yates(1952)}]{yates1952}
\bibinfo{author}{R.~C. Yates}, \bibinfo{title}{Curves and Their Properties},
  \bibinfo{publisher}{National Council of Teachers of Mathematics Inc.},
  \bibinfo{year}{1952}.

\end{thebibliography}

%\section*{Supplementary Material}
%
%Supplementary material that may be helpful in the review process should
%be prepared and provided as a separate electronic file. That file can
%then be transformed into PDF format and submitted along with the
%manuscript and graphic files to the appropriate editorial office.

\end{document}